\documentclass[bibyear]{aa}
\usepackage{graphicx}
\usepackage{txfonts}
\usepackage{placeins}
\usepackage{here}
\usepackage{ulem}
\usepackage{color}

\begin{document}

\title{
  High-angular-resolution ALMA imaging of the inhomogeneous dynamical 
  atmosphere of the asymptotic giant branch star W~Hya
}
\subtitle{
SiO, \mbox{H$_2$O}, \mbox{SO$_{2}$}, SO, HCN, AlO, AlOH, TiO, \mbox{TiO$_{2}$}, and OH lines
}

\author{K.~Ohnaka\inst{1} 
\and
K.~T.~Wong\inst{2}
\and
G.~Weigelt\inst{3} 
\and
K.-H.~Hofmann\inst{3} 
}

\offprints{K.~Ohnaka}

\institute{
  Instituto de Astrof\'isica, Departamento de F\'isica y Astronom\'ia,
  Facultad de Ciencias Exactas, 
  Universidad Andr\'es Bello,
Fern\'andez Concha 700, Las Condes, Santiago, Chile\\
\email{k1.ohnaka@gmail.com}
\and
Department of Physics and Astronomy, Uppsala University,
Box 516, 751 20 Uppsala, Sweden
\and
Max-Planck-Institut f\"{u}r Radioastronomie, 
Auf dem H\"{u}gel 69, 53121 Bonn, Germany
}

\date{Received / Accepted }

\abstract
{}
{
We present high-angular-resolution imaging of the 
asymptotic giant branch 
star W~Hya with the Atacama Large Millimeter/submillimeter Array (ALMA) 
to probe the dynamics and chemistry in the atmosphere and inner wind.
}
{
  \mbox{W~Hya}\ was observed with the longest baselines of ALMA at 250--268~GHz with
  an angular resolution of $\sim$17$\times$20~mas.
}
{
  ALMA's high angular resolution allowed us to resolve the stellar disk 
  of \mbox{W~Hya}\, along with clumpy, irregularly shaped emission extending to 
  $\sim$100~mas. This emission includes a plume in the north-northwest, 
  a tail in the south-southwest, and the extended atmosphere
  elongated in the east-northeast--west-southwest 
  direction, with semimajor and semiminor axes of
  $\sim$70 and 40~mas ($\sim$3.4 and 1.9~\mbox{$R_{\star}$}), respectively. 
  We identified 57 lines, which include 
  SiO, \mbox{H$_2$O}, \mbox{SO$_{2}$}, SO, HCN, AlO, AlOH, TiO, \mbox{TiO$_{2}$}, OH, and 
  some of their isotopologues, 
  with about half of them being in vibrationally excited states. 
  The molecular line images show 
  spatially inhomogeneous molecular formation.  
  Our ALMA data taken at phase 0.53 (minimum light) indicate 
  global, accelerating infall within $\sim$75~mas (3.6~\mbox{$R_{\star}$}) 
  but also outflow at up to $\sim$10~\mbox{km s$^{-1}$}\ in deeper layers. 
While 38 of the detected lines appear in absorption against the
continuum stellar disk as expected, 
we detect nonthermal emission on top of the continuum over the stellar
disk in 19 lines, including SiO, \mbox{H$_2$O}, \mbox{SO$_{2}$}, and AlO. 
The emission of the SiO, AlO, TiO, \mbox{TiO$_{2}$}, SO, and \mbox{SO$_{2}$}\ lines coincides well
with the clumpy dust cloud distribution obtained from contemporaneous visible
polarimetric imaging in addition to \mbox{H$_2$O}\ reported in our previous work. 
This lends support to the idea that SiO, \mbox{H$_2$O}, and AlO are directly
involved in grain nucleation. The overlap of SO/\mbox{SO$_{2}$}\ (possibly also
TiO/\mbox{TiO$_{2}$}) with the dust clouds suggests
the formation of these molecules and dust behind shocks induced by
pulsation and/or convection. 
We detect HCN emission close to the star, 
down to $\sim$30~mas ($\sim$1.4~\mbox{$R_{\star}$}), which is consistent with
shock-induced chemistry. 
}
{}


\titlerunning{ALMA imaging of the complex dynamical atmosphere of the
AGB star W~Hya}
\authorrunning{Ohnaka et al.}
\maketitle

\section{Introduction}
\label{sect_intro}

Low- and intermediate-mass stars experience significant mass loss at the 
asymptotic giant branch (AGB), which plays an important role not only 
in stellar evolution but also in the chemical evolution of galaxies, 
because nuclear-processed material is returned to the interstellar space 
via mass loss. 
It is often postulated that large-amplitude pulsation levitates the material, 
which leads to density enhancement in the cool, upper atmosphere, where dust can form. 
The radiation pressure on the dust grains can then drive the mass loss 
(H\"ofner \& Olofsson \cite{hoefner18}).
Furthermore, the recent 3D models of the dynamical atmosphere of AGB stars show that 
dust formation can occur in low-temperature regions caused by convective 
inhomogeneities in density and temperature (Freytag \& H\"ofner 
\cite{freytag23}).

To clarify the long-standing problem of the mass loss from AGB stars, 
it is indispensable to probe the region within a few \mbox{$R_{\star}$}, 
where the dust forms and the wind accelerates. 
The advance in high-angular observation techniques has made it possible to 
spatially resolve this key region. 
Infrared long-baseline interferometric imaging has revealed inhomogeneous 
structures over the stellar disk and in the atmosphere of AGB stars with 
milliarcsecond angular resolution (e.g., Wittkowski et al. \cite{wittkowski17}; 
Paladini et al. \cite{paladini18}; Ohnaka et al. \cite{ohnaka19}; 
Drevon et al. \cite{drevon22}; Planquart et al. \cite{planquart24}). 
For example, the imaging of the AGB star R~Dor in the 2.3~\mbox{$\mu$m}\ CO lines 
with the AMBER instrument at the Very Large Telescope Interferometer (VLTI) 
shows the irregularly shaped, clumpy atmosphere extending to $\sim$2~\mbox{$R_{\star}$} 
(Ohnaka et al. \cite{ohnaka19}). They also obtained 2D velocity-field maps 
over the surface and atmosphere at different atmospheric heights and revealed 
strong outward acceleration between $\sim$1.5 and 1.8~\mbox{$R_{\star}$}. 
These inhomogeneities in the atmosphere may be the seed of the clumpy cloud 
formation, which has been detected in some AGB stars 
(Ireland et al. \cite{ireland05}; Norris et al. \cite{norris12}; 
Khouri et al. \cite{khouri16}, \cite{khouri18}; 
Ohnaka et al. \cite{ohnaka16}, \cite{ohnaka17};
Adam \& Ohnaka \cite{adam19}).

The Atacama Large Millimeter/submillimeter Array (ALMA) also provides us with 
the spatial resolution needed to resolve the atmosphere and the innermost 
circumstellar environment of nearby cool evolved stars. 
Takigawa et al. (\cite{takigawa17}) imaged the atmosphere of the AGB star 
\mbox{W~Hya}\ in the AlO line at 344~GHz with ALMA. 
Vlemmings et al. (\cite{vlemmings17}) obtained a continuum image of \mbox{W~Hya}\ 
from the same data, which shows a hot spot over the stellar disk. 
The ATOMIUM Large Program observed 17 oxygen-rich AGB stars and red 
supergiants (RSGs) between $\sim$214 and 270~GHz at high angular 
resolutions to study the gas dynamics and chemical properties in the stellar winds 
(Decin et al. \cite{decin20}; Gottlieb et al. \cite{gottlieb22}; 
Wallstr\"om et al. \cite{wallstroem24}). 
Khouri et al. (\cite{khouri24}) presented a detailed analysis of the 
kinematical structure of the extended atmosphere and the innermost circumstellar 
envelope of R~Dor, taking advantage of the ALMA data in the CO lines that 
spatially resolved the stellar disk. 
More recently, Vlemmings et al. (\cite{vlemmings24}) imaged convective cells 
over the surface of R~Dor with ALMA and their time variations within a month. 
Velilla-Prieto et al. (\cite{velilla-prieto23}) showed the formation of different 
molecular species in spatially distinct regions of the clumpy atmosphere of the 
dusty carbon star IRC+10216. 
Asaki et al. (\cite{asaki23}) imaged the compact HCN maser emission 
from the carbon star R~Lep with an angular resolution of down to 5~mas.

In this paper, we present ALMA observations of the AGB star \mbox{W~Hya}\ at spatial resolutions of 17--20~mas at 250--268~GHz. 
Ohnaka et al. (\cite{ohnaka24}; hereafter Paper~I) reported the results on the 268~GHz continuum and two vibrationally excited \mbox{H$_2$O}\ lines from the same data. Here, we present the complete results on all the detected lines. 
In Sect.~\ref{sect_basic}, we provide an overview of the basic properties of
\mbox{W~Hya}\ derived in the literature. Our ALMA observations and data
reduction are described in Sect.~\ref{sect_obs}.
We describe the observational results and interpretation of the data of the
detected spectral lines in Sect.~\ref{sect_res}.
A discussion on the molecule-dust chemistry and dynamics in the atmosphere and the inner wind is presented in Sect.~\ref{sect_discuss}, followed by concluding remarks in Sect.~\ref{sect_concl}.

\section{Basic properties of W~Hya}
\label{sect_basic}

The oxygen-rich AGB star W Hya is one of the AGB stars closest to the Sun,
and therefore, it has been extensively studied from the visible to 
the infrared to the radio (e.g., Zhao-Geisler et al. \cite{zhao-geisler11}; 
Khouri et al. \cite{khouri15} and references therein). 
While it is classified as a semi-regular variable of type a (SRa), 
it shows
a clear periodicity with 389 days (Uttenthaler et al. \cite{uttenthaler11}, 
see also discussion in Nowotny et al. \cite{nowotny10} for the classification
of \mbox{W~Hya}). 

The distances of \mbox{W~Hya}\ measured with different methods show some
discrepancies. 
Vlemmings et al. (\cite{vlemmings03}) measured a distance of 
$98^{+30}_{-18}$~pc based on the OH maser parallax. 
They interpret the errors as due to the variations in the stellar
atmosphere, for example, caused by stellar pulsation. 
Knapp et al. (\cite{knapp03}) obtained a smaller distance of
$78^{+6.5}_{-5.6}$~pc from the
reprocessed Hipparcos parallax. They also derived a period-luminosity
relation based on the reprocessed Hipparcos parallax for a sample of
semiregular variables (\mbox{W~Hya}\ is one of them). 
Recently Andriantsaralaza et al. (\cite{andriantsaralaza22}) derived a
distance of $87^{+11}_{-9}$~pc based on this period-luminosity relation
for semi-regular variables\footnote{The reason for the difference in the
results between Knapp et al. (\cite{knapp03}) and Andriantsaralaza et al.
(\cite{andriantsaralaza22}) is that the period-luminosity
relation for semiregular variables is not tight, showing large scatter. 
This means that the distance derived from the fitted period-luminosity
relation has significant uncertainty, and therefore, the result of 
Andriantsaralaza et al. (\cite{andriantsaralaza22}) deviates from the
value directly derived from the Hipparcos parallax of \mbox{W~Hya}\ itself 
(Knapp et al. \cite{knapp03}).}.
On the other hand, van Leeuwen (\cite{vanleeuwen07}) obtained a parallax
of $9.59 \pm 1.12$~mas based on another reprocessing of the Hipparcos
data, which translates into a distance of $104^{+14}_{-11}$~pc, in agreement
with the result of Vlemmings et al. (\cite{vlemmings03}).
In the present work, we adopted the 98~pc from
Vlemmings et al. (\cite{vlemmings03}) because their method with the 
high-angular-resolution measurements allowed them to separate the parallax
and residual motions due to the variations in the stellar atmosphere. 

The wind terminal velocity has been determined from various 
observations primarily in the far-infrared and radio. 
For example, Khouri et al. (\cite{khouri14b}) derived 7.5~\mbox{km s$^{-1}$}\ by
the model fitting to the CO line profiles, while
Hoai et al. (\cite{hoai22}) obtained, based on the analysis of ALMA data,
a lower value of 5~\mbox{km s$^{-1}$}, which is outside the parameter range of the
model fitting of Khouri et al. (\cite{khouri14b}). 
Hoai et al. (\cite{hoai22}) also pointed out that the CO masers
reported by Vlemmings et al. (\cite{vlemmings21}) appear at a 
blueshifted velocity of $\sim$5.5~\mbox{km s$^{-1}$}. 
In the present work, we adopted the wind terminal velocity of 5~\mbox{km s$^{-1}$}. 
As for the systemic velocity, Khouri et al. (\cite{khouri14b}) derived
40.4~\mbox{km s$^{-1}$}\ in the local standard of rest (LSR), which is in agreement
with the previous studies (see references therein).
However, they note that the high-$J$ $^{13}$CO lines are better fit
with a systemic velocity of 39.6~\mbox{km s$^{-1}$}. 
Vlemmings et al. (\cite{vlemmings17}) also derived a similar, lower value
of 39.2~\mbox{km s$^{-1}$}.
The systemic velocity of 40.4~\mbox{km s$^{-1}$}\ was adopted in our analysis, because
it agrees with the values derived from various observations.

We adopted 41.4~mas as the star's angular diameter
(i.e., 20.7~mas as the star's angular radius \mbox{$R_{\star}$}), as described in
Paper~I. 
Combined with the distance of 98~pc, this translates into a linear radius
of $3\times10^{13}$~cm (430~\mbox{$R_{\sun}$}).

As described in Sect.~\ref{sect_obs}, our ALMA observations took place at phase 0.53. Ohnaka et al. (\cite{ohnaka17}) derived the bolometric flux of $1.69 \times 10^{-8}$~W~m$^{-2}$ at approximately the same phase of 0.54, which corresponds to a bolometric luminosity of 5070~\mbox{$L_{\sun}$}\ at the adopted distance of 98~pc. Combining the bolometric flux and the angular diameter of 41.4~mas results in an effective temperature of 2330~K. 

\begin{table*}
\caption {
  Spectral windows (spws) of our ALMA observations of \mbox{W~Hya}.
}
\begin{center}
\begin{tabular}{r c c c c c c}\hline
\noalign{\smallskip}
spw & Central frequency & Bandwidth & Velocity resolution & Angular resolution
& RMS & Continuum flux density\\ 
   &     (GHz)         & (MHz)     &  (\mbox{km s$^{-1}$})             &   (mas)           &
mJy/beam & (mJy) \\
\hline
\noalign{\smallskip}
1 & 250.727 & 468.75 & 1.17 & $19.7\times16.5$ & 0.57 & $267.7 \pm 1.3$ \\
2 & 251.826 & 468.75 & 1.16 & $19.3\times16.7$ & 0.52 & $270.0 \pm 1.4$ \\
3 & 252.471 & 468.75 & 1.16 & $19.3\times16.7$ & 0.47 & $271.8 \pm 1.0$ \\
4 & 253.703 & 468.75 & 1.15 & $19.1\times16.5$ & 0.52 & $273.7 \pm 1.4$ \\
5 & 265.886 & 468.75 & 1.10 & $18.4\times15.8$ & 0.59 & $301.5 \pm 1.6$ \\
6 & 266.943 & 468.75 & 1.10 & $18.4\times15.8$ & 0.55 & $305.6 \pm 1.5$ \\
7 & 267.198 & 468.75 & 1.10 & $18.3\times15.8$ & 0.60 & $306.0 \pm 1.5$ \\
8 & 267.720 & 937.50 & 1.09 & $18.2\times15.8$ & 0.63 & $306.2 \pm 1.3$ \\
9 & 268.168 & 937.50 & 1.09 & $18.1\times15.7$ & 0.65 & $307.8 \pm 1.3$ \\
\hline
\end{tabular}
\tablefoot{
  The velocity resolution corresponds to twice the channel width, with the Hanning smoothing at the correlator taken into account. The RMS noise is given for each spectral channel.
}
\end{center}
\label{spwlist}
\end{table*}

\newcounter{linecnt}
\stepcounter{linecnt}

\begin{table*}
\caption {
  Molecular lines identified in our ALMA observations of \mbox{W~Hya}\ in Band 6.
}
\begin{center}
\begin{tabular}{r l c c r r}\hline
\noalign{\smallskip}
\# & Species & Transition & Rest frequency (GHz) & $E_{u}/k$ (K)
& $\log A_{ul}$ (s$^{-1}$) \\ 
\hline
\noalign{\smallskip}
\arabic{linecnt} & \mbox{$^{29}$SiO} & $\varv=3$, $J$ = 6 -- 5 & 251.930123 & 5265.98 &
$-3.067$ \stepcounter{linecnt} \\ 
\arabic{linecnt} &\mbox{$^{30}$SiO}\ * & $\varv=2$, $J$ = 6 -- 5 & 250.727751 & 3521.00 & $-3.079$ \stepcounter{linecnt} \\ 
\arabic{linecnt} &\mbox{$^{29}$SiO}\ * & $\varv=2$, $J$ = 6 -- 5 & 253.703469     & 3541.92 & $-3.064$ \stepcounter{linecnt} \\ 
\arabic{linecnt} &\mbox{$^{30}$SiO} & $\varv=1$, $J$ = 6 -- 5 & 252.471372     & 1790.18 & $-3.076$ \stepcounter{linecnt} \\ 
\arabic{linecnt} &\mbox{Si$^{17}$O}\ * & $\varv=0$, $J$ = 6 -- 5 & 250.744695     &   42.12 & $-3.090$ \stepcounter{linecnt} \\ 
\arabic{linecnt} &\mbox{H$_2$O}\ *   & $\varv_2=2$, $6_{5,2}$ -- $7_{4,3}$ & 268.149117
& 6039.00 & $-4.820$ \stepcounter{linecnt} \\ 
\arabic{linecnt} &\mbox{H$_2$O}   & $\varv_2=2$, $9_{2,8}$ -- $8_{3,5}$ & 250.751793 &
6141.05 & $-5.046$ \stepcounter{linecnt} \\ 
\arabic{linecnt} &\mbox{SO$_{2}$}   & $\varv=0$, 36$_{10,26}$ -- 37$_{9,29}$ & 250.816786
& 857.17 & $-4.472$ \stepcounter{linecnt} \\ 
\arabic{linecnt} &\mbox{SO$_{2}$}\ *   & $\varv=0$, 32$_{4,28}$ -- 31$_{5,27}$ &
252.563893 & 531.10 & $-4.384$ \stepcounter{linecnt} \\ 
\arabic{linecnt} &\mbox{SO$_{2}$}\ *   & $\varv=0$, 30$_{9,21}$ -- 31$_{8,24}$ &
266.943325 & 625.92 & $-4.408$ \stepcounter{linecnt} \\ 
\arabic{linecnt} &\mbox{SO$_{2}$}\ *   & $\varv=0$, 47$_{5,43}$ -- 46$_{6,40}$ &
267.428332 & 1103.31 & $-4.530$ \stepcounter{linecnt} \\ 
\arabic{linecnt} &\mbox{SO$_{2}$}   & $\varv=0$, 13$_{3,11}$ -- 13$_{2,12}$ & 267.537451
& 105.82 & $-3.820$ \stepcounter{linecnt} \\
\arabic{linecnt} &\mbox{SO$_{2}$}   & $\varv=0$, 28$_{4,24}$ -- 28$_{3,25}$ & 267.719840
& 415.88 & $-3.666$ \stepcounter{linecnt} \\ 
\arabic{linecnt} &\mbox{SO$_{2}$}   & $\varv=0$, 63$_{6,58}$ -- 62$_{7,55}$ & 267.192971
& 1950.02 & $-4.799$ \stepcounter{linecnt} \\ 
\arabic{linecnt} &\mbox{SO$_{2}$}\     & $\varv=0$, 38$_{5,33}$ -- 37$_{6,32}$ &
253.935883 & 749.09 & $-4.395$ \stepcounter{linecnt} \\
\arabic{linecnt} &\mbox{SO$_{2}$}\ *    & $\varv=0$, 21$_{3,19}$ -- 22$_{0,22}$ &
253.753448 & 234.70 & $-5.788$ \stepcounter{linecnt} \\

\arabic{linecnt} &\mbox{SO$_{2}$}\ *   & $\varv_2=2$, 13$_{1,13}$ -- 12$_{0,12}$ &
251.738875 & 1571.41 & $-3.760$ \stepcounter{linecnt} \\
\arabic{linecnt} &\mbox{SO$_{2}$}\ *   & $\varv_2=1$, 30$_{4,26}$ -- 30$_{3,27}$ &
266.030566 & 1240.54 & $-3.632$ \stepcounter{linecnt} \\
\arabic{linecnt} &\mbox{SO$_{2}$}   & $\varv_2=1$, 30$_{3,27}$ -- 30$_{2,28}$ &
266.815526 & 1227.53 & $-3.682$ \stepcounter{linecnt} \\
\arabic{linecnt} &\mbox{SO$_{2}$}\ *   & $\varv_2=1$, 11$_{3,9}$ -- 11$_{2,10}$ &
268.169791 & 844.26 & $-3.812$ \stepcounter{linecnt} \\ 
\arabic{linecnt} &\mbox{SO$_{2}$}\ *   & $\varv_1=1$, 28$_{4,24}$ -- 28$_{3,25}$ &
268.617160 & 2071.15 & $-3.662$ \stepcounter{linecnt} \\ 
\arabic{linecnt} &\mbox{SO$_{2}$}   & $\varv=0$, 67$_{10,58}$ -- 66$_{11,55}$ &
267.334780 & 2345.14 & $-4.329$ \stepcounter{linecnt} \\ 
\arabic{linecnt} &\mbox{SO$_{2}$}   & $\varv=0$, 66$_{16,50}$ -- 67$_{15,53}$ &
267.356251 & 2655.42 & $-4.366$ \stepcounter{linecnt} \\ 
\arabic{linecnt} &\mbox{SO$_{2}$}\ *   & $\varv_1=1$, 10$_{5,5}$ -- 11$_{4,8}$ &
250.545521 & 1768.80 & $-4.647$ \stepcounter{linecnt} \\ 
\arabic{linecnt} &\mbox{SO$_{2}$}\ *   & $\varv_1=1$, 8$_{3,5}$ -- 8$_{2,6}$ & 251.650159
& 1712.13 & $-3.917$ \stepcounter{linecnt} \\ 
\arabic{linecnt} &\mbox{SO$_{2}$}   & $\varv_1=1$, 13$_{3,11}$ -- 13$_{2,12}$ &
267.803703 & 1762.51 & $-3.817$ \stepcounter{linecnt} \\ 
\arabic{linecnt} &\mbox{SO$_{2}$}\ *   & $\varv_2=1$, 45$_{5,41}$ -- 44$_{6,38}$ &
252.408567 & 1763.60 & $-4.500$ \stepcounter{linecnt} \\ 
\arabic{linecnt} &\mbox{SO$_{2}$}\ *   & $\varv_2=1$, 15$_{6,10}$ -- 16$_{5,11}$ &
267.006774 & 963.79 & $-4.470$ \stepcounter{linecnt} \\ 
\arabic{linecnt} &\mbox{SO$_{2}$}   & $\varv_2=1$, 31$_{9,23}$ -- 32$_{8,24}$ &
267.091925 & 1430.70 & $-4.372$ \stepcounter{linecnt} \\ 
\arabic{linecnt} &\mbox{SO$_{2}$}\ *   & $\varv_3=1$, 14$_{3,12}$ -- 14$_{2,13}$ &
268.242240 & 2077.83 & $-3.804$ \stepcounter{linecnt} \\ 
\arabic{linecnt} &\mbox{SO$_{2}$}\     & $\varv_2=2$, 3$_{3,1}$ -- 3$_{2,2}$ & 268.024200
& 1517.97 & $-4.124$ \stepcounter{linecnt} \\ 

\arabic{linecnt} &\mbox{SO$_{2}$}\     & $\varv_2=2$, 4$_{3,1}$ -- 4$_{2,2}$ & 267.619458
& 1521.64 & $-3.980$ \stepcounter{linecnt} \\ 
\arabic{linecnt} &\mbox{SO$_{2}$}\     & $\varv_3=1$, 46$_{6,41}$ -- 45$_{7,38}$ &
251.851027 & 3042.64 &$-4.420$ \stepcounter{linecnt} \\ 

\arabic{linecnt} &\mbox{SO$_{2}$}\ *   & $\varv_3=1$, 6$_{3,4}$ -- 6$_{2,5}$ & 253.654459
& 2000.78 & $-3.937$ \stepcounter{linecnt} \\ 

\arabic{linecnt} &\mbox{SO$_{2}$}\ & $\varv_3=1$, 14$_{6,9}$--15$_{5,10}$ & 268.035143 &
2143.16 & $-4.494$ \stepcounter{linecnt} \\ 

\arabic{linecnt} &\mbox{$^{34}$SO$_{2}$} & $\varv=0$, 32$_{4,28}$ -- 32$_{3,29}$ & 251.758329
& 530.31 & $-3.709$ \stepcounter{linecnt} \\ 
\arabic{linecnt} &\mbox{$^{34}$SO$_{2}$} & $\varv=0$, 15$_{3,13}$ -- 15$_{2,14}$ & 267.871060 & 131.76 &$-3.821$ \stepcounter{linecnt} \\ 
\arabic{linecnt} &\mbox{$^{34}$SO$_{2}$} & $\varv=0$, 9$_{5,5}$ -- 10$_{4,6}$ & 252.615371 & 100.51 &$-4.700$ \stepcounter{linecnt} \\ 
\arabic{linecnt} &\mbox{$^{34}$SO$_{2}$} & $\varv=0$, 11$_{3,9}$ -- 11$_{2,10}$ & 253.936316 & 82.05 &$-3.893$ \stepcounter{linecnt} \\ 

\arabic{linecnt} &$^{33}$SO$_2$ & $\varv=0$, 5$_{3,3}$ -- 5$_{2,4}$ & 251.702816 & 36.00 &$-4.028$ \stepcounter{linecnt} \\ 

\arabic{linecnt} &SO     & $\varv=0$, $N_J$ = 6$_5$ -- 5$_4$ & 251.825770 &
50.66 & $-3.716$ \stepcounter{linecnt} \\ 
\arabic{linecnt} &SO     & $\varv=0$, $N_J$ = 4$_3$ -- 3$_4$ & 267.197746 &
28.68 & $-6.148$ \stepcounter{linecnt} \\ 
\arabic{linecnt} &AlO *    & $N$ = 7 -- 6              & 267.938239 & 51.44 &
$-2.687$ \stepcounter{linecnt} \\ 
\arabic{linecnt} &AlOH & $J$ = 8 -- 7                & 251.794759 & 54.38 &
$-4.024$ \stepcounter{linecnt} \\ 
\arabic{linecnt} &HCN    & $\varv=0$, $J$ = 3 -- 2 & 265.886434 & 25.52 & $-3.077$ \stepcounter{linecnt} \\ 
\arabic{linecnt} &HCN    & $\varv_2=1^{1e}$, $J$ = 3 -- 2 & 265.852709 & 1049.89 & $-3.142$ \stepcounter{linecnt} \\ 
\arabic{linecnt} &HCN    & $\varv_2=1^{1f}$, $J$ = 3 -- 2 & 267.199283 & 1050.02 & $-3.135$ \stepcounter{linecnt} \\ 
\arabic{linecnt} &TiO   & $\varv=1$, $^{3}\Delta_1$, $J$ = 8 -- 7 & 251.802917
& 1491.68 & $-3.018$ \stepcounter{linecnt} \\ 
\arabic{linecnt} &$^{50}$TiO   & $\varv=0$, $^{3}\Delta_2$, $J$ = 8 -- 7 &
253.591920 & 192.22 & $-3.030$ \stepcounter{linecnt} \\ 
\arabic{linecnt} &$^{49}$TiO   & $\varv=0$, $^{3}\Delta_1$, $J$ = 8 -- 7 &
251.957928 & 52.92 & $-3.077$ \stepcounter{linecnt} \\ 

\arabic{linecnt} &\mbox{TiO$_{2}$} & $8_{6,2}$--$8_{5,3}$ & 251.866957 & 65.77 & $-2.940$ \stepcounter{linecnt} \\ 
\arabic{linecnt} &\mbox{TiO$_{2}$} & $7_{6,2}$--$7_{5,3}$ & 251.977193 & 59.95 & $-3.041$ \stepcounter{linecnt} \\ 
\arabic{linecnt} &\mbox{TiO$_{2}$} & $26_{3,23}$--$26_{2,24}$ & 250.939052 & 272.67 & $-2.825$ \stepcounter{linecnt} \\ 
\arabic{linecnt} &\mbox{TiO$_{2}$} & $9_{6,4}$--$9_{5,5}$ &  251.708056 & 72.31 & $-2.881$ \stepcounter{linecnt} \\ 
\arabic{linecnt} &\mbox{TiO$_{2}$} & $24_{2,22}$--$24_{1,23}$ & 265.770503  & 224.99 & $-2.894$ \stepcounter{linecnt} \\ 

\arabic{linecnt} &OH & $\varv=0$, $N_J$ = $18_{35/2}$, $F$ =
$18^{+}$--$18^{-}$ & 265.734659 & 8859.61 & $-5.859$ \stepcounter{linecnt} \\ 
\arabic{linecnt} &OH & $\varv=0$, $N_J$ = $18_{35/2}$, $F$ =
$17^{+}$--$17^{-}$ & 265.765323 & 8859.61 & $-5.859$ \stepcounter{linecnt} \\ 
\hline
\end{tabular}
\tablefoot{
  The lines marked with an asterisk (*) show emission excess on top of the continuum over the stellar disk. 
}
\end{center}
\label{linelist}
\end{table*}

\section{Observations and data reduction}
\label{sect_obs}

Our ALMA observations took place between June 7, 2019, and June 8, 2019, from 22:52 to 04:42 (UTC) in four consecutive execution blocks (EBs) in Cycle 6 with the C43-10 configuration of the 12-m array (Program ID: 2018.1.01239.S, P.I.: K.~Ohnaka). The variability phase of W Hya at the time of our ALMA observations was 0.53 at minimum light. Each EB consists of 53--55 target scans with a scan length up to 54.4~s. The on-source time of each EB is about 45 min, and the total on-source time of the project is about 3~hours.
The quasar J1337$-$1257 was observed as the bandpass and absolute flux calibrator, J1342$-$2900 as the phase-referencing calibrator, and J1351$-$2912 as the check source.

The observations were carried out under very good atmospheric conditions with a precipitable water vapor (PWV) of about 0.55~mm. W Hya was observed from an elevation of ${\sim}59\degr$ through its transit at ${\sim}85\degr$ until ${\sim}44\degr$. This resulted in a broad and nonoverlapping hour angle range, and hence a good $u\varv$ coverage. The shortest and longest baselines are 83.1~m and 16.2~km, respectively, which results in an angular resolution of 16--20~mas and a largest recoverable scale of 190 mas at the observed frequencies.

As listed in Table~\ref{spwlist}, the spectral setups consist of nine spectral windows (spws) centered at 250.7, 251.8, 252.5, 253.7, 265.9, 266.9, 267.2, 267.7, and 268.2~GHz. The bandwidth of the first seven windows is {468.75}~MHz, while that of the windows at 267.7 and 268.2~GHz is {937.50}~MHz.
The four highest-frequency windows (spws 6--9) partially overlap together.
The channel width in each spectral window is 488.3~kHz.
The velocity resolution of 1.1--1.2~km~s$^{-1}$ corresponds to twice the spectral channel width with the Hanning smoothing at the correlator. 

The visibility data were calibrated with the Common Astronomy Software Applications CASA (Casa Team \cite{casa22}), version 5.6.1-8, following the standard steps of the ALMA Cycle 7 pipeline. Among 44 antennas in the array, DA54 and DA61 were completely flagged due to bad amplitudes. After the pipeline calibration, we self-calibrated the data and manually reconstructed the images. We first identified spectral channels containing line emission or absorption from an initial set of image cubes and produced the continuum visibility data from the line-free channels. The continuum images of W~Hya were reconstructed, which provided initial models for self-calibration, which was done in two iterations. The first iteration was for phase alone with the solution interval being the scan length (${\lesssim}54$~s). The second iteration was for both the amplitude and phase using the entire EB (${\sim}45$~min) as the solution interval.

The continuum and spectral line images of W~Hya were reconstructed from the self-calibrated visibility data. We adopted a pixel scale of 3~mas and the robust weighting with a robustness parameter of 0.5 in the CASA task \texttt{tclean}.
For the vibrationally excited HCN line ($\varv_2=1^{1e}$, $J$ = 3 -- 2) at 265.853~GHz (Sect.~\ref{subsect_res_hcn_265}) and AlOH ($J$ = 8 -- 7) line at 251.795~GHz (Sect.~\ref{subsect_res_alo_aloh}), we reconstructed the images with the natural weighting to enhance the weak extended emission. 
We also imaged the continuum data in each spectral window using the multifrequency synthesis (MFS) method. Because the millimeter continuum of W Hya is spatially resolved, we deconvolved the images with the multi-scale algorithm (Cornwell \cite{cornwell08}) for size scales of 1 (point sources), 5, and 15 pixels and with a small-scale bias parameter of 0.7.
The typical RMS noise in the continuum MFS images is ${\sim}40$~$\mu$Jy/beam. The restoring beams for the continuum images are about 19$\times$17~mas in the lower sideband (spws 1--4) and 18$\times$16~mas in the upper sideband (spws 5--9) as listed in Table~\ref{spwlist}.

Except for HCN $\varv=0$, $J=3$--2 at 265.886 GHz and SO $\varv=0$,
$N_J=6_5$--$5_4$ at 251.826 GHz, the emission of most spectral lines is confined
within a radius of ${\sim}0{\farcs}2$ from the center of the continuum 
emission. 
The RMS noise in the spectral line cubes is typically
${\sim}0.5$--$0.6$~mJy/beam over the channel width of 488~kHz
(Table~\ref{spwlist}). As discussed in
Sect.~\ref{subsect_res_29sio_253}, spw 4 covers the $^{29}$SiO $\varv=2$
$J$ = 6--5 line, which
exhibits strong maser emission in W~Hya. Due to the dynamic range limit,
the RMS noise is much higher in the channels containing the strongest
emission, up to 2.7~mJy/beam. The maximum imaging dynamic range is close to
6800.

  In order to determine the position of the stellar disk center,
  we fit the visibility data at the line-free spectral channels in each
  spectral window before self-calibration with a uniform ellipse using 
  \texttt{uvmultifit} (Mart\'i-Vidal et al. \cite{marti-vidal14}).
  The continuum disk center averaged over the different spectral
  windows is 
  ($\alpha$, $\delta$) = (13:49:01.92528, $-28$:22:04.69895)
  in the International Celestial Reference System (ICRS). 
  The images presented in this paper are centered at this disk center
  position. 

  We identified the spectral line detections mainly using the
  Splatalogue
service\footnote{\url{https://splatalogue.online/\#/advanced}}, the Cologne
Database for Molecular
Spectroscopy\footnote{\url{https://cdms.astro.uni-koeln.de/}}
(CDMS, Endres et al. \cite{endres16}; M\"uller et al. \cite{mueller01};
\cite{mueller05}),
the Jet Propulsion
Laboratory (JPL) catalog\footnote{\url{https://spec.jpl.nasa.gov/}}
(Pickett et al. \cite{pickett98}), and the HITRAN
database\footnote{\url{https://hitran.org/}} (Gordon et al. \cite{gordon22}).
Table~\ref{linelist}
lists the 57 molecular lines identified in the present work, which include
$^{29}$SiO, $^{30}$SiO, Si$^{17}$O, H$_2$O, SO$_2$, $^{34}$SO$_2$,
$^{33}$SO$_2$, SO, AlO, AlOH, HCN, TiO, $^{49}$TiO, $^{50}$TiO, TiO$_2$, and OH.

\begin{figure*}
\centering
  \resizebox{17cm}{!}{\rotatebox{0}{\includegraphics{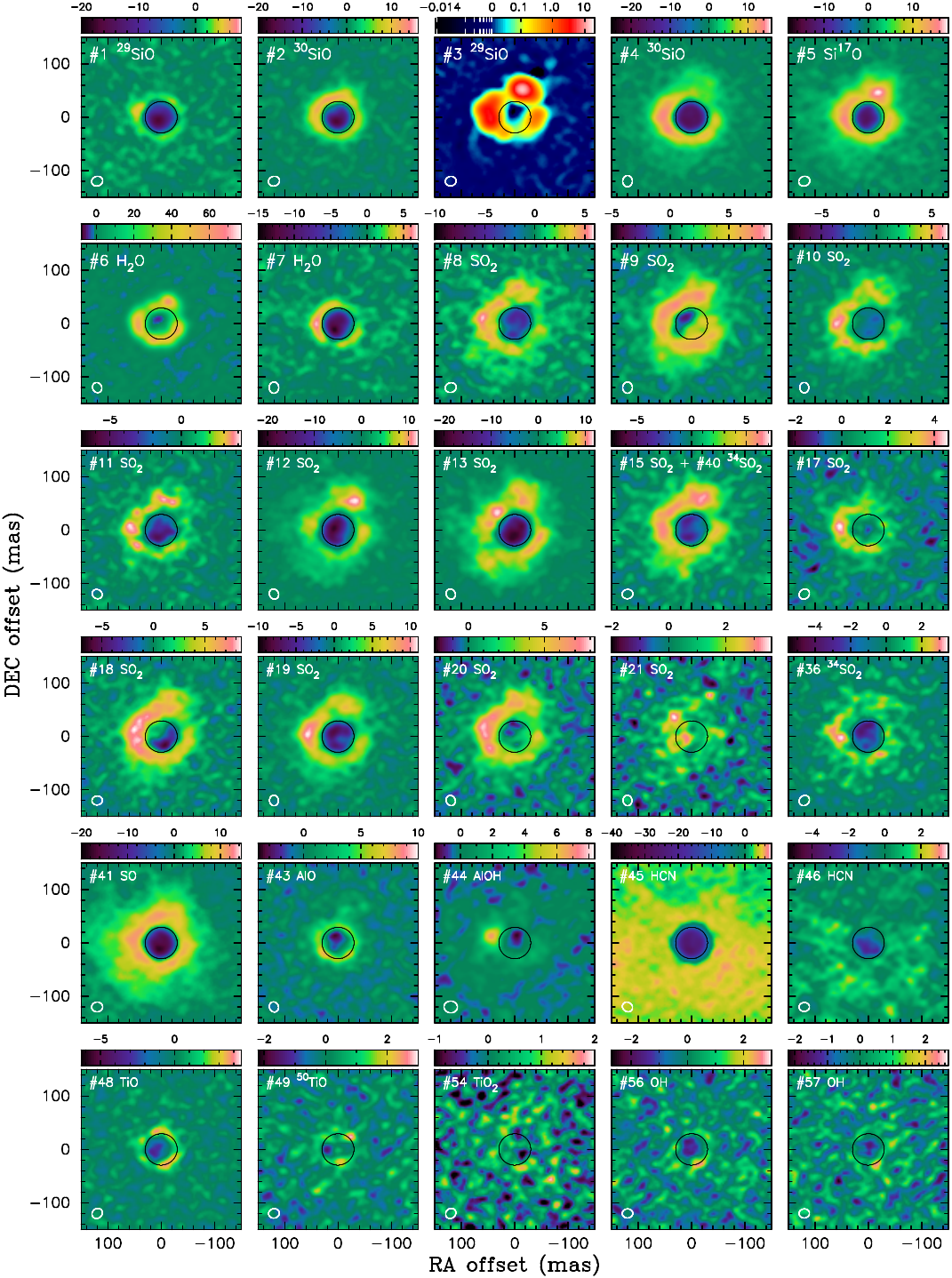}}}
\caption{
  Continuum-subtracted maps of \mbox{W~Hya}\ observed at the systemic velocity 
  in the different molecular lines presented in the main text.
  The images of two \mbox{H$_2$O}\ lines (\#6 and \#7) are from Paper~I. 
  The color scale shown above each panel corresponds to mJy/beam except for
  \#3, where it is given in Jy/beam.
  The black circles represent the ellipse fit to the continuum image.
  The identification number of each line in Table~\ref{linelist} is 
  shown in the upper left corner. The restoring beam size 
  is shown in the lower left corner of each panel.
  North is up, and east is to the left. 
}
\label{centralmap_all}
\end{figure*}

Figure~\ref{centralmap_all} gives an overview of the continuum-subtracted
images of the different molecular lines at the systemic velocity,
which cover most of the identified
species including two \mbox{H$_2$O}\ lines reported in Paper~I 
(the channel maps and/or integrated intensity maps of all
the detected lines will be presented below and in Appendices on
Zenodo \url{https://doi.org/10.5281/zenodo.17118092}).
ALMA's high angular resolution allows us to resolve
the stellar disk and highly inhomogeneous atmosphere and innermost
circumstellar envelope. The different molecular lines show distinct
morphology, indicating differences in their formation and excitation. 

\begin{figure}
\resizebox{\hsize}{!}{\rotatebox{0}{\includegraphics{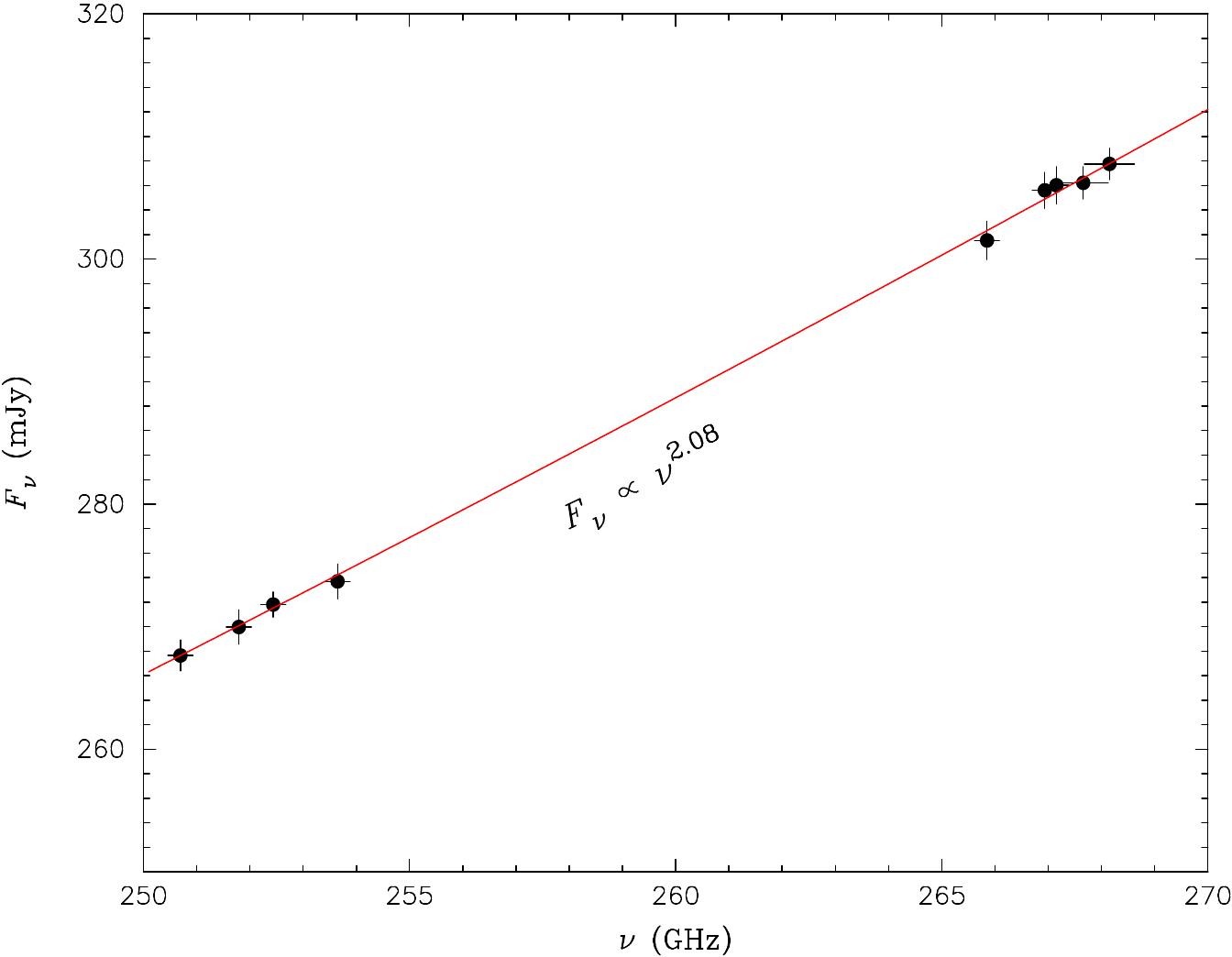}}}
\caption{
  Continuum spectrum of \mbox{W~Hya}\ at 250--268~GHz. The black dots represent the
  observed continuum fluxes, while the solid red line represents a power-law
  fit with $F_{\nu} \propto \nu^{2.08}$. The horizontal bars show the
  bandwidth of the spectral windows, not errors.
}
\label{cont_flux_spec}
\end{figure}

\section{Results}
\label{sect_res}

\subsection{Continuum}
\label{subsect_res_cont}

In Paper I, we presented the continuum image restored
with a beam of $18\times16$~mas in the 268.2~GHz spectral window (spw 9), 
which can be fit with a uniform elliptical disk 
with a major and minor axis of $59.1 \pm 0.3$~mas and $57.7 \pm 0.2$~mas
with a position angle of $16 \pm 23$\degr. 
The continuum images obtained in other spectral windows are 
similar to the one at 268.2~GHz as shown in Fig. A.1. 
The aforementioned uniform elliptical disk fitting to the continuum data
in other spectral windows 
results in major and minor axes of 58--60~mas, as found at 268.2~GHz.
In the present work, 
we use the geometrical mean of the major and minor axis of 
59.3~mas derived in Paper I 
as the millimeter continuum angular diameter (i.e., 29.6~mas as
the continuum angular radius \mbox{$R_{\rm cont}$}). 

Table~\ref{spwlist} lists the flux densities integrated over the continuum
images in nine spectral windows. 
The flux densities are in broad agreement with the measurement by
Dehaes et al. (\cite{dehaes07}), who obtained 
$280.0 \pm 17.2$~mJy at 250~GHz with a full width at half maximum 
(FWHM) bandwidth of 90~GHz in May 2003
(phase 0.65). 
As Fig.~\ref{cont_flux_spec} shows, the continuum flux measurements 
can be well fit with a power law, $F_{\nu} \propto \nu^{2.08 \pm 0.03}$.
This is close to the 
power-law index of 1.86 expected from the radio-photosphere model,
where the opacity source is mainly optically thick 
free-free emission (Reid \& Menten \cite{reid97}, \cite{reid07};
Matthews et al. \cite{matthews15}). 

The millimeter continuum angular size at 250--268~GHz 
is 1.4 times larger than the star's angular diameter of 41.4~mas
adopted in the present work (Sect.~\ref{sect_basic}). 
Our continuum angular size is in broad agreement with the
$(69 \pm 10) \times (46 \pm 7)$ mas obtained at 43~GHz by
Reid \& Menten (\cite{reid07}), but it is larger than the
$56.5 \times 51.0$~mas
obtained at 338~GHz by Vlemmings et al. (\cite{vlemmings17}) at a
variability phase of $\sim$0.3.
Given that the free-free opacity decreases with the frequency, 
the difference in the continuum angular size can be explained
if the continuum at 251--268~GHz forms
at layers slightly farther away from the star than that at 338~GHz. 
However, it is also possible that the phase dependence of the angular
size is responsible for the observed difference in the continuum angular
size as discussed by Matthews et al. (\cite{matthews15}) and 
recently modeled by Bojnordi Arbab et al. (\cite{bojnordi_arbab24}).

All continuum images obtained at 250.7--268.2~GHz show a slight offset of the
intensity peak to the northwest (NW) by 5--7~mas
(0.2--0.3~\mbox{$R_{\star}$}) from
the center of the stellar disk obtained by the aforementioned uniform
elliptical disk fitting, 
but the asymmetry in the intensity is
1\% at most, as described for the 268.2~GHz continuum image in Paper~I.  
Although the contemporaneous visible polarimetric imaging reveals
clumpy dust cloud formation within $\sim$2~\mbox{$R_{\star}$}\ (Paper~I), 
the asymmetry in the millimeter continuum images 
cannot be attributed to a dust clump for the following reason. 
Ohnaka et al. (\cite{ohnaka17}) derived a dust optical depth of 0.8 
at 0.55~\mbox{$\mu$m}\ 
from the visible polarimetric imaging of \mbox{W~Hya}\ at minimum light.
This translates into a dust optical
depth at 268~GHz ($\tau_{\rm mm}^{\rm dust}$) of $10^{-7}$,
if we adopt the optical constants of forsterite (\mbox{Mg$_2$SiO$_4$}) from
J\"ager et al. (\cite{jaeger03}) and a grain size of 0.1~\mbox{$\mu$m}. 
Approximating the stellar continuum intensity $I_{\star}$ with the blackbody
radiation at 2150~K (see below), the dust thermal emission $I_{\rm dust}$ is
estimated as
$I_{\star} e^{-\tau_{\rm mm}^{\rm dust}} + B_{\nu}(T_{\rm dust})
  (1-e^{-\tau_{\rm mm}^{\rm dust}})$.
Assuming a condensation temperature of 1500~K for $T_{\rm dust}$,
the difference between the stellar continuum intensity and dust emission
intensity is $\sim \!\! 10^{-7}$ of the stellar continuum intensity,
too small to account for the observed contrast of 1\% of the asymmetry.

The observed smooth continuum images is in marked contrast to 
the detection of a hot spot over the stellar disk of \mbox{W~Hya}\ in the
338~GHz continuum reported by Vlemmings et al. (\cite{vlemmings17}), 
although our spatial resolution of 20$\times$17~mas is comparable to their 
restoring beam size of 17~mas. 
They derived a brightness temperature as high as $> \! 5.3 \times 10^4$~K. 
The presence of hot gas may be temporarily variable, 
given that the ALMA observations of Vlemmings et al. (\cite{vlemmings17})
took place in December 2015 ($\sim$3.5 years $\approx$ 
3.3 pulsation cycles before our observations).
However, Hoai et al. (\cite{hoai22}), imaging the 338 GHz continuum from the
same dataset, did not reproduce such a hot spot.
Detailed investigation of the imaging and model fitting of the 338 GHz
continuum image would be necessary, which is, however, beyond the scope of
the present paper. 

The uniform elliptical disk fitting in nine spectral
windows results in brightness temperatures of 2130--2170~K over the
millimeter continuum stellar disk. 
Assuming that the continuum is formed by the optically thick free-free
emission, the derived brightness temperatures correspond to the gas
temperature in the millimeter-continuum-forming layers. 
Given the effective temperature of 2330~K (Sect.~\ref{sect_basic}), 
the gas temperature falls off only slightly from the 
photosphere to the millimeter continuum-forming layers at $\sim$1.4~\mbox{$R_{\star}$}. 
Vlemmings et al. (\cite{vlemmings17}) derived a
brightness temperature of $2495 \pm 255$~K averaged over the entire
stellar disk observed at 338~GHz at phase 0.3,
while their continuum image (their Fig.~1) shows
that the brightness temperature over the stellar disk outside the hot spot
is lower, 1800--2300~K.
Reid \& Menten (\cite{reid07}) imaged \mbox{W~Hya}\ in the 43~GHz continuum
at phase 0.25 and derived a brightness temperature of $2380 \pm 550$~K. 
The radio photosphere model of Reid \& Menten (\cite{reid97}) predicts
the brightness temperature to increase with the frequency.
However, the large uncertainties, inhomogeneities over the stellar disk 
as well as the phase dependence of the atmosphere make it difficult to
compare with the prediction of the radio photosphere model. 
Contemporaneous observations with comparable angular resolutions will be
useful to address this point. 

\begin{figure*}
\centering
\resizebox{17cm}{!}{\rotatebox{0}{\includegraphics{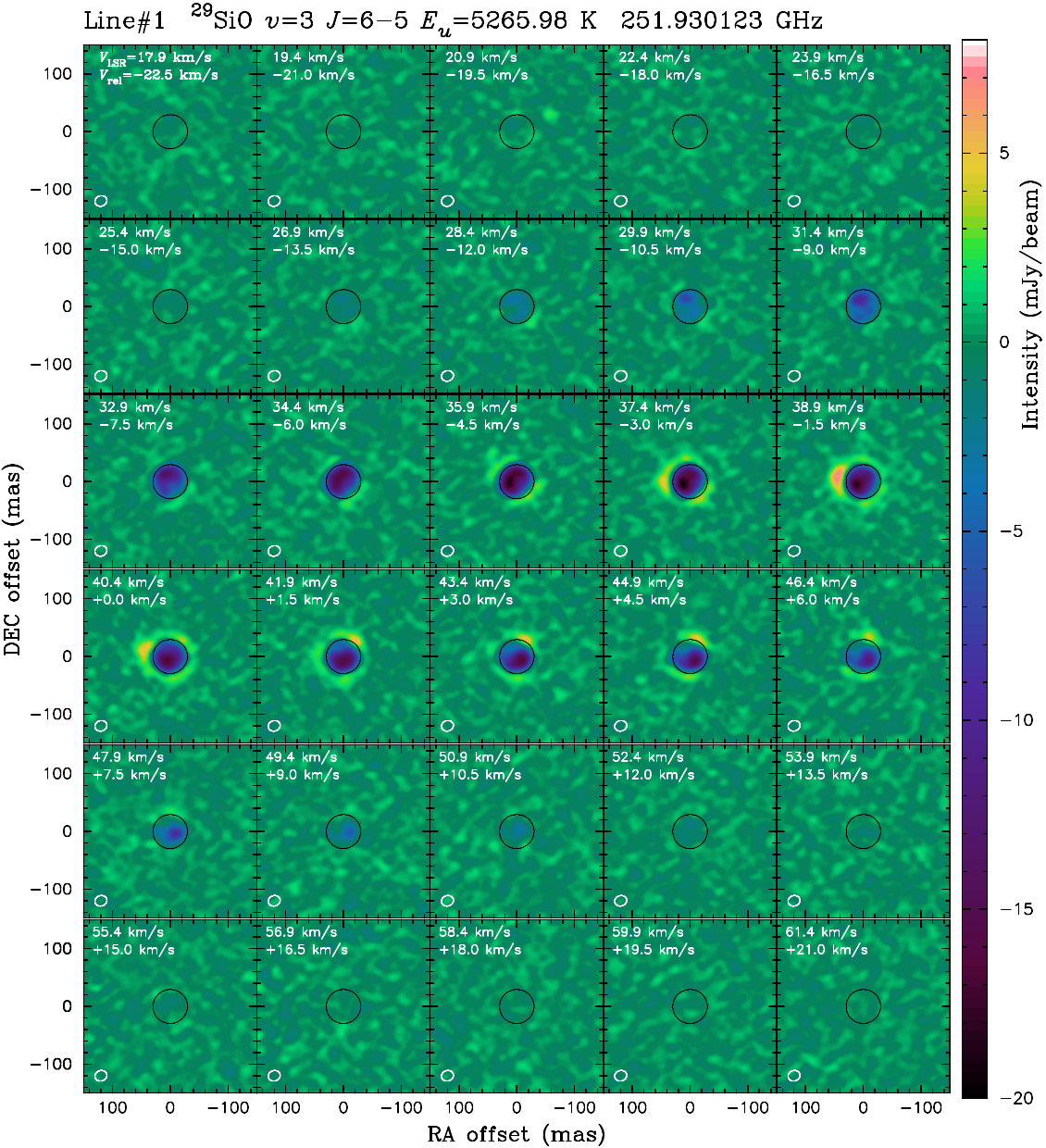}}}
\caption{
  Continuum-subtracted channel maps of \mbox{W~Hya}\ observed in the
  \mbox{$^{29}$SiO}\ $\varv=3$ $J$ = 6 -- 5 line at 251.930123~GHz. 
  The black circles represent the ellipse fit to the continuum image.
  In the upper left corner of each panel, 
  the LSR velocity and the relative
  velocity \mbox{$V_{\rm rel}$} \ = \mbox{$V_{\rm LSR}$} - \mbox{$V_{\rm sys}$}, \mbox{$V_{\rm sys}$} = 40.4~\mbox{km s$^{-1}$} are shown.
  The restoring beam size 
  is shown in the lower left corner of each panel.
  North is up, and east is to the left. 
} 
\label{channelmap_29sio_251}
\end{figure*}

\begin{figure*}[h]
\centering
\resizebox{17cm}{!}{\rotatebox{0}{\includegraphics{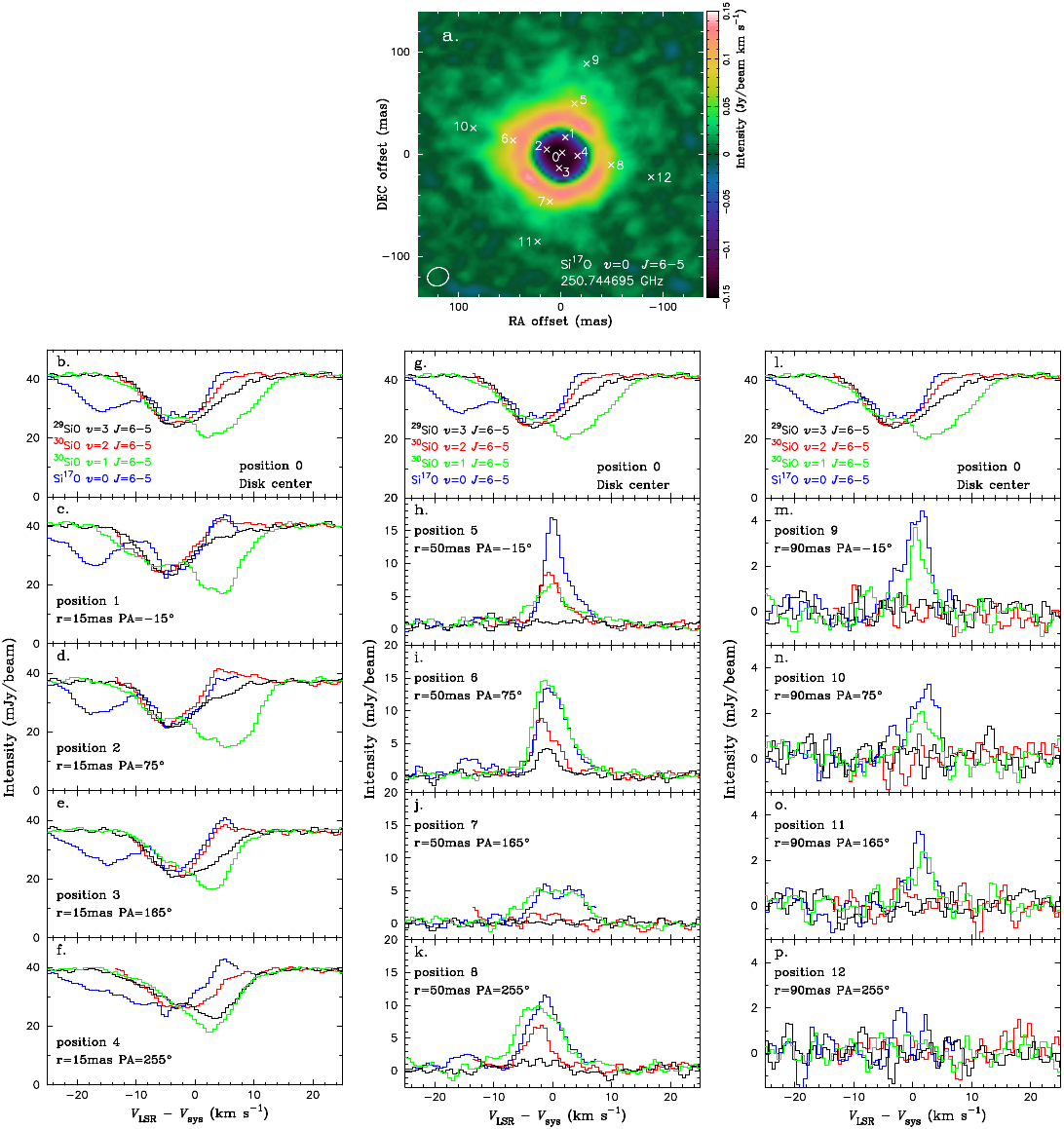}}}
\caption{
  Spatially resolved spectra of four SiO lines detected in our ALMA
  observations. 
  {\bf a:} Continuum-subtracted intensity map of the \mbox{Si$^{17}$O}\ line
  ($\varv=0$, $J$ = 6 -- 5) integrated from \mbox{$V_{\rm rel}$}\ = $-10$ to 10~\mbox{km s$^{-1}$}. 
  The crosses and 
  numbers represent the positions where the spatially resolved spectra
  shown in panels b--p were derived. 
  {\bf b}--{\bf f:} Spatially resolved spectra observed at five positions
  over the stellar disk. 
  The black, red, green, and blue lines represent the spectra of
  the \mbox{$^{29}$SiO}\ $\varv=3$ $J$ = 6 -- 5, \mbox{$^{30}$SiO}\ $\varv=2$ $J$ = 6 -- 5,
  \mbox{$^{30}$SiO}\ $\varv=1$ $J$ = 6 -- 5, and \mbox{Si$^{17}$O}\ $\varv=0$ $J$ = 6 -- 5 lines,
  respectively. The spectra were obtained from the data cube with
  the continuum emission. 
  {\bf g}--{\bf k:} SiO spectra obtained at positions 5--8 at
  a radial distance of 50~mas from the stellar disk center. 
  These spectra were extracted from the continuum-subtracted data cube. 
  The spectra measured at the disk center are shown in panel {\bf g}
  to facilitate comparison between the absorption and emission spectra.
  {\bf l}--{\bf p:} SiO spectra obtained at positions 9--12 at
  a radial distance of 90~mas from the stellar disk center, shown
  in the same manner as in panels {\bf g}--{\bf k}. 
}
\label{whya_sio_combined}
\end{figure*}

\begin{figure*}
\centering
\resizebox{17cm}{!}{\rotatebox{0}{\includegraphics{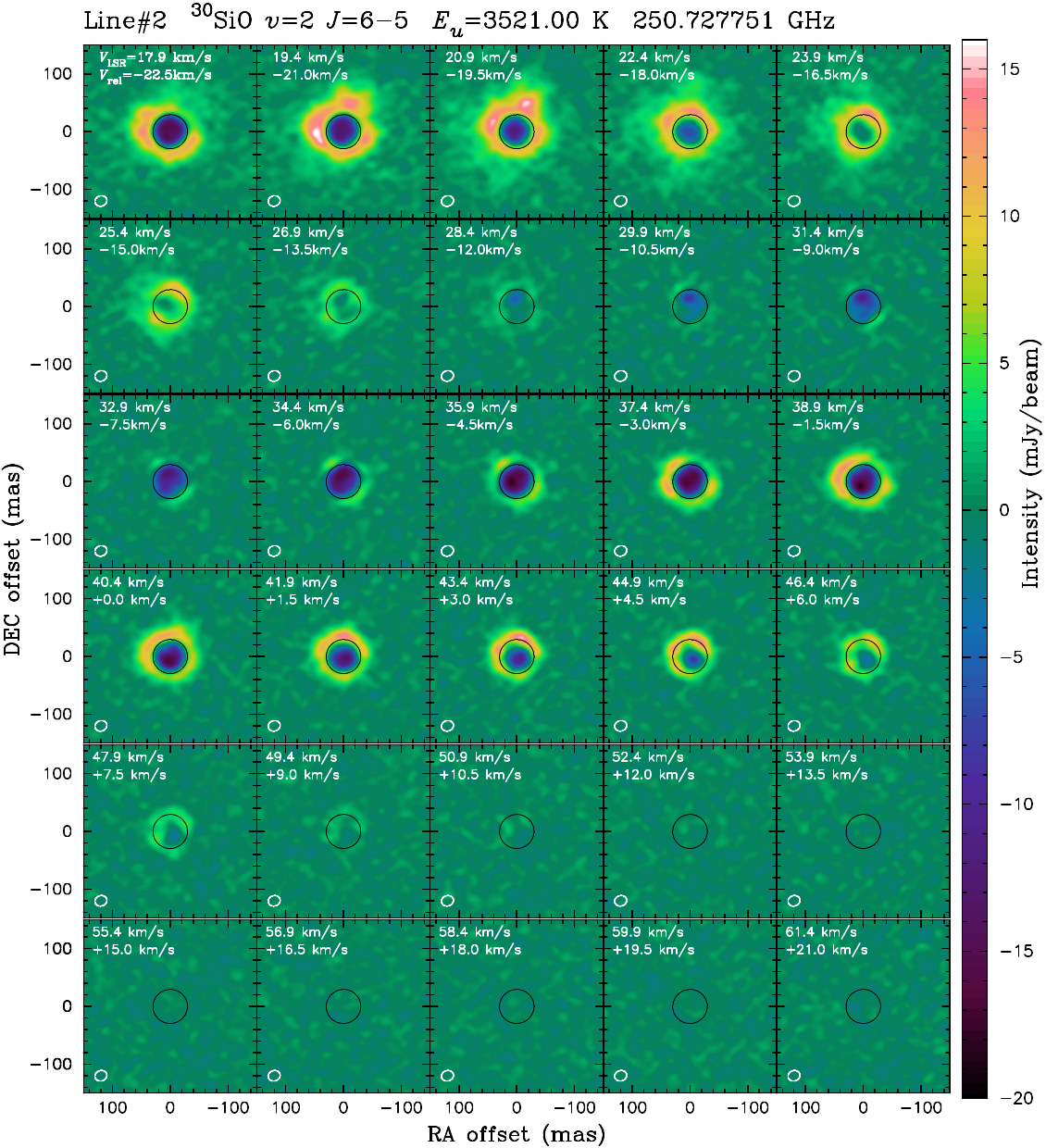}}}
\caption{
  Continuum-subtracted channel maps of \mbox{W~Hya}\ obtained in the \mbox{$^{30}$SiO}\
  $\varv=2$ $J$ = 6 -- 5 line at 250.727751~GHz, shown in the same manner as
  Fig.~\ref{channelmap_29sio_251}. The maps at $\mbox{$V_{\rm rel}$} \le -10.5$~\mbox{km s$^{-1}$}\ are
  severely affected by the \mbox{Si$^{17}$O}\ line at 250.745~GHz
  (presented in Fig.~\ref{channelmap_si17o_250}). 
}
\label{channelmap_30sio_250}
\end{figure*}

\subsection{\mbox{$^{29}$SiO}\ $\varv=3$, $J$ = 6 -- 5 at 251.930~GHz}
\label{subsect_res_29sio_251}

Figure~\ref{channelmap_29sio_251} shows the continuum-subtracted channel
maps obtained for the \mbox{$^{29}$SiO}\ line ($\varv=3$, $J$ = 6 -- 5) at 251.930~GHz  
with a restoring beam size of $21 \times 18$~mas. 
In each panel, the velocity in the LSR as well as the relative velocity with
respect to the systemic velocity $\mbox{$V_{\rm rel}$} = V_{\rm LSR} - V_{\rm sys}$ is
shown. 
The images show compact emission extending to a radius of $\sim$40~mas 
($\sim$1.3~\mbox{$R_{\rm cont}$}\ = 1.9~\mbox{$R_{\star}$}), which means that the \mbox{$^{29}$SiO}\ line probes
the region close to the star because of its high upper level energy of 
5266~K. 
Furthermore, 
thanks to the spatial resolution of $\sim$20~mas and the large angular size 
of \mbox{W~Hya}, we can see absorption over the stellar disk,
which is 
expected because the gas temperature is generally lower than the
continuum brightness temperature, and it decreases with the radial distance.
Along the line of sight to the stellar disk, the cooler
layers of \mbox{$^{29}$SiO}\ line formation absorb the background continuum,
resulting in an absorption spectrum. The redshifted and
blueshifted velocities of the absorption indicate infalling and outflowing
motions, respectively.

Our ALMA images reveal that the absorption over the stellar disk is
inhomogeneous. 
The images obtained at velocities from \mbox{$V_{\rm rel}$}\ = $1.5$ to $7.5$~\mbox{km s$^{-1}$}\ 
show prominent absorption in the southwest (SW) region of the stellar disk,
revealing an infalling gas clump or cell. 
On the other hand, the images obtained at \mbox{$V_{\rm rel}$}\ = $-9$ to $\sim \! 0$~\mbox{km s$^{-1}$}\
show that absorption is particularly deep in the
northeast (NE)--east (E)--southeast (SE) regions. 

To see the variation of the absorption over the stellar disk, 
we extracted spatially resolved spectra from the data cube without
the continuum subtraction at five different positions over the stellar disk:
at the center of the fitted elliptical disk and
four positions with a radial offset of 15~mas (0.5~\mbox{$R_{\rm cont}$}) at position angles
of $-15$\degr, 75\degr, 165\degr, and 255\degr, which 
we refer to as position 0, 1, 2, 3, and 4, respectively, as labeled in 
Fig.~\ref{whya_sio_combined}a.  
The four position angles were selected to probe the plume, tail, and
extended, elongated atmosphere, which will be described in the next
subsections.

Figure~\ref{whya_sio_combined} (left column, black lines) reveals that 
the spectra of \mbox{$^{29}$SiO}\ obtained at all five positions
show broad absorption extending approximately from \mbox{$V_{\rm rel}$}\ =
$-16$ to $10$~\mbox{km s$^{-1}$}, 
which indicates the presence of outflowing and infalling layers 
along the line of sight over the stellar disk. We present a simple
model to explain the observed data in Sect.~\ref{subsect_res_sio_model}. 
If the current mass of \mbox{W~Hya}\ is assumed to be 1~\mbox{$M_{\sun}$}\ based on the
results of Khouri et al. (\cite{khouri14a}) and Danilovich et al.
(\cite{danilovich17}), 
the escape velocity at 40~mas (1.9~\mbox{$R_{\star}$}\ = 3.9~au at the distance of 98~pc) 
is 21~\mbox{km s$^{-1}$}, which is higher than the outflow velocity seen in
the \mbox{$^{29}$SiO}\ absorption line. 
Therefore, the material within $\sim$2~\mbox{$R_{\star}$}\ is still gravitationally
bound if only marginally. 
The observed velocity range of the \mbox{$^{29}$SiO}\ line 
is in broad agreement -- albeit somewhat smaller --
with the velocities from $-20$ to $20$~\mbox{km s$^{-1}$}\ seen in the CO $\varv = 1$
$J = 3 - 2$ line toward \mbox{W~Hya}\ (Vlemmings et al. \cite{vlemmings17}),
given the difference in the upper level energy and possible time variations. 
The broad line profile is also observed in the first-overtone
($\varv = 2 - 0$) SiO lines near
4~\mbox{$\mu$m}\ for a sample of Mira-type and semi-regular variables 
(Lebzelter et al. \cite{lebzelter01}).
They show that the 4~\mbox{$\mu$m}\ SiO lines are broad, approximately covering
from $-20$ to $20$~\mbox{km s$^{-1}$}\ with respect to the systemic velocity,
comparable to the velocity width of the millimeter \mbox{$^{29}$SiO}\ line.

\subsection{\mbox{$^{30}$SiO}\ $\varv=2$, J = 6 -- 5 at 250.728~GHz}
\label{subsect_res_30sio_250}

Figure~\ref{channelmap_30sio_250} shows the continuum-subtracted channel maps
of the \mbox{$^{30}$SiO}\ line ($\varv=2$, $J$ = 6 -- 5) at 250.728~GHz.
The images at velocities more blueshifted than $\mbox{$V_{\rm rel}$} \approx \! -10.5$~\mbox{km s$^{-1}$}\
are severely contaminated by the adjacent \mbox{Si$^{17}$O}\ line at 250.745~GHz
(see Sect.~\ref{subsect_res_si17o_250}). 
The images obtained near the systemic velocity show emission elongated in
the east-northeast (ENE)--west-southwest (WSW) direction with a semimajor
and semiminor axis of
$\sim$45~mas (1.5~\mbox{$R_{\rm cont}$}\ = 2.2~\mbox{$R_{\star}$}) and
$\sim$40~mas (1.3~\mbox{$R_{\rm cont}$}\ = 1.9~\mbox{$R_{\star}$}), respectively. 
The elongated emission is similar to that seen in the 
\mbox{$^{29}$SiO}\ line ($\varv=3$ $J$ = 6 -- 5) presented in
Fig.~\ref{channelmap_29sio_251}. However,
the emission is slightly more extended and more prominent, 
because the upper level energy of 3520~K of the \mbox{$^{30}$SiO}\ ($\varv=2$) line is
lower than that of the \mbox{$^{29}$SiO}\ ($\varv=3$) line (5266~K). 

The channel maps obtained at \mbox{$V_{\rm rel}$}\ = $-9$ to $-6$~\mbox{km s$^{-1}$}\ show prominent 
blueshifted absorption in the northern region of the
stellar disk. Strong absorption appears in the SE region of the
stellar disk at \mbox{$V_{\rm rel}$}\ = $-4.5$ to $0$~\mbox{km s$^{-1}$}. These features are seen in the
channel maps of the \mbox{$^{29}$SiO}\ line described above.
On the other hand, the channel maps obtained at \mbox{$V_{\rm rel}$}\ = $1.5$ to $9$~\mbox{km s$^{-1}$}\
show that the bright ring-like emission extends inward over the stellar disk,
different from that of the \mbox{$^{29}$SiO}\ line.  
The "leak" of the emission off the stellar limb into the stellar disk
can occur due to the finite beam size (e.g, Wong et al. \cite{wong16}).
However, the channel maps at some other velocity channels (e.g., \mbox{$V_{\rm rel}$}\ =
$-3.0$ and $-1.5$~\mbox{km s$^{-1}$}) do not show emission over the stellar disk
in spite of the strong emission just outside the stellar limb,
particularly in the east. 
Therefore, the emission detected over the stellar disk at the redshifted 
velocities cannot be attributed to such a leak due to the finite beam size. 

The spatially resolved spectra (Fig.~\ref{whya_sio_combined}, left column,
red lines) show that the blue part of the absorption of the \mbox{$^{30}$SiO}\ line
between \mbox{$V_{\rm rel}$}\ = $-10$ and 0~\mbox{km s$^{-1}$}\ is 
nearly identical to that of the \mbox{$^{29}$SiO}\ line (black lines in the figure).
As will be discussed in the next subsections, the spectra
of the \mbox{$^{30}$SiO}\ $\varv=1$ $J$ = 6 -- 5 line 
(and \mbox{Si$^{17}$O}\ $\varv=0$ $J$ = 6 -- 5 line to some extent) in this blueshifted
velocity range are identical.
This suggests that the blue part of these lines is optically thick,
corresponding to the same brightness temperature. In an optically thick case,
this means that these lines originate from the gas at the same kinetic
temperature at the same radial range.

However, the red part shows noticeably shallower absorption 
or even emission at some velocities due to the aforementioned
emission over the stellar disk. For the same reason, at position 4,
the deepest absorption at the redshifted velocity of $\sim$5~\mbox{km s$^{-1}$}\ seen
in the spectrum of the \mbox{$^{29}$SiO}\ line is also absent. 
The emission over the stellar disk manifests itself as a bump at \mbox{$V_{\rm rel}$}\ =
$3$ to $10$~\mbox{km s$^{-1}$}\ in the spectrum at position 2 (PA = 75\degr)
and also at position 1 (PA = $-15$\degr) to a lesser extent. 
The redshifted emission over the stellar disk may seem to indicate the
presence of 
gas hotter than the 2100--2200~K of the continuum-forming layer. 
However,
if we assume that different molecular species have the same kinetic
temperature and they are in local thermodynamical equilibrium (LTE),
such hot gas would give rise to emission over the stellar disk
in the \mbox{$^{29}$SiO}\ $\varv=3$ line as well as the \mbox{$^{30}$SiO}\ $\varv=1$ line
(Sect.~\ref{subsect_res_29sio_251}), which is not observed.
If a line forms in non-LTE, its excitation is determined not only by
the collision but also by the radiative pumping, and therefore, 
the line can appear in absorption or emission. 
In Sect.~\ref{subsect_res_si17o_250}, we present
the interpretation that the redshifted emission over the stellar disk is
of the nonthermal origin -- suprathermal
(excitation temperature $>$ kinetic temperature)
or maser action (excitation temperature $<$ 0).

\begin{figure*}
\centering
\resizebox{17cm}{!}{\rotatebox{0}{\includegraphics{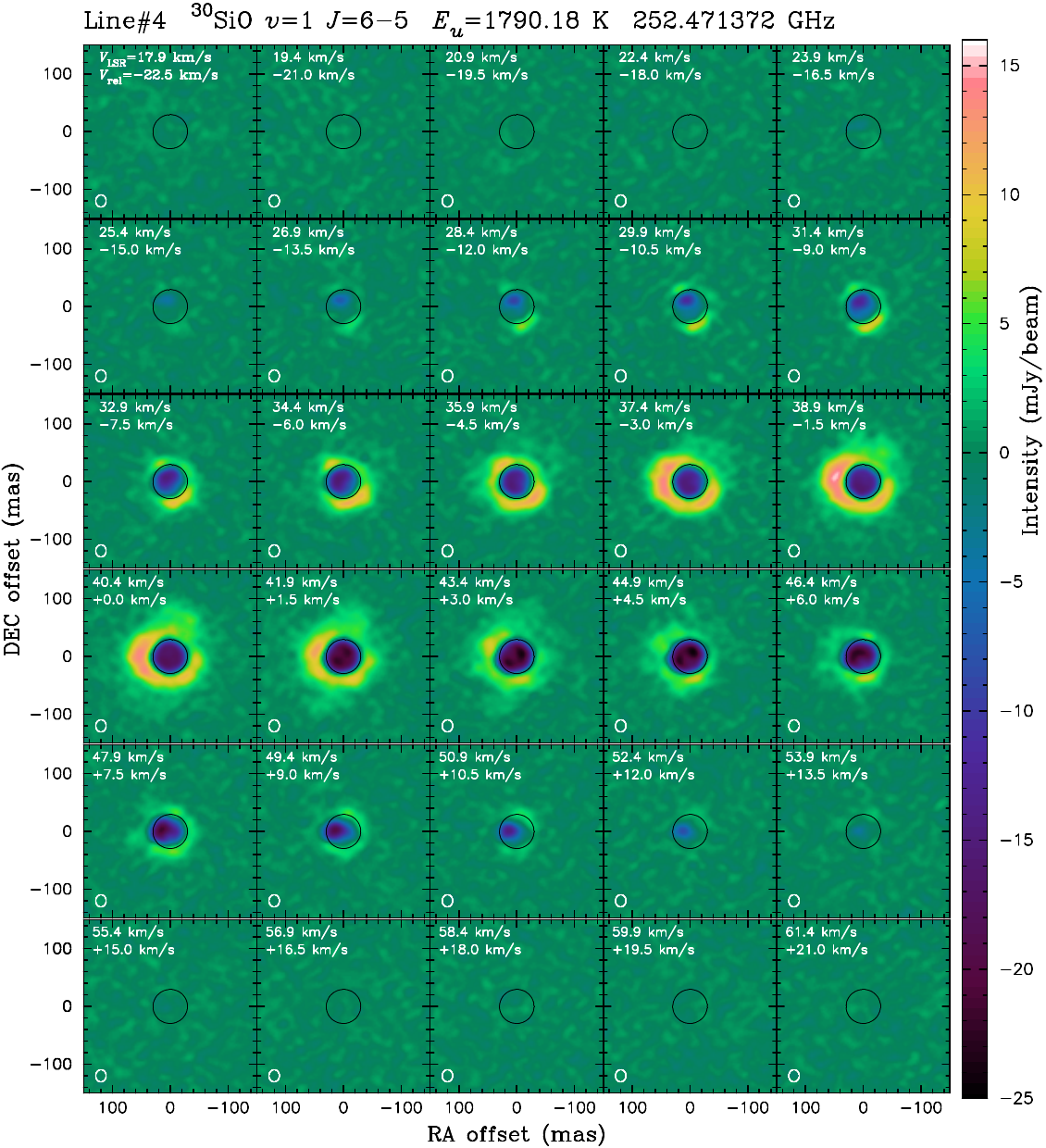}}}
\caption{
  Continuum-subtracted channel maps of \mbox{W~Hya}\ obtained in the \mbox{$^{30}$SiO}\
  $\varv=1$ $J$ = 6 -- 5 line at 252.471372~GHz, shown in the same manner as
  Fig.~\ref{channelmap_29sio_251}.
}
\label{channelmap_30sio_252}
\end{figure*}

\begin{figure}
\resizebox{\hsize}{!}{\rotatebox{0}{\includegraphics{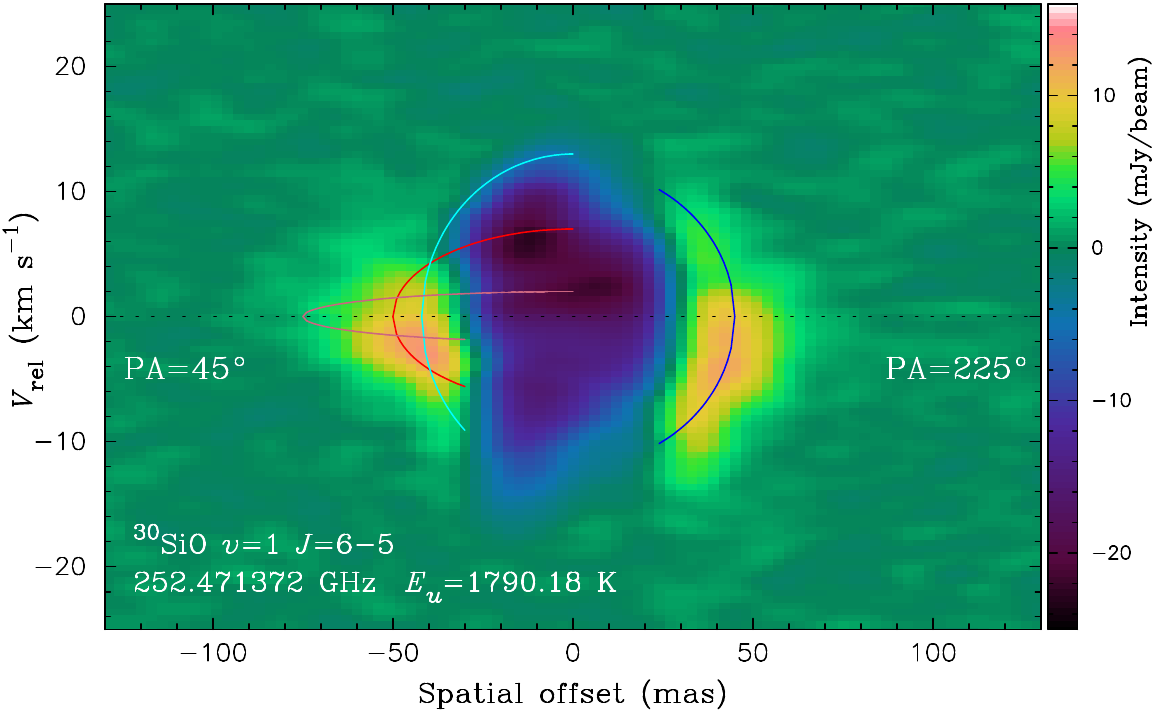}}}
\caption{
  Position-velocity diagram in the direction from PA = 45\degr\ to 225\degr\
 obtained from the continuum-subtracted channel maps of the \mbox{$^{30}$SiO}\ line
  ($\varv=1$, $J$ = 6 -- 5).
  The pink, red, and light blue curves at the negative spatial offsets
  (on the side of PA = 45\degr) represent the traces expected from
  shells infalling at velocities of 2~\mbox{km s$^{-1}$}\ at a radius of 75~mas,
  7~\mbox{km s$^{-1}$}\ at 50~mas, and
  13~\mbox{km s$^{-1}$}\ at 42~mas, respectively.
  The blue curve on the PA = 225\degr\ side shows the trace from an infalling
  or outflowing (although infalling is more plausible) shell with a velocity
  of 12~\mbox{km s$^{-1}$}\ at a radius of 45~mas. See Sect.~\ref{subsect_res_30sio_252}. 
}
\label{pvmap_30sio_252_angle135}
\end{figure}

\subsection{\mbox{$^{30}$SiO}\ $\varv = 1$, $J$ = 6 -- 5 at 252.471~GHz}
\label{subsect_res_30sio_252}

The continuum-subtracted channel maps of the \mbox{$^{30}$SiO}\ $\varv = 1$ 
$J$ = 6 -- 5 line at 252.471~GHz, shown in Fig.~\ref{channelmap_30sio_252}, 
reveal a complex outer atmosphere extending out to a
radius of $\sim$100~mas (i.e., $\sim$3.3~\mbox{$R_{\rm cont}$}\ = 4.8~\mbox{$R_{\star}$}).
It is much more prominent and more extended
than in the \mbox{$^{29}$SiO}\ $\varv=3$ and \mbox{$^{30}$SiO}\ $\varv=2$ lines,
because the upper level energy of 1790~K is lower than those of the two
SiO lines presented above.
Three prominent features can be recognized in Fig.~\ref{channelmap_30sio_252}:
a plume in the north-northwest (NNW), a tail in the south-southeast (SSE),
and the extended atmosphere elongated in the ENE--WSW direction. 

We extracted spatially resolved spectra at the same five positions
over the stellar disk as in the case of the \mbox{$^{29}$SiO}\ and \mbox{$^{30}$SiO}\ lines.
Spectra were also extracted off the limb of the star, 
at angular radii of 50~mas (1.7~\mbox{$R_{\rm cont}$}\ = 2.4~\mbox{$R_{\star}$}) and 
90~mas (3~\mbox{$R_{\rm cont}$}\ = 4.3~\mbox{$R_{\star}$}), 
at the same four position angles of $-15$\degr, 75\degr, 165\degr, and
255\degr\ (Fig.~\ref{whya_sio_combined}, middle and right columns).
The spectra obtained at PA = $-15$\degr\ and 165\degr\ correspond to the
plume and tail, respectively, while those obtained at PA = 75\degr\ and
255\degr\ probe the extended atmosphere.

The spectra extracted over the stellar disk (positions 0 to 4),
plotted in Fig.~\ref{whya_sio_combined} (left column, green lines),
show blueshifted absorption extending to $\mbox{$V_{\rm rel}$} \approx -15$~\mbox{km s$^{-1}$}, which is 
nearly identical to that of the lines of 
\mbox{$^{29}$SiO}\ $\varv=3$ and \mbox{$^{30}$SiO}\ $\varv=2$, because these SiO lines are
optically thick in their blue part 
as explained in Sect.~\ref{subsect_res_30sio_250}. 
On the other hand, 
the spectra of the \mbox{$^{30}$SiO}\ $\varv=1$ line exhibit 
very deep redshifted absorption with the strongest absorption at \mbox{$V_{\rm rel}$} =
3--7~\mbox{km s$^{-1}$}, extending to $\mbox{$V_{\rm rel}$} \approx +15$~\mbox{km s$^{-1}$}\ except for position 4 at 
PA = 225\degr.
This deep redshifted absorption is not seen
in other SiO lines described above.

In order to estimate where the redshifted deep absorption originates,
we examined the $p$-$V$ diagram extracted 
in the direction at PA = 45\degr --225\degr\ with a slit width of 5~mas
from the \mbox{$^{30}$SiO}\ $\varv=1$ data. 
We selected this direction because the redshifted absorption appears
to be the strongest in the NE (PA $\approx$ 45\degr) and the weakest
in the southwest (SW, PA $\approx$ 225\degr). 
The resulting $p$-$V$ diagram is shown in
Fig.~\ref{pvmap_30sio_252_angle135}. 
Assuming that the most prominent emission and the deepest absorption
on the NE side trace a 
coherent kinematic structure in the form of a partial spherical shell,
we may derive the extent of different shells by tracing various patterns
in the $p$-$V$ diagram.
First, the deepest absorption over the stellar disk and the brightest
emission off the stellar limb on the NE side can be fit with an infall
velocity of 7~\mbox{km s$^{-1}$}\ and a radius of 50~mas (1.7~\mbox{$R_{\rm cont}$}\ = 2.4~\mbox{$R_{\star}$}) 
as shown with the red line in
Fig.~\ref{pvmap_30sio_252_angle135}. 
Second, the most redshifted absorption over the stellar disk on the NE side 
can be traced by the infall velocity of up to 13~\mbox{km s$^{-1}$}\ and
a radius of 42~mas (1.4~\mbox{$R_{\rm cont}$}\ = 2.0~\mbox{$R_{\star}$})
as shown with the light blue line. 
Third, the most extended emission on the NE side near the systemic velocity 
can be fit with an infall
velocity of 2~\mbox{km s$^{-1}$}\ and a radius of 75~mas (2.5~\mbox{$R_{\rm cont}$}\ = 3.6~\mbox{$R_{\star}$})
as shown with the pink line. 
Fourth, the emission off the limb of the star on the SW side 
can be fit with a partial shell with a radius
of 45~mas (1.5~\mbox{$R_{\rm cont}$}\ = 2.2~\mbox{$R_{\star}$}) radially
expanding or infalling at 12~\mbox{km s$^{-1}$}\ as show by the blue line.

This interpretation means an accelerating infall on the NE side (PA = 45\degr) from $\sim$2~\mbox{km s$^{-1}$} at 3.6~\mbox{$R_{\star}$} to $\sim$7~\mbox{km s$^{-1}$} at 2.4~\mbox{$R_{\star}$} and then to $\sim$13~\mbox{km s$^{-1}$} at 2.0~\mbox{$R_{\star}$}. The strong redshifted absorption observed over the stellar disk suggests the infall from PA = $-15$\degr to $\sim$165\degr. Also, the infalling material seen off the limb of the stellar disk can lead to redshifted self-absorption due to colder material on the near side. The emission spectrum of the \mbox{$^{30}$SiO} \ $\varv=1$ line extracted on the NE side at 50~mas $\approx$ 1.7~\mbox{$R_{\rm cont}$} (Fig.~\ref{whya_sio_combined}i, position 6, green line) indeed shows that the red part is weaker or suppressed compared to the blue part. The same trend is also seen in the \mbox{$^{30}$SiO} \ $\varv=2$ (red line) and \mbox{Si$^{17}$O} \ $\varv=0$ (blue line in Fig.~\ref{whya_sio_combined}i, described in Sect.~\ref{subsect_res_si17o_250} below) lines. In light of these signatures of self-absorption in the emission spectra, it is possible that self-absorption contributes to the absorption spectra of the \mbox{$^{30}$SiO} \ $\varv=1$, \mbox{$^{30}$SiO} \ $\varv=2$, and \mbox{Si$^{17}$O} \ $\varv=0$ lines over the stellar disk. In the case of the \mbox{$^{29}$SiO} \ $\varv=3$ line, self-absorption cannot be confirmed, because while the blue part is very weak at position 6 (Fig.~\ref{whya_sio_combined}i, black line), the emission off the stellar limb is compact, and no emission is detected at other positions.

The fitting on the SW side in the $p$-$V$ diagram in
Fig.~\ref{pvmap_30sio_252_angle135} does not allow us to 
distinguish between infall and expansion. 
However, the emission spectrum of the \mbox{$^{30}$SiO}\ $\varv=1$ line at
position 8 (Fig.~\ref{whya_sio_combined}k, green line)
shows that the red part is
weak or suppressed -- the same signature of self-absorption due to
the infalling material on the near side. 
In addition, the vibrationally excited \mbox{H$_2$O}\ line at 268~GHz
($\varv_2=2$, $6_{5,2}$--$7_{4,3}$) reported in Paper~I shows mostly 
infall in the SW region at $\sim$2~\mbox{$R_{\rm cont}$}\ = 2.9~\mbox{$R_{\star}$}.
Therefore, it is more plausible
that the material in the SW region is also infalling. 

\begin{figure*}
\centering
\resizebox{17cm}{!}{\rotatebox{0}{\includegraphics{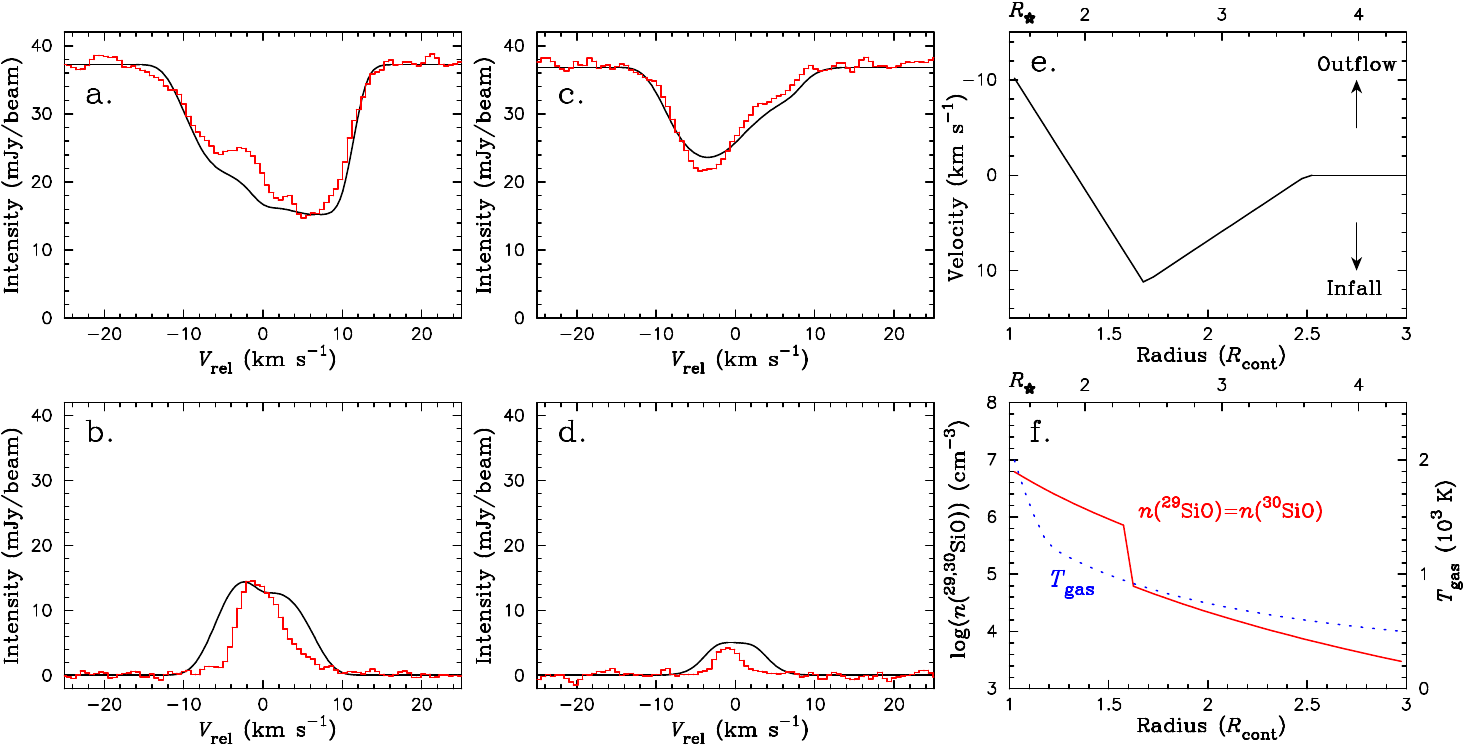}}}
\caption{
Spherical LTE model for the \mbox{$^{30}$SiO}\ $\varv=1$ $J$ = 6 -- 5 and
\mbox{$^{29}$SiO}\ $\varv=3$ $J$ = 6 -- 5 lines.
Panels {\bf a} and {\bf b} show a comparison between the model 
(black) and the spectra of the \mbox{$^{30}$SiO}\ $\varv=1$ line (red) observed
over the stellar disk and off the limb of the disk (positions 2 and 6
in Fig.~\ref{whya_sio_combined}), respectively. 
Panels {\bf c} and {\bf d} show a comparison for the \mbox{$^{29}$SiO}\ $\varv=3$ line
in the same manner.
The velocity profile of the model is shown in panel {\bf e}. 
Positive and negative velocities correspond to infall and outflow,
respectively. 
Panel {\bf f} shows the $^{29}$SiO number density (assumed to be equal to
that of $^{30}$SiO) on the left ordinate and the gas temperature on the
right ordinate.
In panels {\bf e} and {\bf f}, the radius is shown in the units of \mbox{$R_{\rm cont}$}\
(below) and \mbox{$R_{\star}$}\ (above).
}
\label{linemodelplot}
\end{figure*}

\subsection{Modeling of the \mbox{$^{30}$SiO}\ $\varv = 1$ and \mbox{$^{29}$SiO}\ $\varv = 3$ 
lines}
\label{subsect_res_sio_model}

We constructed a spherical model in LTE to examine whether the above 
picture can account for the observed spectra of the \mbox{$^{30}$SiO}\ $\varv=1$
line as well as the \mbox{$^{29}$SiO}\ $\varv=3$ line. We assumed LTE primarily
for simplicity, because full non-LTE radiative transfer modeling is
beyond the scope of this paper. Neither of the two lines 
shows nonthermal (suprathermal or maser) emission over the stellar disk,
which would be evidence of non-LTE. 
We refrain from modeling
the \mbox{$^{30}$SiO}\ $\varv=2$ (Sect.~\ref{subsect_res_30sio_250}) and
\mbox{Si$^{17}$O}\ $\varv=0$
lines (below in Sect.~\ref{subsect_res_si17o_250}) in the present work,
because they exhibit nonthermal emission over the stellar disk. 

The velocity $V$ (positive and negative $V$ corresponds to infall
and outflow, respectively) was assumed to increase from 0~\mbox{km s$^{-1}$}\ 
at 2.5~\mbox{$R_{\rm cont}$}\ to an infall velocity of $V_{\rm infall}^{\rm max}$
down at $R_{\rm infall}^{\rm max}$ with the decreasing radius, 
representing an accelerating infall toward the star.
On the other hand, the images of the \mbox{$^{29}$SiO}\ line at the negative
velocities show predominant blueshifted absorption over the stellar
disk, and the emission off the stellar limb extends only to $\sim$40~mas
($\sim$1.3~\mbox{$R_{\rm cont}$}), which suggests outward motion in the layers close
to the star. 
We assumed that the velocity
$V$ monotonically increases from the most negative value
$V_{\rm  outward}^{\rm max}$ at \mbox{$R_{\rm cont}$}\ to $V_{\rm infall}^{\rm max}$ at
$R_{\rm infall}^{\rm max}$, beyond which the velocity  
follows the aforementioned infall (see Fig.~\ref{linemodelplot}e). 

We assumed a power-law density profile for the $^{29}$SiO and
$^{30}$SiO number densities ($n(^{29}{\rm SiO})$ and $n(^{30}{\rm SiO})$),
adopting an isotope ratio of $^{29}$Si/$^{30}$Si = 1
based on the observed value of $0.99 \pm 0.05$ (Peng et al. \cite{peng13}).
The gas temperature ($T_{\rm gas}$) was approximated with a broken power-law
profile for the following reason.
We showed in Paper~I that the strong nonthermal emission of the
vibrationally excited \mbox{H$_2$O}\ line at 268~GHz over the stellar disk
suggests maser action, which requires a gas temperature lower than
$\sim$900~K and an \mbox{H$_2$O}\ density higher than $\sim \! 10^{4}$~\mbox{cm$^{-3}$},
based on the maser model of Gray et al. (\cite{gray16}). 
Given that the strong, spotty \mbox{H$_2$O}\ emission is seen at angular radii of
40--60~mas (2--3~\mbox{$R_{\star}$}\ $\approx$ 1.3--2~\mbox{$R_{\rm cont}$}), the temperature steeply
decreases from $\sim$2150~K in the continuum-forming layer at \mbox{$R_{\rm cont}$}\ to
$\la$900~K at 1.3--2~\mbox{$R_{\rm cont}$}. However, if such a steep gradient is
extended to larger radii, it leads to
an unrealistic temperature lower than $\sim$100~K already at 3.3~\mbox{$R_{\rm cont}$}\
(5~\mbox{$R_{\star}$}), and therefore, the temperature gradient should be shallower
beyond some radius.

The monochromatic intensity at a projected angular distance $p$ from the
stellar disk center (i.e., position observed in the plane of the sky)
was obtained as 
\[
I_{\nu}(p) = B_{\nu}(T_{\rm cont}) \, e^{-\tau_{\nu,p}^{\rm max}} \,
{\rm circ}(\mbox{$R_{\rm cont}$})
+ \!\!
\int B_{\nu}(T_{\rm gas}(r))\, e^{-\tau_{\nu,p}} \, d\tau_{\nu,p}\, ,
\]
where $B_{\nu}$ denotes the Planck function, and the function
${\rm circ}(\mbox{$R_{\rm cont}$})$ takes a value of 1 for $p \le \mbox{$R_{\rm cont}$}$ and 0
elsewhere. The integration is carried out along the line of sight at $p$,
and $\tau_{\nu,p}$ represents the optical depth along that line of sight. 
For $p \le \mbox{$R_{\rm cont}$}$, $\tau_{\nu,p}^{\rm max}$ corresponds to the optical
depth measured from the observer to the deepest layer.
We assumed a Gaussian line profile with a FWHM of a turbulent velocity of
2~\mbox{km s$^{-1}$}\ (Hoai et al. \cite{hoai22}) with the velocity field taken into
account in the observer's frame (e.g., Mihalas \cite{mihalas78}).

We focused on reproducing the spatially resolved spectra of the \mbox{$^{29}$SiO}\
$\varv=3$ and \mbox{$^{30}$SiO}\ $\varv=1$ lines extracted at
position 2 over the disk and position 6 off the stellar limb used in
the above $p$-$V$ diagram analysis. 
Figure~\ref{linemodelplot} shows the
best-fit model and comparison with the observed spectra.
The absorption and emission spectra of both lines are reasonably
reproduced, given the simplifications in our model and the complexity of
the object. 
The model is characterized with a gas temperature profile
$T_{\rm gas} \propto r^{-3}$ at $r \le 1.2$~\mbox{$R_{\rm cont}$}\ and
$T_{\rm gas} \propto r^{-1}$ at $r > 1.2$~\mbox{$R_{\rm cont}$}. 
The density profile is steep,
$n(^{29}{\rm SiO}) =  n(^{30}{\rm SiO}) \propto r^{-5}$. 
We found it necessary to introduce a step-like decrease   
in the SiO number density by a factor of 10 at 1.6~\mbox{$R_{\rm cont}$}\ 
to reproduce the observed \mbox{$^{30}$SiO}\ $\varv=1$ line for the following reason. 
A $^{29}$SiO (and $^{30}$SiO) number density of
$\sim 5\times10^6$~\mbox{cm$^{-3}$}\ at 1~\mbox{$R_{\rm cont}$}\ is needed so that the blue
part of the lines becomes optically thick as explained above. However,
this makes the emission of the \mbox{$^{29}$SiO}\ line too strong off the stellar
limb and the absorption of the \mbox{$^{30}$SiO}\ line too deep over the stellar
disk. The step-like decrease in the SiO number density makes the
\mbox{$^{30}$SiO}\ absorption less pronounced over the stellar disk and 
suppresses the \mbox{$^{29}$SiO}\ emission off the limb, while the emission of the
\mbox{$^{30}$SiO}\ line can still be seen because this line is intrinsically stronger
due to its lower $E_u$.

The derived velocity profile (Fig.~\ref{linemodelplot}e) shows
an accelerating infall toward the star,
starting from 0~\mbox{km s$^{-1}$}\ at 2.5~\mbox{$R_{\rm cont}$}\
(3.6~\mbox{$R_{\star}$}) and reaching $\sim$11~\mbox{km s$^{-1}$}\ at 1.7~\mbox{$R_{\rm cont}$}\ (2.5~\mbox{$R_{\star}$}). 
Then the infall decelerates at smaller radii and turns to outflow 
at 1.3~\mbox{$R_{\rm cont}$}\ ($\sim$2~\mbox{$R_{\star}$}). The outflow motion becomes stronger at
smaller radii and reaches $V = -10$~\mbox{km s$^{-1}$}\ at the deepest layer.
The infall at $>$1.7~\mbox{$R_{\rm cont}$}\ gives rise to the deep,
redshifted absorption over the stellar disk as observed,
while the steep velocity gradient ranging from $-10$ to $11$~\mbox{km s$^{-1}$}\ in the
inner region ($<$1.7~\mbox{$R_{\rm cont}$}) accounts for the observed broad absorption
profiles. 
The deep redshifted absorption does not appear in 
the spectrum of the \mbox{$^{29}$SiO}\ line. 
This is because the \mbox{$^{30}$SiO}\ $\varv=1$ line has a lower upper level energy,
and therefore, it is excited over a large range of the radial distance 
(Fig.~\ref{channelmap_30sio_252}), including the region where the infall
reaches $V \approx 10$~\mbox{km s$^{-1}$} (1.7~\mbox{$R_{\rm cont}$}\ $\approx$ 50~mas). 
The highly excited \mbox{$^{29}$SiO}\ $\varv=3$ line is confined to the innermost
region (Fig.~\ref{channelmap_29sio_251}), and therefore, it traces only
the motion in the deep layers. 
Our model is also in reasonable agreement with the infall at up to 
$\sim$15~\mbox{km s$^{-1}$}\ within
2--3~\mbox{$R_{\star}$}\ inferred from the nonthermal \mbox{H$_2$O}\ emission at 268~GHz in
Paper~I.

It is worth noting that the location of the deceleration of the infall
(toward the star) 
approximately coincides with that of the step-like change in the
SiO number density, although they were treated as independent parameters
of the fitting. 
The deceleration of the infall and the presence of the outflow 
in the layers below $\sim$1.7~\mbox{$R_{\rm cont}$}\ (2.5~\mbox{$R_{\star}$}) suggest that 
the density may increase in this region due to the compression of
the layers moving in the opposite directions. 
We also note that 
the gas temperature remains below $\sim$1000~K at $\ga$1.3~\mbox{$R_{\rm cont}$}\ 
($\ga$1.9~\mbox{$R_{\star}$}) due to the steep temperature gradient at the smallest
radii. 
The density enhancement and the low gas temperature 
provide a favorable condition for dust
formation and the radiative pumping of the 268~GHz \mbox{H$_2$O}\ maser as
discussed in Paper~I. 
In fact, the clumpy dust clouds start to form at $\sim$40~mas
($\sim$1.3~\mbox{$R_{\rm cont}$}\ = 1.9~\mbox{$R_{\star}$}).
The likely maser emission of the 268~GHz \mbox{H$_2$O}\ line is seen between
$\sim$40~mas (1.3~\mbox{$R_{\rm cont}$}\ = 1.9~\mbox{$R_{\star}$}) and 60~mas (2~\mbox{$R_{\rm cont}$}\ = 2.9~\mbox{$R_{\star}$}). 
These locations of the dust formation and \mbox{H$_2$O}\ emission 
coincide with the region of the expected density enhancement and the
gas temperatures below $\sim$1000~K in our model.

Our modeling suggests the
change in the SiO number density by a factor of $10 \pm 5$. 
The pulsation-driven dynamical model shows density jumps by a factor of
$\ga$10 within a few \mbox{$R_{\star}$}\ (H\"ofner et al. \cite{hoefner16}, 
\cite{hoefner22}) at the shock fronts. 
The 3D convective models also develop shocks with 
density jumps by a factor of $\ga$10 in the regions where infall
and outflow collide (H\"ofner \& Freytag \cite{hoefner19};
Freytag \& H\"ofner \cite{freytag23}). 
Therefore, the density jump suggested by our model can correspond to
the shocks generated by pulsation and/or convective motion.
However, the step-like change in the SiO number density can also be
interpreted as due to the depletion of Si onto dust grains.
Our current data and model do not allow us to distinguish between
two cases, density enhancement or Si depletion. It is possible that
both effects give rise to the step-like change in the SiO number
density. Observations of multiple SiO lines (including $^{28}$SiO lines) 
at different excitation
energies are necessary to examine the Si depletion onto dust grains. 

To obtain the fractional abundance of \mbox{$^{29}$SiO}\ and \mbox{$^{30}$SiO}\ with respect to
H$_2$, we estimated the H$_2$ number density as follows.
The 3D dynamical models (H\"ofner \& Freytag
\cite{hoefner19}; Freytag \& H\"ofner \cite{freytag23}) show 
densities of $10^{-12}$--$10^{-11}$~g~cm$^{-3}$ at $\sim$1~\mbox{$R_{\rm cont}$}\
$\approx$1.5~\mbox{$R_{\star}$}\ when spherically (i.e., directionally) averaged.
However, the models also produce local cells with densities of 
up to $\sim \!\! 10^{-10}$~g~cm$^{-3}$ at $\sim$1.5~\mbox{$R_{\star}$}.
As discussed above, the clumpy dust cloud formation and the 268~GHz \mbox{H$_2$O}\
emission suggest the presence of (infalling) dense, cool gas clumps. 
Therefore, if we adopt the locally enhanced density of
$\sim \!\! 10^{-10}$~g~cm$^{-3}$ from the 3D models, the H$_2$ number density is
estimated to be $2.3\times 10^{13}$~\mbox{cm$^{-3}$}. This translates into 
fractional \mbox{$^{29}$SiO}\ and \mbox{$^{30}$SiO}\ abundances of $2.2\times 10^{-7}$ at
1~\mbox{$R_{\rm cont}$}.
If we assume $^{28}$Si/$^{29}$Si $\ga$ 10 and
$^{28}$Si/$^{30}$Si $\ga$ 10 based on the results on other AGB stars
(Tsuji et al. \cite{tsuji94}; Decin et al. \cite{decin10};
De Beck \& Olofsson \cite{debeck18}), the (total) SiO abundance is
expected to be $\ga 2.2 \times 10^{-6}$.

We can compare this value with the SiO abundance expected
from the Si abundance. 
If we assume the solar Si abundance ($\log \varepsilon_{\rm Si} = 7.57$,
Deshmukh et al. \cite{deshmukh22}) for \mbox{W~Hya}, and all Si is
locked up in SiO, the maximum fractional SiO abundance with respect to
H$_2$ is $7.4 \times 10^{-5}$. 
Therefore, the SiO abundance estimated from our modeling does not
contradict the Si abundance. However, given the simplifications in the
model and the uncertainties in the silicon isotope ratios and the
estimate of the H$_2$ number density, 
we cannot draw a definitive conclusion about the fraction of Si
locked up in SiO or in dust grains. 

It should also be kept in mind that our model is only for estimating the
physical properties
of the circumstellar environment within $\sim$2.5~\mbox{$R_{\rm cont}$}, and therefore,
the steep density gradient is expected to become shallower and approach
$\propto r^{-2}$ at some radius beyond the region considered in our model.
Also, the steep temperature gradient corresponds to the dense, cool
pockets inferred from the likely maser emission of the 268~GHz \mbox{H$_2$O}\ line
and do not necessarily represent the global temperature profile on
a larger spatial scale such as $T \propto r^{-0.65}$ derived by
Khouri et al. (\cite{khouri14b}).

\begin{figure*}
\centering
\resizebox{17cm}{!}{\rotatebox{0}{\includegraphics{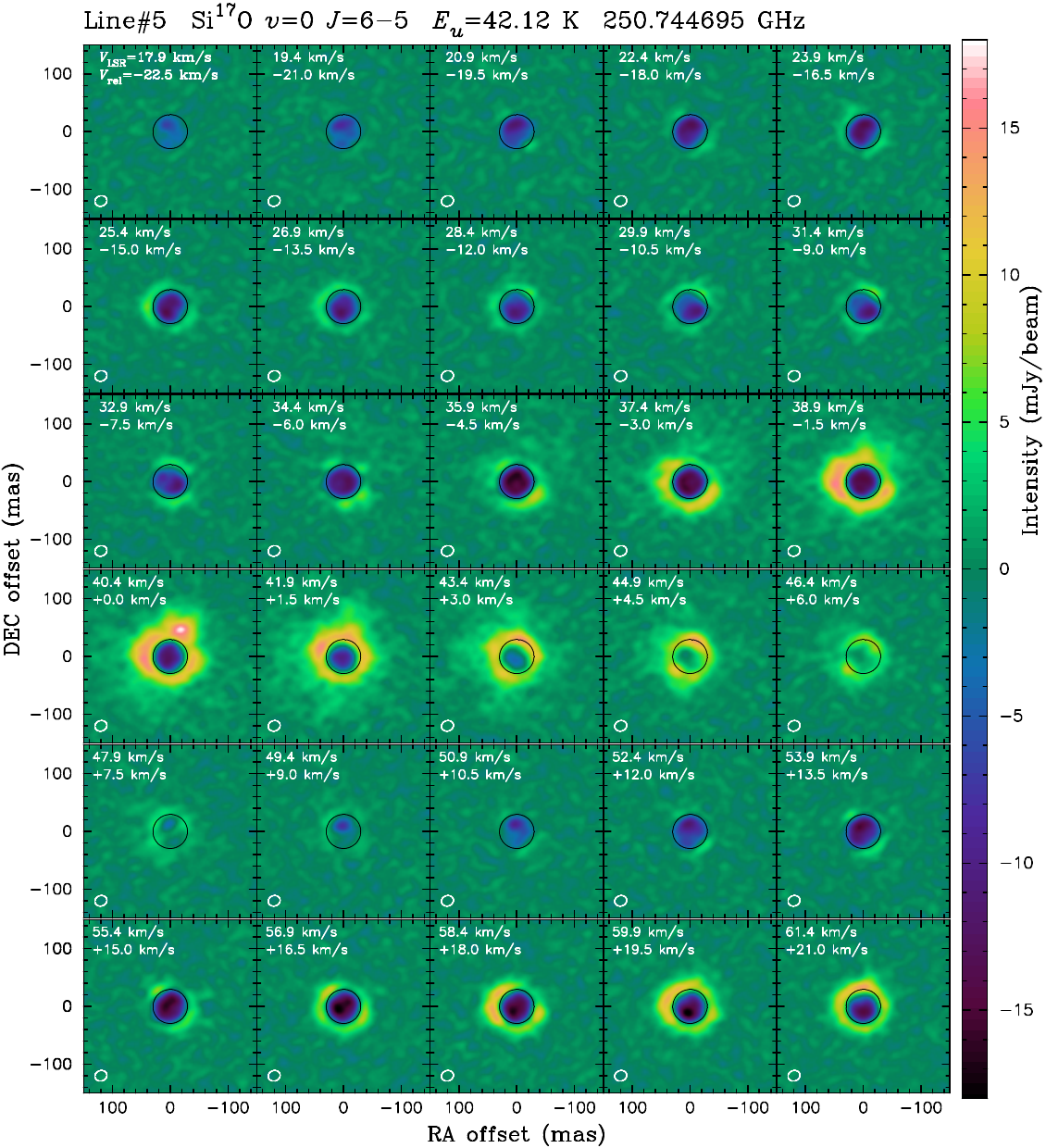}}}
\caption{
  Continuum-subtracted channel maps of \mbox{W~Hya}\ obtained in the \mbox{Si$^{17}$O}\
  $\varv=0$  $J$ = 6 -- 5 line at 250.744695~GHz, shown in the same manner as
  Fig.~\ref{channelmap_29sio_251}.
  The signals in the images at $\mbox{$V_{\rm rel}$} \ge 9$~\mbox{km s$^{-1}$}\ are due to the
  \mbox{$^{30}$SiO}\ $\varv=2$ $J$ = 6 -- 5 line. 
}
\label{channelmap_si17o_250}
\end{figure*}

\subsection{\mbox{Si$^{17}$O}\ $\varv = 0$, $J$ = 6 -- 5 at 250.745~GHz}
\label{subsect_res_si17o_250}

Figure~\ref{channelmap_si17o_250} shows the channel maps obtained for the
\mbox{Si$^{17}$O}\ line ($\varv = 0$, $J$ = 6 -- 5) at 250.745~GHz. 
The maps at velocities more redshifted than \mbox{$V_{\rm rel}$}\ = 9.0~\mbox{km s$^{-1}$}\ are dominated
by the \mbox{$^{30}$SiO}\ $\varv=2$ line described in Sect.~\ref{subsect_res_30sio_250}.
The \mbox{Si$^{17}$O}\ maps reveal the
same structures as found in the \mbox{$^{30}$SiO}\ $\varv=1$ line -- the NNW plume,
the SSE tail, and the extended atmosphere
elongated in the ENE--WSW direction. 
Given the upper level energy of mere 42~K, much lower than
the other SiO lines discussed above, this \mbox{Si$^{17}$O}\ line can be excited
across a very broad radial range, extending to cooler regions at larger
distances from the star.
The images obtained from \mbox{$V_{\rm rel}$}\ = 0 to 3~\mbox{km s$^{-1}$}\ show irregularly shaped
emission extending out to a radius of $\sim$120~mas (= 4~\mbox{$R_{\rm cont}$}\ =
5.8~\mbox{$R_{\star}$}).
Given the largest recoverable scale of 190~mas (i.e., a radius of 95~mas),
it is possible that some extended \mbox{Si$^{17}$O}\ emission is missing. 
It is worth noting that 
the channel maps at \mbox{$V_{\rm rel}$}\ = 1.5--6~\mbox{km s$^{-1}$}\ show that the emission 
goes inside the stellar limb, in a manner similar to the \mbox{$^{30}$SiO}\ $\varv=2$
line at 250.728~GHz discussed in Sect.~\ref{subsect_res_30sio_250}. 
The emission from the \mbox{Si$^{17}$O}\ line covers a larger fraction of
the stellar disk than the \mbox{$^{30}$SiO}\ $\varv=2$ line. 

The spectra obtained over the stellar disk (Fig.~\ref{whya_sio_combined},
left column, blue lines) show broad absorption, with the deepest absorption
centered at \mbox{$V_{\rm rel}$}\ = $-4$ to $-5$~\mbox{km s$^{-1}$}. 
In addition, the spectra obtained at position 0, 1, and 4 
show a narrow absorption feature at
$\mbox{$V_{\rm rel}$} \approx -5$~\mbox{km s$^{-1}$}, which is due to the absorption
by the material in the wind that has reached the terminal velocity. 
Takigawa et al. (\cite{takigawa17}) and Hoai et al. (\cite{hoai22}) also 
detected narrow features at $\sim \! -5$~\mbox{km s$^{-1}$}\ originating
from the wind at the terminal velocity in the \mbox{$^{29}$SiO}\ $\varv=0$
$J$ = 8 -- 7 line and in the CO $\varv=0$ $J$ = 3 -- 2 line (this latter
in emission, which Vlemmings et al. \cite{vlemmings21} interpret as
maser emission\footnote{Paper~I mistakenly mentions that Vlemmings et al.
(\cite{vlemmings21}) interpreted the CO $\varv=1$ $J$ = 3 -- 2 as masers.
It is the CO $\varv=0$ $J$ = 3 -- 2 line that 
Vlemmings et al. (\cite{vlemmings21}) present as maser emission.}).
The emission covering the stellar disk at \mbox{$V_{\rm rel}$}\ = 1.5--6~\mbox{km s$^{-1}$}\
appears as a bump at 4--5~\mbox{km s$^{-1}$}\ in the spectra except at position 0,
although not definitive at position 2.

Surprisingly, the deep redshifted absorption seen in the
\mbox{$^{30}$SiO}\ $\varv=1$ line is not observed in \mbox{Si$^{17}$O}\ $\varv=0$,
although we expect the cool \mbox{Si$^{17}$O}-bearing gas at outer radii to
produce similar, if not stronger, absorption.
This suggests that there is anomalous emission over the stellar disk.
Assuming that, under thermal excitation, the underlying \mbox{Si$^{17}$O}\ absorption
is as strong as that of the \mbox{$^{30}$SiO}\ ($\varv=1$), 
the excess \mbox{Si$^{17}$O}\ emission is approximately 20 mJy/beam to fill in the deep absorption
and result in the emission bump at \mbox{$V_{\rm rel}$}\ = 4 to 5~\mbox{km s$^{-1}$}. 
The emission intensity is comparable to the continuum intensity of
$\sim$40~mJy/beam of the stellar disk. 
However, similar emission over the stellar disk is not observed in the 
\mbox{$^{30}$SiO}\ $\varv=1$ line, indicating that
the \mbox{Si$^{17}$O}\ emission is nonthermal -- suprathermal or maser.
The nonthermal emission is redshifted by 4 to 5~\mbox{km s$^{-1}$}, as in the case of
the \mbox{$^{30}$SiO}\ $\varv=2$ line (Sect.~\ref{subsect_res_30sio_250}) and 
the likely maser emission of the $\varv_2=2$ \mbox{H$_2$O}\ line at 268~GHz reported in
Paper~I. 

To better visualize the spatial distribution of the suprathermal or maser 
emission, we created a set of difference maps 
by subtracting the channel maps of the \mbox{$^{30}$SiO}\ $\varv=1$ line
from those of the \mbox{Si$^{17}$O}\ $\varv=0$ line.
Figure~B.1 shows the resulting difference channel maps.
Only the maps between \mbox{$V_{\rm rel}$}\ = $-4.5$ and 7.5~\mbox{km s$^{-1}$}\ are presented,
because the data at velocities more blueshifted than
\mbox{$V_{\rm rel}$}\ = $-4.5$~\mbox{km s$^{-1}$}\ and more redshifted than
$7.5$~\mbox{km s$^{-1}$}\ are affected by the blend of the \mbox{H$_2$O}\
$\varv_2=2$ line (Sect.~\ref{subsect_res_h2o_250})
and the \mbox{$^{30}$SiO}\ $\varv=2$ line (Sect.~\ref{subsect_res_30sio_250}),
respectively. The figure reveals that the emission is covering
the stellar disk and slightly outside the limb at \mbox{$V_{\rm rel}$}\ = 0 to 7.5~\mbox{km s$^{-1}$}.
At the systemic velocity, the excess emission is
seen in the NNW, which corresponds to the strong emission in the NNW plume.
The images at \mbox{$V_{\rm rel}$}\ = 3.0 to 6.0~\mbox{km s$^{-1}$}\ show bright emission
in the western region of the stellar disk. 

The large extension of the \mbox{Si$^{17}$O}\ emission allows us to probe the
dynamics at larger distances than with other SiO lines.  
The spatially resolved spectra extracted off the limb of the
stellar disk at 90~mas
(Figs.~\ref{whya_sio_combined}m--\ref{whya_sio_combined}o, blue lines)
show that the blue part is noticeably weaker than the red part, which
can be interpreted as due to self-absorption in the outflow,
as in the case of the
HCN line discussed in Sect.~\ref{subsect_res_hcn_265}. 
The same trend is seen in the \mbox{$^{30}$SiO}\ $\varv=1$ line
(Figs.~\ref{whya_sio_combined}m--\ref{whya_sio_combined}o, green lines),
although the intensity is lower. 
These results suggest a global outflow at 90~mas (3~\mbox{$R_{\rm cont}$}\ = 4.3~\mbox{$R_{\star}$}).

We also find very diffuse emission that is difficult to recognize in the individual channel maps.
Figure~B.2 shows spectra derived by integrating the intensity in different annular regions. 
The emission clearly appears in the spectra from the annular regions up to an outer radius of 900~mas (30~\mbox{$R_{\rm cont}$} = 43~\mbox{$R_{\star}$}), ranging from \mbox{$V_{\rm rel}$}\ = $-10$ to $+10$~\mbox{km s$^{-1}$}. 
The spectrum obtained from the annular regions from 900~mas to 1200~mas (Fig.~B.2d) shows possible emission, but it cannot be considered to be definitive given its low S/N.
Moreover, given the largest recoverable scale of 190~mas, weaker and/or smoother emission should be missing, and the detected diffuse emission probably represents only a fraction of the real extended emission.
We do not find such diffuse emission in other SiO lines and other molecular lines described below except for the SO and HCN lines. 

We note that there is another absorption feature centered at \mbox{$V_{\rm rel}$}\ = $-16$ to $-18$~\mbox{km s$^{-1}$}\ in the spectra of the \mbox{Si$^{17}$O}\ $\varv=0$ line extracted over the stellar disk (Figs.~\ref{whya_sio_combined}b--\ref{whya_sio_combined}f), which is not seen in any other SiO lines. 
In the spectrum taken at PA = 225\degr, this absorption is less pronounced, and it blends with the absorption centered at \mbox{$V_{\rm rel}$}\ = $-4$ to $-5$~\mbox{km s$^{-1}$}, forming a single, very broad absorption trough. 
Paper~I tentatively identified it as the vibrationally excited \mbox{H$_2$O}\ line ($\varv_2 = 2$, $9_{2,8}$--$8_{3,5}$). We discuss this issue in more detail in Sect.~\ref{subsect_res_h2o_250}. 

\begin{figure*}
\centering
  \resizebox{17cm}{!}{\rotatebox{0}{\includegraphics{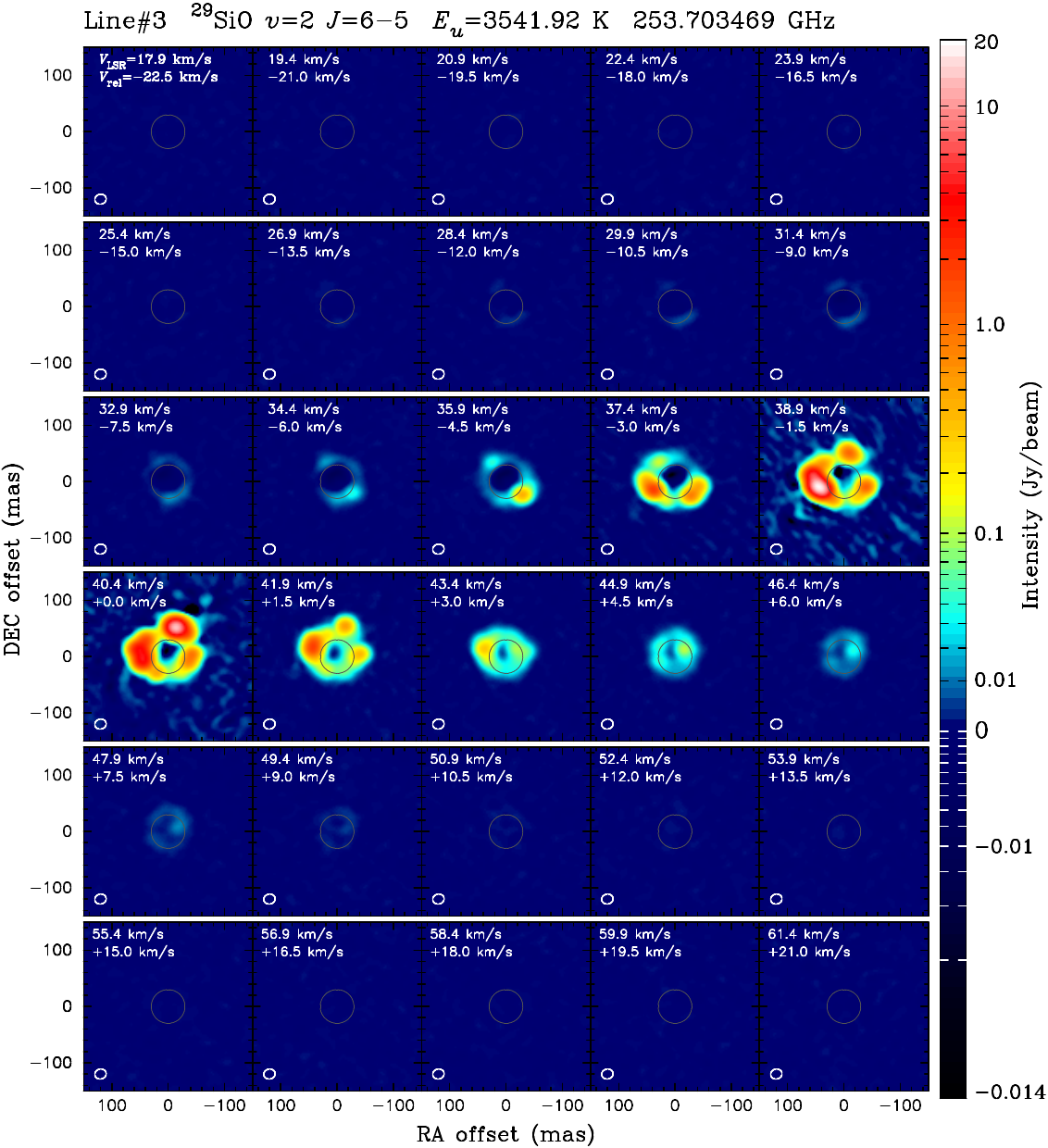}}}
\caption{
  Continuum-subtracted channel maps of \mbox{W~Hya}\ obtained for the \mbox{$^{29}$SiO}\ $\varv=2$ $J$ = 6 -- 5 line at 253.703469 GHz, shown in the same manner as Fig.~\ref{channelmap_29sio_251}, except that the unit of the color scale is different from other channel maps to show the strong maser emission and weaker emission over the stellar disk.
}
\label{channelmap_29sio_253}
\end{figure*}

\subsection{\mbox{$^{29}$SiO}\ $\varv = 2$, $J$ = 6 -- 5 at 253.703~GHz}
\label{subsect_res_29sio_253}

We detect strong masers in the \mbox{$^{29}$SiO}\ $\varv=2$ $J$ = 6 -- 5
line at 253.703469~GHz. The channel maps of the continuum-subtracted images, 
shown in Fig.~\ref{channelmap_29sio_253}, reveal three major maser clouds 
in the east, NNW, and WSW, where the peak intensity reaches $\sim$18~Jy/beam 
at \mbox{$V_{\rm rel}$}\ = $-1.5$ to 0~\mbox{km s$^{-1}$}.
As mentioned in Sect.~\ref{sect_obs}, 
the images at the spectral channels with the strong maser emission 
are limited by the dynamic range, resulting in RMS noise of 0.8, 2.0, 2.7,
and 0.7~mJy/beam at \mbox{$V_{\rm rel}$}\ = $-3.0$, $-1.5$, 0, and 1.5~\mbox{km s$^{-1}$}, respectively, 
while it is $\sim$0.5~mJy/beam at other spectral channels.
The peak intensity of 18~Jy/beam corresponds to a brightness temperature of
$10^6$~K. 
The three major maser clouds are located at a radius of $\sim$50~mas
($\sim$2.4~\mbox{$R_{\star}$}).
Reid \& Menten (\cite{reid07}) and Imai et al. (\cite{imai10}) show that
the 43~GHz $^{28}$SiO maser spots toward \mbox{W~Hya}\ are distributed
in a partial ring-like structure at radii of 40--50~mas,
which is in qualitative agreement with our ALMA data. 
The \mbox{$^{29}$SiO}\ masers imaged with ALMA probably consist of a number of
individual spots such as those revealed by radio observations at milliarcsecond
resolution (e.g., Imai et al. \cite{imai10}).

Figures~B.3g--B.3j
show the spatially resolved spectra extracted at four different positions over
the maser clouds as marked in Fig.~B.3a. 
The spectra (green lines) 
are narrow and centered approximately at the systemic velocity. 
Therefore, the masers are primarily tangentially
amplified as in the case for the 43~GHz SiO masers, suggesting that 
the \mbox{$^{29}$SiO}\ masers detected in our ALMA
data likely originate from the same radial range as the 43~GHz masers. 
However, a closer look at the line profiles extracted from the maser
clouds off the limb of the stellar disk reveals the presence
of broad components (the blue lines in Figs.~B.3g--B.3j). 

While the channel maps at \mbox{$V_{\rm rel}$}\ = $-10.5$ to $-1.5$~\mbox{km s$^{-1}$}\
show absorption over the stellar disk, emission is detected 
over the stellar disk between \mbox{$V_{\rm rel}$}\ = $-3$ to 9.0~\mbox{km s$^{-1}$}\
(i.e., excess emission on top of the continuum).
In particular, enhanced emission is seen
in the western region of the stellar disk at \mbox{$V_{\rm rel}$}\ = 3.0 to 7.5~\mbox{km s$^{-1}$}.
These western blobs are close to, but do not exactly coincide with, the 
excess emission in the difference maps of
\mbox{Si$^{17}$O}\ (Fig.~\ref{channelmap_si17o_250}) and the channel maps of 
\mbox{SO$_{2}$}\ line \#9 (Sect.~\ref{subsect_res_so2}, Fig.~\ref{channelmap_so2_252}). 
The spatially resolved spectra extracted over the stellar disk
(Figs.~B.3b--B.3f) 
show that the emission is broad, ranging from \mbox{$V_{\rm rel}$}\ = $-3$~\mbox{km s$^{-1}$}\ to
$\sim$10~\mbox{km s$^{-1}$}, and it reaches 20--60~mJy/beam at \mbox{$V_{\rm rel}$}\ = 0--4~\mbox{km s$^{-1}$}. 
This indicates overall infall of the material in front of the star
within $\sim$50~mas, which is consistent with our modeling of the
SiO lines (Fig.~\ref{linemodelplot}e).
Also plotted in Figs.~B.3b--B.3f
(orange lines) are the spectra of the vibrationally excited \mbox{H$_2$O}\ line
($\varv_2=2$, $6_{5,2}$--$7_{4,3}$) at 268~GHz (reported in Paper~I), 
extracted at the same positions as the \mbox{$^{29}$SiO}\ masers.
The spectra over the stellar disk show that 
both lines show strong, broad emission at the redshifted velocities
up to 8--12~\mbox{km s$^{-1}$}, while the emission is weak or it turns to absorption
at the blueshifted velocities.

The emission excess over the stellar disk of the \mbox{H$_2$O}\ and SiO lines 
cannot be accounted for by the tangential maser amplification. 
As discussed in Paper~I, the density and gas temperature fulfill the
conditions for the radiative pumping of the \mbox{H$_2$O}\ masers, and therefore,
the line may easily turn to masers. 
In the case of the \mbox{$^{29}$SiO}\ maser, which was first discovered in the 
RSG VY~CMa by Cernicharo \& Bujarrabal (\cite{cernicharo92}), 
Herpin \& Baudry (\cite{herpin00}) show that it can be
produced by the line overlaps between infrared ro-vibrational transitions
of \mbox{$^{29}$SiO}\ and $^{28}$SiO/\mbox{$^{30}$SiO}.
Humphreys et al. (\cite{humphreys02}) detected occasional 43~GHz SiO maser
spots over the stellar disk of the Mira star TX~Cam.
Their models including pulsation-induced velocity variations in the
atmosphere 
indeed predict occasional appearance of maser spots over the stellar disk
by the radial amplification, in addition to the predominantly tangentially
amplified masers outside the limb of the stellar disk. 
Therefore, the emission excess over the stellar disk 
suggests that the radial amplification is effective to some extent 
in the 253~GHz \mbox{$^{29}$SiO}\ line. 
Alternatively, it is also possible that the broad emission over the surface is 
suprathermal but not masers, while the intense emission off the stellar
disk limb is tangentially amplified masers.

\begin{table*}
\caption {
  Upper and lower level energies and the rest frequency
  of the \mbox{H$_2$O}\ $\varv_2=2$ $9_{2,8}$--$8_{3,5}$ line from different
  databases. 
}
\begin{center}
\begin{tabular}{l l l l}\hline
\noalign{\smallskip}
Reference & $E_u$ (cm$^{-1}$) & $E_l$ (cm$^{-1}$) & Rest frequency (GHz) \\ 
\hline
\noalign{\smallskip}
JPL: Yu et al. (\cite{yu12}) & 4268.24065 & 4259.87647 &
250.7517934 \\
& ($\pm 0.840 \times 10^{-3}$)  & ($\pm 0.870 \times 10^{-3}$)  &
($\pm 0.0362551$)  \\ 
JPL: Lanquetin et al. (\cite{lanquetin01}) & 4268.24144  & 
4259.87647  & 250.7754917 \\
& ($\pm 0.840 \times 10^{-3}$) & ($\pm 0.870 \times 10^{-3}$) &
($\pm 0.0362551$)\\
Furtenbacher et al. (\cite{furtenbacher20a}) & 4268.2407664560  & 
4259.8764187989  & 250.756834  \\
& ($\pm 4.021\times 10^{-5}$) & ($\pm 1.002\times 10^{-5}$) & ($\pm 0.001242$) \\
Furtenbacher et al. (\cite{furtenbacher20b}) & 4268.2408304879 &
4259.8764338705 & 250.758302  \\
& ($\pm 2.868 \times 10^{-5}$) & ($\pm 1.137 \times 10^{-5}$) &
($\pm 0.000925$) \\
\hline
\end{tabular}
\tablefoot{
The uncertainty in the rest frequency listed in the JPL
  Catalog (Yu et al. \cite{yu12}) is $\pm 0.0001264$. The uncertainty
  listed in the table is calculated from the sum of the uncertainty in each
  energy level in quadrature.   
}
\end{center}
\label{table_h2o_250}
\end{table*}

\subsection{Tentative identification of the vibrationally excited \mbox{H$_2$O}\
  line ($\varv_2 = 2$, $9_{2,8}$ -- $8_{3,5}$)}
\label{subsect_res_h2o_250}

The spectra of the \mbox{Si$^{17}$O}\ $\varv=0$ line extracted over the stellar disk
show an absorption feature at \mbox{$V_{\rm rel}$}\ = $-16$ to $-18$~\mbox{km s$^{-1}$}\ with respect to
the \mbox{Si$^{17}$O}\ line (Figs.~\ref{whya_sio_combined}b--\ref{whya_sio_combined}f). 
In Paper~I, we tentatively attributed this absorption 
to the vibrationally excited \mbox{H$_2$O}\ line 
($\varv_2 = 2$, $9_{2,8}$--$8_{3,5}$, $E_u = 6141$~K), at a rest frequency of
250.756834~GHz (Furtenbacher et al. \cite{furtenbacher20a}) or
250.758302~GHz (Furtenbacher et al. \cite{furtenbacher20b}).
We describe here details of this tentative identification. 

While the frequency of the 268~GHz \mbox{H$_2$O}\ line is very accurately
measured in the laboratory by Pearson et al. (\cite{pearson91})
with an error of 0.15~MHz as quoted in the JPL catalog,
the transition $\varv_2 = 2$, $9_{2,8}$--$8_{3,5}$ has never been measured
in the laboratory, and there are noticeable differences among
various calculations. Table~\ref{table_h2o_250} lists a few selected 
examples. The latest
JPL catalog for vibrationally excited \mbox{H$_2$O}\ reports a rest frequency of
250.7517934$\pm$0.0362551~GHz for this transition
(first row in Table~\ref{table_h2o_250}, Yu et al. \cite{yu12}).
If this frequency is adopted,
not only the absorption observed over the stellar disk but also
the emission off the disk limb is blueshifted by $\sim$6~\mbox{km s$^{-1}$}.
Given that the emission off the limb is expected to be roughly centered
around the systemic velocity in the case of globally spherical motion,
the systematic blueshift of the emission off the limb is difficult to
interpret. 

Recently, Furtenbacher et al. (\cite{furtenbacher20a}, b) produced the W2020
database of validated experimental transitions and empirical energy levels of
\mbox{H$_2$O}. If we adopt the empirical energies as shown in Table 3, the line
frequencies of the 251 GHz transition are  
$250.756834 \pm 0.001242$~GHz and $250.758302 \pm 0.000925$~GHz,
respectively. With these values, both the
observed absorption and emission appear approximately at the systemic
velocity as seen in Fig.~2 of Paper~I. 
As Table~\ref{table_h2o_250} shows, the differences in the
frequency originate from the energy of the upper state ($\varv_2=2$,
$J_{K_a K_c}=9_{2,8}$),
which is less constrained by experimental transitions than the lower
state. Moreover, the upper level energy is considered slightly less reliable
in the updated W2020 database (Furtenbacher et al. \cite{furtenbacher20b})
than its original version.
Our tentative identification of the \mbox{H$_2$O}\ $\varv_2 = 2$,
$9_{2,8}$--$8_{3,5}$ line suggests that the upper state
energies from Furtenbacher et al. (\cite{furtenbacher20a}, b)
agree with the observation.

We also considered the possibility that the absorption blueshifted by
16~\mbox{km s$^{-1}$}\ with respect to the \mbox{Si$^{17}$O}\ line rest frequency 
originates from \mbox{Si$^{17}$O}\ itself in an outflow, instead of the \mbox{H$_2$O}\ line.
However, we deem it to be unlikely, because 
the other SiO lines do not show such salient absorption blueshifted
by 16--18~\mbox{km s$^{-1}$}.
Another possibility is that it can be explained by 
a clumpy cloud outflowing at 16--18~\mbox{km s$^{-1}$}\ just covering the
stellar disk in front of the star in cool, far regions where the other SiO
lines with much higher upper level energies cannot be excited. 
However, such a scenario seems to be too fortuitous.

We also checked whether this absorption is due to molecules
other than \mbox{H$_2$O}\ and \mbox{Si$^{17}$O}.
On the Splatalogue line list, there is an $^{17}$OH line
($\nu_{\rm rest}$ = 250.757411~GHz),
which corresponds to a velocity shift of $-15.2$~\mbox{km s$^{-1}$}\
with respect to the \mbox{Si$^{17}$O}\ line. 
However, it is unlikely that the absorption at issue 
is attributed to $^{17}$OH for the following reason.
There is another $^{17}$OH line ($\nu_{\rm rest}$ = 250.742102~GHz),
which corresponds to a velocity shift of $3.1$~\mbox{km s$^{-1}$}\ with respect to the
\mbox{Si$^{17}$O}\ line. The upper level energy
of this line is the same as the one at $-15.2$~\mbox{km s$^{-1}$}\ but
the Einstein coefficient $A_{ul}$ is 48 times larger. Therefore,
we expect to see absorption at $3.1$~\mbox{km s$^{-1}$}\ 
stronger than that at $-15.2$~\mbox{km s$^{-1}$}.
However, the observed spectra show no absorption at $3.1$~\mbox{km s$^{-1}$}, 
which makes it very unlikely that the absorption centered at \mbox{$V_{\rm rel}$}\ =
$-16$~\mbox{km s$^{-1}$}\ is due to $^{17}$OH. 
Therefore, the vibrationally excited \mbox{H$_2$O}\ line ($\varv_2 = 2$,
$9_{2,8}$--$8_{3,5}$) is the most likely candidate for the absorption
feature.

\begin{figure*}
\centering
\resizebox{17cm}{!}{\rotatebox{0}{\includegraphics{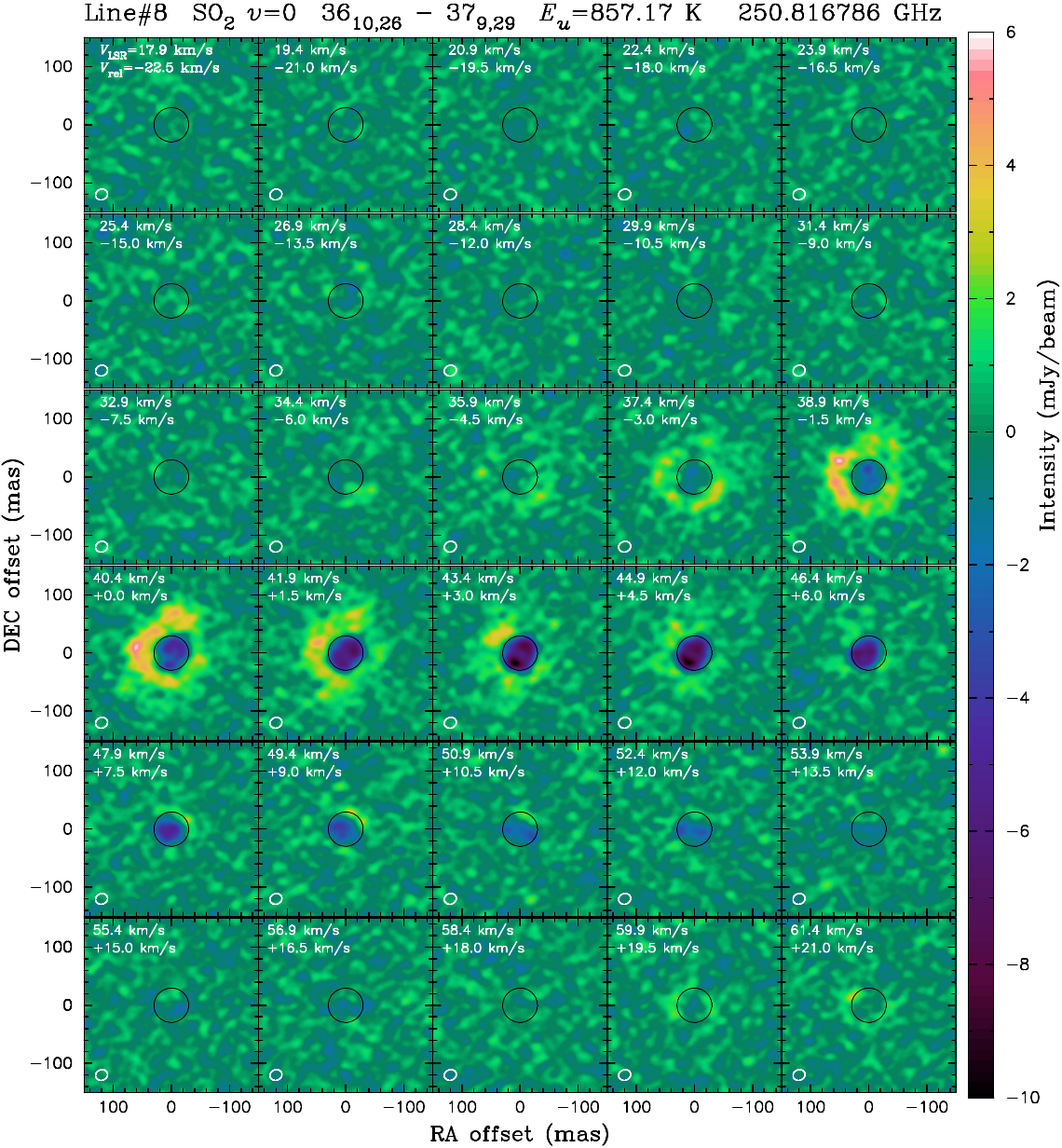}}}
\caption{
  Continuum-subtracted channel maps of \mbox{W~Hya}\ obtained in the \mbox{SO$_{2}$}\ line
  ($\varv=0$ $J_{K_a,K_c} = 36_{10,26} - 37_{9,29}$) at 250.816786~GHz, 
  shown in the same manner as Fig.~\ref{channelmap_29sio_251}.
}
\label{channelmap_so2_250}
\end{figure*}

\begin{figure*}
\centering
\resizebox{17cm}{!}{\rotatebox{0}{\includegraphics{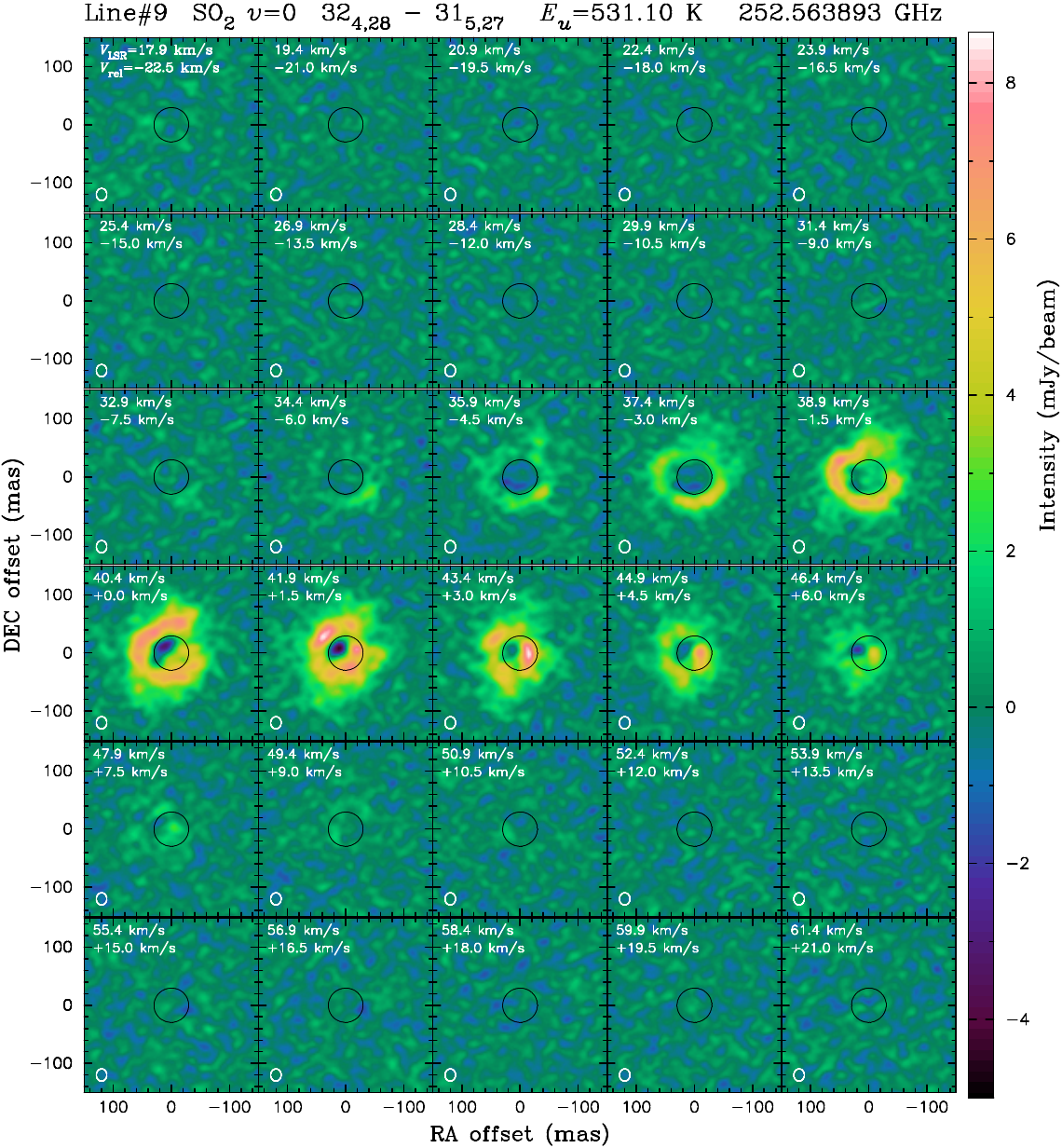}}}
\caption{
  Continuum-subtracted channel maps of \mbox{W~Hya}\ obtained in the \mbox{SO$_{2}$}\ line
  ($\varv=0$ $J_{K_a,K_c} = 32_{4,28} - 31_{5,27}$) at 252.563893~GHz, 
  shown in the same manner as Fig.~\ref{channelmap_29sio_251}.
}
\label{channelmap_so2_252}
\end{figure*}

\subsection{\mbox{SO$_{2}$}\ lines}
\label{subsect_res_so2}

Figures~\ref{channelmap_so2_250} and \ref{channelmap_so2_252} show
the channel maps of two representative \mbox{SO$_{2}$}\ lines in the ground
vibrational state $\varv =0$. 
The remaining \mbox{SO$_{2}$}\ lines identified in the current work are
presented in Figs.~C.1--C.19. 
For very weak \mbox{SO$_{2}$}\ lines, it is difficult to recognize signals in the
channel maps. However, the signatures of the \mbox{SO$_{2}$}\ lines 
could be detected in the integrated intensity maps, which are shown in
Fig.~C.20. 
The extended emission of the \mbox{SO$_{2}$}\ lines seen in the channel maps 
is similar to that observed
in the lines of \mbox{$^{30}$SiO}\ $\varv=1$ and \mbox{Si$^{17}$O}\ $\varv=0$. The NNW plume
extends to a radius of $\sim$100~mas ($\sim$3.3~\mbox{$R_{\rm cont}$}\ = 4.8~\mbox{$R_{\star}$}),
while the SSE tail can be seen up to a radius of $\sim$120~mas
($\sim$4~\mbox{$R_{\rm cont}$}\ = 5.8~\mbox{$R_{\star}$}). The extended atmosphere elongated in
the ENE--WSW direction has approximately the same angular size as
seen in the \mbox{$^{30}$SiO}\ $\varv=1$ and \mbox{Si$^{17}$O}\ $\varv=0$ lines. 

It is worth noting that the channel maps of the \mbox{SO$_{2}$}\ line \#9 
(Fig.~\ref{channelmap_so2_252}) reveal clear 
emission over the stellar disk at \mbox{$V_{\rm rel}$}\ = $-1.5$ to 7.5~\mbox{km s$^{-1}$}.
In particular, there is a salient emission spot in the western half
of the stellar disk, which appears to be the strongest at $3$~\mbox{km s$^{-1}$}\ and
reaches $\sim$8.5~mJy/beam above the continuum or 2588~K in brightness
temperature.  
Only in the NE region of the stellar disk does the line appear in absorption
as expected for thermal excitation.
The western emission is also close to the excess emission seen in the lines
of \mbox{Si$^{17}$O}\ (Sect.~\ref{subsect_res_si17o_250},
Fig.~\ref{channelmap_si17o_250}) and \mbox{$^{29}$SiO}\
(Sect.~\ref{subsect_res_29sio_253}, Fig.~\ref{channelmap_29sio_253}). 
The emission over the stellar disk on top of the continuum is seen in
the channel maps of other detected \mbox{SO$_{2}$}\ lines, but not always located in
the same regions of the disk. 

We also checked whether the \mbox{SO$_{2}$}\ lines 
in the ground and vibrationally excited states detected with low S/N 
show emission or absorption over the stellar disk by measuring the flux 
within a radius of 30~mas from the continuum-subtracted data. 
We also measured the flux in an annular region with an inner
and outer radius of 30 and 90~mas, respectively, to confirm the
detection of the emission. 
Figure~C.21 shows that the emission from the annular
region is indeed detected in all cases.
Only three lines (panels a, c, and k) show absorption over the stellar
disk (the blueshifted absorption in the $\varv=0$ $63_{6,58}$--$62_{7,55}$
line in panel g is due to the blend of the SO $\varv=0$ $N_J$ = $4_3$--$3_4$
line).
Six lines show emission (panels d, f, h, j, l, and o), and 
there are three lines that show both emission and absorption on the stellar
disk (panels m, n, and p).
The remaining lines (panels b, e, and i) show neither clear absorption nor
emission. 
As in the case of the \mbox{$^{30}$SiO}\ $\varv$ = 2 and \mbox{Si$^{17}$O}\ $\varv$ = 0 lines,
the emission on top of the continuum over the stellar disk is considered
to be of nonthermal origin, suprathermal or maser emission.
Our results reveal that the majority of the detected \mbox{SO$_{2}$}\ lines
exhibit nonthermal emission over the stellar disk.

Danilovich et al. (\cite{danilovich16}, \cite{danilovich20}) report the
detection of a number of \mbox{SO$_{2}$}\ lines toward a small sample of AGB
stars. More recently, Wallstr\"om et al. (\cite{wallstroem24}) present 
the detection of many
vibrationally excited \mbox{SO$_{2}$}\ lines in a larger sample of AGB stars
and RSGs. While the stellar disks were not spatially resolved by the
observations of Danilovich et al. (\cite{danilovich16}), their modeling of the
\mbox{W~Hya}\ data shows \mbox{SO$_{2}$}\ forms close to the star, down to $2
\times 10^{14}$~cm (= 6.7~\mbox{$R_{\star}$}) or even closer. The ALMA
observations of Danilovich et al. (\cite{danilovich20}) of the low mass-loss
rate (a few $\times \, 10^{-7}$~\mbox{$M_{\sun}$~yr$^{-1}$}) AGB star R~Dor,
whose mass-loss rate is low and similar to that of \mbox{W~Hya}, show that
\mbox{SO$_{2}$}\ forms close to the star, with its intensity peaking within
the continuum emission. Our high-angular-resolution ALMA images of
\mbox{W~Hya}\ confirm that \mbox{SO$_{2}$}\ indeed forms very close to the
star, down to $\sim$2~\mbox{$R_{\star}$}.

Also, Danilovich et al. (\cite{danilovich16}) present likely maser emission
of the vibrationally excited \mbox{SO$_{2}$}\ ($\varv_2=1$
$25_{4,22}$--$26_{1,25}$) line at 279.497~GHz in R~Dor. Vlemmings et
al. (\cite{vlemmings17}) report the detection of a vibrationally excited
\mbox{SO$_{2}$}\ line at 342.436~GHz ($\varv_2=1$, $23_{3,21}$--$23_{2,22}$,
$E_{u}$ = 1021~K) toward \mbox{W~Hya}\ and tentatively identify another
\mbox{SO$_{2}$}\ line at 345.017~GHz ($\varv_2 = 2$, $27_{3,25}$--$28_{0,28}$,
$E_{u}$ = 1832~K). Interestingly, the line at 345.017~GHz detected by
Vlemmings et al. (\cite{vlemmings17}) appears in emission over the stellar
disk with a FWHM of $7.3 \pm 0.6$~\mbox{km s$^{-1}$}, just as many of the
\mbox{SO$_{2}$}\ lines detected in our ALMA observations.

We carried out a population diagram analysis to estimate the excitation
temperature and column density of \mbox{SO$_{2}$}, assuming LTE and that the
lines are optically thin. While we detect 27 \mbox{SO$_{2}$}\ lines, many of
them show nonthermal emission over the stellar disk. Therefore, we selected
only the \mbox{SO$_{2}$}\ lines that show no emission on top of the continuum
over the stellar disk at any frequencies across the lines, which are the lines
\#8, \#12, \#13, \#19, \#22, and \#26 (we excluded the line \#33 shown in
panel k in Fig.~C.21 due to possible blend of the adjacent strong SO line
\#41). We measured the line flux of the emission off the stellar disk by
integrating across the line profile in two annular regions between a radius of
30 and 60~mas (i.e., between 1 and 2~\mbox{$R_{\rm cont}$}\ or $\sim$1.5 and
3.0~\mbox{$R_{\star}$}) and between 60 and 100~mas (between 2 and
3.3~\mbox{$R_{\rm cont}$} = $\sim$3 and 5~\mbox{$R_{\star}$}, approximately up
to the largest recoverable scale) to derive the properties in the innermost
and intermediate regions, respectively. The error in the absolute flux
calibration was assumed to be 10\% (ALMA Technical Handbook, Cortes et
al. \cite{cortes25}).

\begin{figure}
\resizebox{\hsize}{!}{\rotatebox{0}{\includegraphics{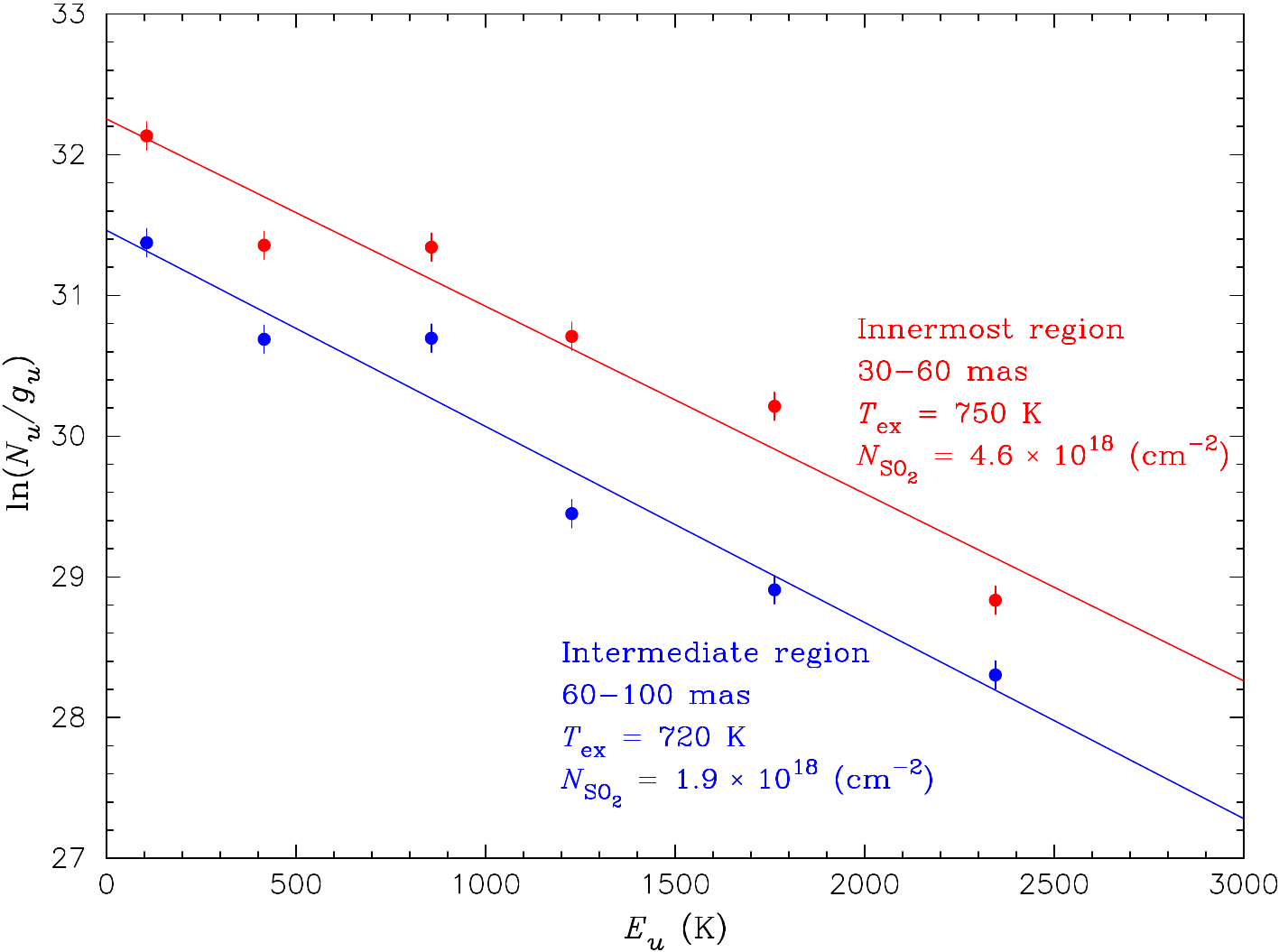}}}
\caption{
  Population diagram for the \mbox{SO$_{2}$}\ lines without signatures of
  nonthermal emission in the innermost and intermediate regions defined as
  annular areas between 30 and 60~mas ($\sim$1 and 2~\mbox{$R_{\rm cont}$})
  and between 60 and 100~mas ($\sim$2 and 3.3~\mbox{$R_{\rm cont}$}),
  respectively. The red and blue dots represent the measurements in the
  innermost and intermediate regions, respectively. The solid red and blue
  lines represent the fits in the respective regions. The excitation
  temperatures ($T_{\rm ex}$) and column densities ($N_{\rm SO_2}$) derived in
  two regions are also shown.
}
\label{pop_diagram_so2}
\end{figure}

  Figure~\ref{pop_diagram_so2} shows that the measurements are reasonably
  fit
  with an excitation temperature ($T_{\rm ex}$) of 750~K and a column density 
  (\mbox{$N_{\rm so_2}$}) of $4.6 \times 10^{18}$~\mbox{cm$^{-2}$}\ in the innermost region,
  while the data in the intermediate region result in \mbox{$T_{\rm ex}$}\ = 720~K and
  \mbox{$N_{\rm so_2}$}\ = $1.9 \times 10^{18}$~\mbox{cm$^{-2}$}. The uncertainties in the
  \mbox{$T_{\rm ex}$}\ and \mbox{$N_{\rm so_2}$}\ are $\pm$100~K and 46\% in the inner region and
  $\pm$90~K and 42\% in the intermediate region, respectively. 
  The fit in either region suggests that LTE
  is fairly valid for the selected lines.
  We also calculated the optical depth of the lines using Eq. (27) of
  Goldsmith \& Langer (\cite{goldsmith99}) and confirmed that the
  optical depth is $\la$0.7 for the selected lines, lending support to
  the population diagram analysis.

  To obtain the fractional abundance of \mbox{SO$_{2}$}\ with respect to H$_2$,
  we estimated the H$_2$ column density as follows.
  The pulsation+dust-driven wind models of Bladh et al. (\cite{bladh19}) show
  mass densities of $10^{-12}$--$10^{-11}$~g~cm$^{-3}$ at $\sim$1~\mbox{$R_{\rm cont}$}\
  $\approx$1.5~\mbox{$R_{\star}$}, which translate into H$_2$ number densities of
  $3\times 10^{11}$--$3\times 10^{12}$~\mbox{cm$^{-3}$}. Assuming a density
  profile $\propto r^{-2}$, we computed the H$_2$ column density along
  each line of sight and then the value averaged over two annular areas. 
  The averaged H$_2$ column density is
  $2.8\times10^{25}$--$2.8\times10^{26}$~\mbox{cm$^{-2}$}\ and 
  $1.6\times10^{25}$--$1.6\times10^{26}$~\mbox{cm$^{-2}$}\ in the innermost and
  intermediate regions, respectively. 
  Because the density profile
  in the wind acceleration region can be steeper than $\propto r^{-2}$,
  the H$_2$ column density can be lower. This sets lower limits of
  $\sim \! 2\times10^{-8}$ (innermost region) and 
  $\sim \! 1\times10^{-8}$ (intermediate region) 
  on the fractional \mbox{SO$_{2}$}\ abundance with respect to H$_2$.  

  The chemical equilibrium model of Ag\'undez et al. (\cite{agundez20})
  predicts a fractional \mbox{SO$_{2}$}\ abundance lower than a few $\times 10^{-10}$
  at 2--5~\mbox{$R_{\star}$}\ in an O-rich case, which is 50--100 times 
  lower than the lower limits derived above. 
  We note that if we adopt the temperature and pressure profiles of
  Ag\'undez et al. (\cite{agundez20})\footnote{Because the pressure and
  temperature profiles
  are shown only to 10~\mbox{$R_{\star}$}\ in Ag\'undez et al. (\cite{agundez20}),
  we extrapolated up to 150~\mbox{$R_{\star}$}\ with power-law decreases.},
  the H$_2$ column density averaged
  over the innermost annular area is $\sim \! 8 \times 10^{25}$~\mbox{cm$^{-2}$}, 
  which does not improve the disagreement.

 The nonequilibrium chemistry model of Gobrecht et al. (\cite{gobrecht16})
 predicts the formation of SO in weak shocks at low post-shock temperatures at
 $\ga$3~\mbox{$R_{\star}$}\ through S + OH $\rightarrow$ SO + H and SH + O
 $\rightarrow$ SO + H, and the subsequent formation of
 \mbox{SO$_{2}$}\ through SO + OH $\rightarrow$ \mbox{SO$_{2}$}\ + H. The
 predicted \mbox{SO$_{2}$}\ fractional abundance remains a few $\times
 10^{-9}$ at 3~\mbox{$R_{\star}$}\ during most of the phase, which is lower
 than our observationally estimated values, but it reaches $\sim \! 4 \times
 10^{-8}$ at $\sim$4~\mbox{$R_{\star}$}\ just before a shock passage. Because
 their model is not specifically constructed for \mbox{W~Hya}, this may be
 regarded as fair agreement with our observationally estimated value of $\ga 1
 \times 10^{-8}$ in the intermediate region. Still, given the phase dependence
 of the abundance, multi-epoch observations and full non-LTE radiative
 transfer modeling are necessary to test the nonequilibrium chemistry models.

\subsection{\mbox{$^{34}$SO$_{2}$}\ lines}
\label{subsect_res_34so2}

We detect four \mbox{$^{34}$SO$_{2}$}\ lines in the ground vibrational state,
and the channel maps are shown in Figs.~C.22--C.24 and Fig.~C.7 (this last
line is blended with an \mbox{SO$_{2}$}\ line). The morphology of the emission
is similar to the \mbox{SO$_{2}$}\ lines. The emission appears to be clumpy,
partially due to lower S/N than in other stronger \mbox{SO$_{2}$}\ lines. The
$32_{4,28}$--$32_{3,29}$ line appears in absorption over the stellar disk. In
other two lines, the S/N is not high enough to recognize the morphology of the
extended emission, although the channel maps at $-1.5$ to $+1.5$~\mbox{km
  s$^{-1}$}\ of the $15_{3,13}$--$15_{2,14}$ line (Fig.~C.23) seem to hint the
similar morphology as other \mbox{SO$_{2}$}\ lines. In the case of the
$9_{5,5}$--$10_{4,6}$ line (Fig.C.24), the emission appears at $-6$ to
0~\mbox{km s$^{-1}$}, indicating a blueshift of the
\mbox{$^{34}$SO$_{2}$}\ line. We note that there is a TiO$_2$ line
($27_{6,22}$--$27_{5,23}$, $E_u$ = 317~K) at 252.617051~GHz, which corresponds
to a velocity shift of $-2$~\mbox{km s$^{-1}$}\ with respect to the
\mbox{$^{34}$SO$_{2}$}\ line. However, as we discuss in
Sect.~\ref{subsect_res_tio_tio2}, TiO$_2$ emission is generally very weak, and
the contribution of this TiO$_2$ line is likely negligible compared to that of
the \mbox{$^{34}$SO$_{2}$}\ line.

\begin{figure*}
\centering
\resizebox{17cm}{!}{\rotatebox{0}{\includegraphics{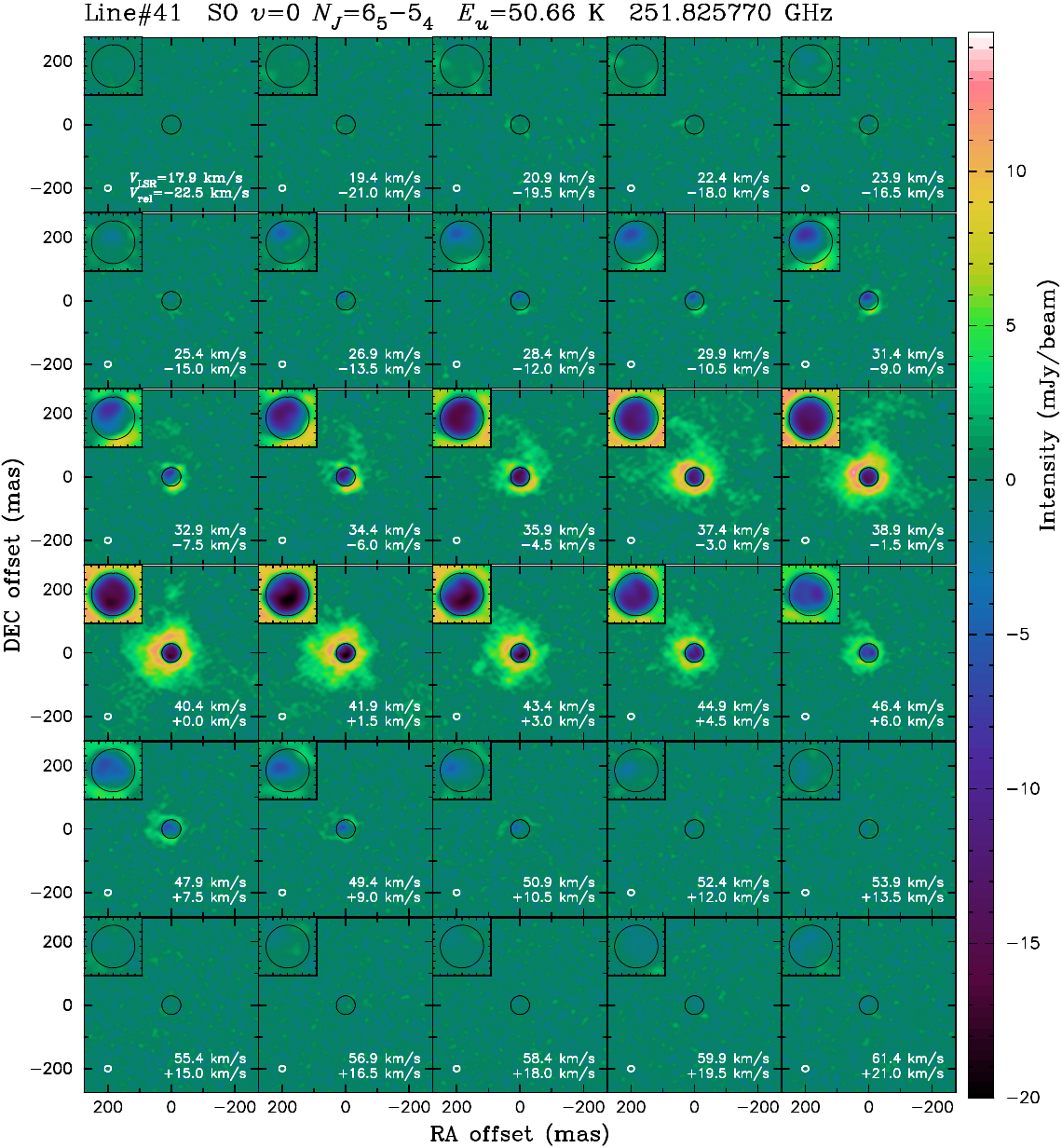}}}
\caption{
  Continuum-subtracted channel maps of \mbox{W~Hya}\ obtained in the SO line ($\varv=0$, $N_J = 6_{5} - 5_{4}$) at 251.825770~GHz, shown in the same manner as Fig.~\ref{channelmap_29sio_251} but for a field of view of 550$\times$550~mas. In addition, the insets show the enlarged view (80$\times$80~mas) of the signal over the stellar disk.
}
\label{channelmap_so_251}
\end{figure*}

\subsection{SO lines}
\label{subsect_res_so_251}

We detect two SO lines,  
$N_J = 6_5$--$5_4$ at 251.825770~GHz and $N_J = 4_3$--$3_4$ at 267.197746~GHz. 
We describe here only the first SO line (line \#41), because the other line is
of low S/N, and it is also blended with neighboring \mbox{SO$_{2}$}\ and
HCN lines (the channel maps of this weak SO line are 
shown in Fig.~C.25). 
Figure~\ref{channelmap_so_251} shows the channel maps obtained for the
$N_J = 6_5$--$5_4$ SO line (note that the field of view of 550$\times$550~mas
is larger than that of the SiO, \mbox{H$_2$O}, and \mbox{SO$_{2}$}\ lines presented above). 
We can recognize the NNW plume and the SSE tail,
and the images taken near the systemic velocity show
the extended atmosphere elongated in the ENE--WSW direction. 

However, the SO emission is more extended than that of the 
\mbox{SO$_{2}$}\ lines.
The NNW plume also extends 
farther to form an arm that winds and extends to the NE between \mbox{$V_{\rm rel}$}\ =
$-7.5$ to $-1.5$~\mbox{km s$^{-1}$}. This can be 
an outflow in front of the plane of the sky or an 
infall behind the plane of the sky. The arm extends relatively far
from the star to a radius of $\sim$150~mas ($\sim$7.2~\mbox{$R_{\star}$}). 
We deem an outflow to be more plausible, because
the \mbox{Si$^{17}$O}\ $\varv=0$ emission indicates a global outflow
already at 90~mas ($\sim$4.3~\mbox{$R_{\star}$}) as discussed in
Sect.~\ref{subsect_res_si17o_250}. 
The SO emission extends beyond 
the largest recoverable radius of $\sim$95~mas (corresponding to a half of
the largest recoverable scale of 190~mas). 
Nevertheless, emission at a radius larger than 95~mas can still be detected
as far as it is sufficiently compact in a form of plumes or filaments.
Therefore, the northern arm is likely real, 
while smooth extended emission may not be fully recovered.

Figure~C.26 (left column) shows 
the spatially resolved spectra extracted at the same five positions
over the stellar disk as for the SiO lines (as labeled in panel a). 
The spectra show narrow absorption features at \mbox{$V_{\rm rel}$}\ = $-5$~\mbox{km s$^{-1}$},
as seen in \mbox{Si$^{17}$O}, 
which originates from the wind that has reached the terminal velocity. 
The broad, overall absorption shows the deepest points at 
\mbox{$V_{\rm rel}$}\ = 2--3~\mbox{km s$^{-1}$}.
The absorption at positions 0, 1, and 2 is particularly broad, 
extending from \mbox{$V_{\rm rel}$}\ = $-20$ to 15~\mbox{km s$^{-1}$}.
On the other hand, 
the spectra extracted outside the stellar disk
(Figs.~C.26g--C.26j) 
show emission only to about $\pm$10~\mbox{km s$^{-1}$}, not up to $\pm$20~\mbox{km s$^{-1}$}. 
This implies that the high-velocity absorption wings originate from the region
close to the star as in the case of 
the high excitation \mbox{$^{29}$SiO}\ $\varv=3$ line that 
originates from the region within $\sim$2~\mbox{$R_{\star}$}. 
As mentioned in Sect.~\ref{subsect_res_29sio_251}, the escape velocity at
1.9~\mbox{$R_{\star}$}\ is 21~\mbox{km s$^{-1}$}. The most blueshifted wing of the SO line extending
to \mbox{$V_{\rm rel}$}\ $\approx \! -20$~\mbox{km s$^{-1}$}\ suggests the presence of material
almost at the escape velocity at $\sim$2~\mbox{$R_{\star}$}. 
The detection of SO close to the star is consistent with the analysis of 
\mbox{W~Hya}\ by Danilovich et al. (\cite{danilovich16}), who show that SO forms
$2 \times 10^{14}$~cm (= 6.7~\mbox{$R_{\star}$}) or even closer.

Danilovich et al. (\cite{danilovich20}) also report that the 
\mbox{SO$_{2}$}\ and
SO emission trace the same structures in the low mass-loss rate AGB star
R~Dor.
Our ALMA observations show that this seems to be the case 
for the NNW plume, SSE tail, and the elongated atmosphere 
of \mbox{W~Hya}, which can be seen in both \mbox{SO$_{2}$}\ and SO. 
We compare the channel map at the systemic
velocity of the SO line and \mbox{SO$_{2}$}\ line \#12, because this latter line
has an upper level energy of 106~K, which is the closest to that of
the SO line (51~K).
Figure~C.27 shows the azimuthally
averaged 1D intensity profiles of the SO and \mbox{SO$_{2}$}\ lines. 
The SO emission is more extended primarily due to its lower upper level
energy. Nevertheless, there is a hint that the intensity peak of the
\mbox{SO$_{2}$}\ line is slightly farther away than that of the SO line,
in spite of the higher upper level energy of the \mbox{SO$_{2}$}\ line,
suggesting that the \mbox{SO$_{2}$}\ abundance may peak slightly outside the SO
abundance.
We note that the full range of the excitation energy of \mbox{SO$_{2}$}\ is not
probed in the present work 
unlike the study of Danilovich et al. (\cite{danilovich20}),
and therefore, we do not have observational constraints farther out. 
High-angular-resolution ALMA imaging dedicated to \mbox{SO$_{2}$}\ and SO lines
(e.g., in Band 7), combined with a compact configuration or a single-dish
telescope, are necessary to investigate the detailed \mbox{SO$_{2}$}\ and SO distribution
at beyond $\sim$5~\mbox{$R_{\star}$}. 

Because we detect only two SO lines, and the $N_J$ = $4_3$--$3_4$ line
(\#42) is of low S/N and blended with the \mbox{SO$_{2}$}\ and HCN lines, we cannot
apply the population diagram analysis. We estimated the SO column density
in the same annular regions as used for the \mbox{SO$_{2}$}\ lines
(Sect.~\ref{subsect_res_so2}) from the $N_J$ = $6_5$--$5_4$ line (\#41)
alone by assuming the excitation temperatures derived from the
\mbox{SO$_{2}$}\ lines.
The resulting area-averaged SO column densities are
$7.0 \times 10^{17}$~\mbox{cm$^{-2}$}\ in the innermost region
between 30 and 60~mas (between 1 and 2~\mbox{$R_{\rm cont}$}\ $\approx$ 1.5--3~\mbox{$R_{\star}$})
and $3.0 \times 10^{17}$~\mbox{cm$^{-2}$}\ in the intermediate region
between 60 and 100~mas (between 2 and 3.3~\mbox{$R_{\rm cont}$}\ $\approx$ 3--5~\mbox{$R_{\star}$}). 
Adopting higher or lower excitation temperatures of 1000~K and 500~K
leads to an uncertainty in the SO column density by a factor of $\sim$2. 
These column densities translate into fractional SO
abundances of $\ga 3 \times 10^{-9}$ and $\ga \! 2 \times 10^{-9}$ in the
innermost and intermediate regions, respectively, if the H$_2$ column
densities estimated in Sect.~\ref{subsect_res_so2} are adopted.
It should also be kept in mind that we may
be missing the extended, smooth emission as mentioned above, which can
lead to even higher SO abundances.

The derived lower limits on the SO
abundance are in broad agreement with the chemical equilibrium model
of Ag\'undez et al. (\cite{agundez20}), which predicts abundances of
$10^{-8}$--$10^{-7}$ within $\sim$5~\mbox{$R_{\star}$}. 
The nonequilibrium model predicts the SO abundance of a few
$\times 10^{-8}$--$10^{-7}$ within $\sim$5~\mbox{$R_{\star}$}\ (Gobrecht et al.
\cite{gobrecht16}), which is also consistent with the observational lower
limits. 
As mentioned in Sect.~\ref{subsect_res_so2}, SO forms in weak shocks,
followed by the subsequent formation of \mbox{SO$_{2}$}\ at slightly larger radii 
in the nonequilibrium chemistry model
(see Fig.~4 of Gobrecht et al. \cite{gobrecht16}).
This is qualitatively consistent
with the 1D intensity peak positions of the SO and \mbox{SO$_{2}$}\ lines shown
in Fig.~C.27. 
However, Gobrecht et al. (\cite{gobrecht16}) mention that \mbox{SO$_{2}$}\ does not 
form efficiently by the reactions of SO and OH in their model,
which is why the predicted
\mbox{SO$_{2}$}\ abundance remains lower than the observed values in the literature
(see their Table~5).
Still, the observed overlap of the SO and \mbox{SO$_{2}$}\ emission suggests some 
chemical reactions that connect \mbox{SO$_{2}$}\ and SO. 

Vlemmings et al. (\cite{vlemmings11}) detected an SO line $5_5$--$4_4$ in
W~Hya at an angular resolution of $\sim$1\farcs5$\times$1\farcs0 
and found a slight offset of 0\farcs29 between the integrated red- and
blueshifted velocities (their Fig.~7). This led them to consider the possible
presence of a bipolar outflow with its axis approximately in the N-S
direction or a rotating disk aligned in the N-S direction, 
although they did not exclude that it might be due to complex kinematical
structures. Our channel maps obtained at a higher angular resolution
do not show signatures of a bipolar outflow or a rotating disk and
indicate that the emission within the largest recoverable scale
(i.e., within a radius of $\sim$95~mas = $\sim$4.6~\mbox{$R_{\star}$}) is primarily
influenced by the complex kinematical structures.
However, as presented below, the SO emission is much more extended
than the largest recoverable scale, and therefore, we cannot examine
the kinematical structures on larger spatial scales.

We also detect diffuse, patchy SO emission extending out to a radius of
$\sim$1\arcsec\ ($\sim$33~\mbox{$R_{\rm cont}$}\ = 48~\mbox{$R_{\star}$}),
as shown in Fig.~C.28. 
Because it is more extended than the largest recoverable scale of 190~mas
of our ALMA observations, the detected structures represent high intensity
regions embedded in more diffuse emission with lower intensities. 
The complex structures seen within 250~mas may give rise to these 
arc-like structures as they disperse at large distances from the star. 
Arc-like structures are detected in various AGB stars, and
they are often caused by binary companions 
(e.g., Maercker et al. \cite{maercker12}; Decin et al. \cite{decin20};
Siebert et al. \cite{siebert22}). 
However, 
there is no signature of a binary companion for \mbox{W~Hya}\ reported in the
literature. 
On the other hand, clumpy structures are observed
on the surface of some AGB stars
(e.g., \mbox{Vlemmings} et al. \cite{vlemmings19}, \cite{vlemmings24}; 
Velilla-Prieto et al. \cite{velilla-prieto23}),
probably associated with convective cells of sizes comparable to the
stellar radius.
The 3D dynamical models of Freytag \& H\"ofner (\cite{freytag23}) show
that the convective cells expand as the material is accelerated and form a
collection of partial arcs at radii greater than 3--4~\mbox{$R_{\star}$}. 
Therefore, the arc-like structures of the SO line detected up to
$\sim$1\arcsec\ toward \mbox{W~Hya}\ can be a consequence of this process.

\begin{figure*}
\centering
\resizebox{17cm}{!}{\rotatebox{0}{\includegraphics{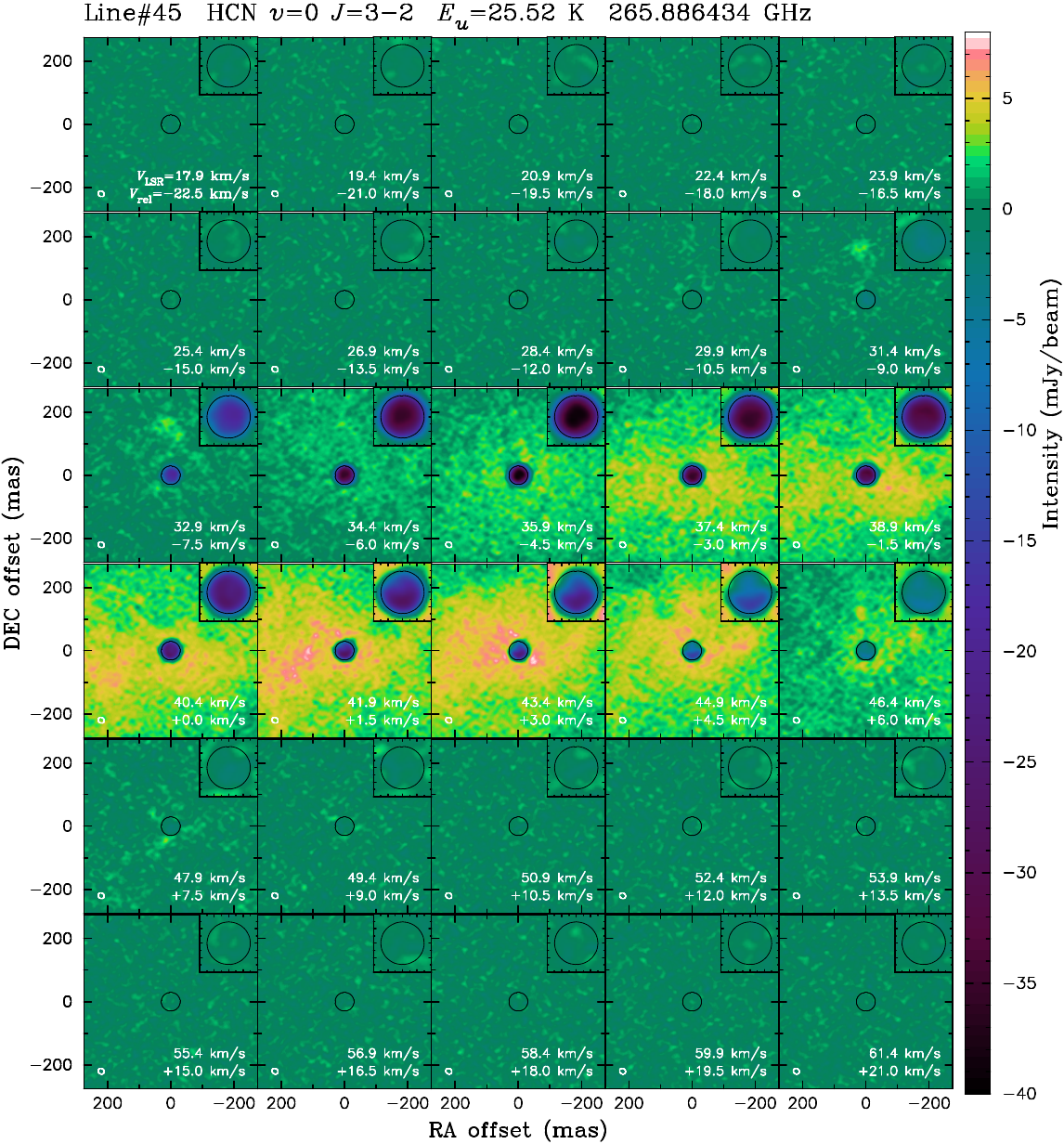}}}
\caption{
  Channel maps of \mbox{W~Hya}\ obtained in the HCN ($\varv=0$, $J$ = 3 -- 2)
  line at 265.886434~GHz,
  shown in the same manner as Fig.~\ref{channelmap_29sio_251} but 
  in a larger field of view of 550$\times$550~mas.  
  The insets show the enlarged view (80$\times$80~mas) of the signal over
  the stellar disk.
}
\label{channelmap_hcn_265}
\end{figure*}

\subsection{HCN lines}
\label{subsect_res_hcn_265}

\begin{figure}
\resizebox{\hsize}{!}{\rotatebox{0}{\includegraphics{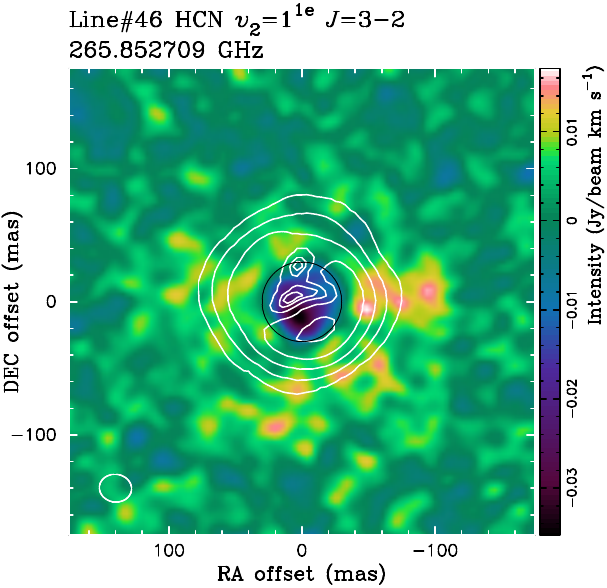}}}
\caption{
  Color map: Intensity map of the HCN ($\varv_2=1^{1e}$, $J$ = 3 -- 2) line at 265.852709~GHz integrated from \mbox{$V_{\rm rel}$}\ = $-10.5$ to 10.5~\mbox{km s$^{-1}$}. Contours: Polarized intensity obtained at 820~nm contemporaneously with our ALMA observations (Paper~I). The contours are drawn on a logarithmic scale at 10\%, 17\%, 30\%, 52\%, and 90\% of the peak value. The black circle represents the millimeter continuum stellar disk. The beam size of the HCN intensity map is shown in the lower left corner.
}
\label{integmap_hcn_v2_1_265}
\end{figure}

We detect three HCN lines in our ALMA observations:
$\varv=0$, $J$ = 3 -- 2 at 265.886434~GHz, $\varv_2=1^{1e}$, $J$ = 3 -- 2
at 265.852709~GHz, and $\varv_2=1^{1f}$, $J$ = 3 -- 2 at 267.199283~GHz.
The data of the ground-state transition show detailed morphological and
kinematical structures as described below.
On the other hand, S/N of $\la 5$ of the individual channel maps of
the two vibrationally excited lines is not sufficient to discuss 
such structures, as shown in
Figs.~C.25 and D.1 for completeness.
To better visualize the signals of the $\varv_2=1^{1e}$ line, 
we created an integrated intensity map from the channel maps reconstructed
with the natural weighting (Fig.~\ref{integmap_hcn_v2_1_265}).
The $\varv_2=1^{1f}$ line is blended with SO, and therefore, we do not
discuss it in the present work.

The ground-state HCN line 
consists of six hyperfine structure transitions. 
The three strongest components, $F$ = 4 -- 3, 3 -- 2, and 2 -- 1,
carry most of the intensity of the rotational transition
(Ahrens et al. \cite{ahrens02}), and they are closely spaced within a velocity
difference of 0.35~\mbox{km s$^{-1}$}, which is smaller than the velocity resolution of
our data.
Therefore, we treat HCN $\varv=0$ as a single line in the present work 
and adopt the rest frequency of 265.886434~GHz from CDMS, which also
corresponds to the weighted mean of the hyperfine components.

Figure~\ref{channelmap_hcn_265} shows the channel maps obtained for the
ground-state HCN line\footnote{
  We note that the grainy appearance of the maps near the systemic
  velocity is a result of nonuniform emission and missing short spacings
  that introduce artifacts on the size scales adopted
  in the CLEAN process.}.
Its emission is by far more extended than other SiO, \mbox{SO$_{2}$}, and SO lines. 
As discussed below, the entire HCN emission extends beyond the
field of view of 550$\times$550~mas of the figure, out to 
a radius of $\sim$1\arcsec. 
The extended emission is more prominent at the redshifted velocities than
at the blueshifted velocities. This is also seen in the spectra 
extracted at different positions, shown in Figs.~D.2g--D.2j,
which reveal that the emission is stronger in the red wing than in the blue
wing except for positions 11 and 25 at PA = 165\degr, where the line
profiles are approximately symmetric. 

The observations of the same HCN line toward \mbox{W~Hya}\ with a lower spatial 
resolution of 0\farcs55$\times$0\farcs40 using the Submillimeter Array (SMA) 
also show that the red wing of the emission
is more prominent than the blue wing (Muller et al. \cite{muller08}). 
As these authors explain, the observed asymmetry in the line profile 
is due to the self-absorption caused by colder
HCN gas in the outer region. 
In an expanding envelope, the blueshifted emission originates from the
hemisphere on the near side, while the redshifted emission originates
from the gas behind the plane of sky. Because the HCN emission is optically
thick as shown below,
we see the radiation from the layer with an optical depth of 1
along the line of sight. 
Assuming approximately spherical
expansion, at a given relative velocity, the blueshifted emission originates
from cooler HCN gas (and thus lower brightness temperature) at larger radii 
on the near-side hemisphere due to self-absorption,
while the redshifted emission originates from warmer gas (i.e.,
higher brightness temperature) at smaller radii on the far-side hemisphere.

The HCN spectra obtained at different angular distances at a given position
angle are nearly identical
(Figs.~D.2g--D.2j), which suggests that the line is optically thick. 
The spectrum at position 7 is somewhat different,
presumably because it originates from a bright clump. 
The spectra extracted over the stellar disk are shown in
Figs.~D.2b--D.2f.
Because the line is optically thick, the brightness temperature (shown 
in the right ordinate) allows us to estimate the gas temperature of the
HCN line formation. 
Given that the three strongest hyperfine structure components appear to
be a single component as mentioned above, 
the narrow, deep absorption seen at \mbox{$V_{\rm rel}$}\ = $-5$~\mbox{km s$^{-1}$}\ is attributed to the
wind that has reached its terminal
velocity\footnote{If there were contribution from 
the much weaker hyperfine components $F$ = 3 -- 3 and 2 -- 2, they would
appear at velocity shifts of 1.7 and $-2.4$~\mbox{km s$^{-1}$}\ with respect to the main
components at \mbox{$V_{\rm rel}$}\ = $-5$~\mbox{km s$^{-1}$},
respectively, i.e., at \mbox{$V_{\rm rel}$}\ = $-3.3$ and $-7.4$~\mbox{km s$^{-1}$}. 
However, there is no signature of absorption at these velocities. 
The contributions from the weaker hyperfine components are therefore
insignificant.}.
The brightness temperature at the deepest absorption at $-5$~\mbox{km s$^{-1}$}\ ranges
from $\sim$150 to $\sim$200~K.

We estimated the radius that corresponds to
these brightness temperatures, using the power-law temperature distributions 
derived for the circumstellar envelope of \mbox{W~Hya}.
Muller et al. (\cite{muller08}) derived 
$T_{\rm gas} (r) \, ({\rm K}) = 650 \, (10^{14}/r({\rm cm}))$ 
based on the radial intensity profile of the HCN ($\varv=0$, $J$ = 3 -- 2)
line alone. On the other hand, Khouri et al. (\cite{khouri14b}) obtained 
$T_{\rm gas} (r) \, ({\rm K})= 2500 \, (2.7\times10^{13}/r({\rm cm}))^{0.65}$, 
which was constrained by multiple single-dish CO spectra from $J$ = 1 -- 0 
($E_u$ = 5.5~K) to $J$ = 24 -- 23 (1656~K).
With the model of Muller et al. (\cite{muller08}),
the temperatures of 150--200~K correspond to radii of
(3.2--4.3) $\times$ $10^{14}$~cm (= 11--14~\mbox{$R_{\star}$}).
The escape velocity at these
distances is 8--9~\mbox{km s$^{-1}$}, which is higher than the terminal velocity of 5~\mbox{km s$^{-1}$}\
for a 1~\mbox{$M_{\sun}$}\ star. 
This contradicts that the deepest absorption dip originates from the wind that
has reached the terminal velocity.
On the other hand, if we adopt the temperature profile of Khouri et al.
(\cite{khouri14b}), the observed temperatures of 150--200~K
are reached at (1.3--2.0) $\times$ $10^{15}$~cm (43--67~\mbox{$R_{\star}$}) with a local
escape velocity of 3.7--4.5~\mbox{km s$^{-1}$}, lower than the wind terminal velocity 
(the local escape velocity drops to 5~\mbox{km s$^{-1}$}\ already at 
$\sim \!\!\! 1.06 \times 10^{15}$~cm = 35~\mbox{$R_{\star}$}, corresponding to 0\farcs72).
This lends support to the temperature profile from Khouri et al.
(\cite{khouri14b}). 
The line profiles show another two absorption dips at \mbox{$V_{\rm rel}$}\ = $-2$~\mbox{km s$^{-1}$}\
and 2--3~\mbox{km s$^{-1}$}\ (particularly at positions 0, 3, and 4). 
Both absorption dips indicate possible inhomogeneities along the line of
sight at radii where the gas is still gravitationally bound. 

A carbon-bearing molecule such as HCN is not expected in the oxygen-rich atmosphere of \mbox{W~Hya}\ in chemical equilibrium.
Gobrecht et al. (\cite{gobrecht16}) present the modeling of the formation of HCN in shocks in the dynamical atmosphere.
The aforementioned SMA observations by Muller et al. (\cite{muller08}) show that the HCN emission originates from the region within $\sim$10~\mbox{$R_{\star}$}, suggesting that HCN forms in shocks in the inner region of the circumstellar envelope. 
The optically thick HCN $\varv=0$ line is estimated to originate far away from the star at 43--67~\mbox{$R_{\star}$}, as discussed above, and therefore, it does not set a stricter constraint on the HCN formation region. 
On the other hand, the vibrationally excited ($\varv_2=1^{1e}$) HCN line is much weaker and not optically thick. 
The integrated intensity map of this line (Fig.~\ref{integmap_hcn_v2_1_265}) reveals the presence of HCN close to the star, down to $\sim$30~mas ($\sim$1.4~\mbox{$R_{\star}$}). 
The models of Gobrecht et al. (\cite{gobrecht16}) predict the formation of HCN at 1--2~\mbox{$R_{\star}$}. 
The detection of HCN close to the star lends strong support to the idea that HCN forms in shock-induced chemistry in the dynamical atmosphere.

  Gobrecht et al. (\cite{gobrecht16}) also show that
  the HCN abundance drastically decreases in the region where H$_2$ is
  depleted due to grain nucleation.
  The contours in Fig.~\ref{integmap_hcn_v2_1_265} show the 
  polarized intensity map obtained at 820~nm contemporaneously with our ALMA
  data using VLT/SPHERE-ZIMPOL, as reported in Paper~I. 
  The polarized intensity map shows the formation of clumpy dust clouds
  in the east, SW, and NNW. 
  While the emission of the HCN $\varv_2=1^{1e}$ line is present in the
  dust cloud formation regions, the HCN emission does not exactly
  coincide with the clumpy dust clouds and extends farther out to
  $\sim$100~mas without showing a signature of depletion. 
  This suggests that the dust formation does not completely deplete H$_2$ as
  predicted by the model. 
  However, it should be kept in mind that the models of Gobrecht et al.
  (\cite{gobrecht16}) correspond to IK~Tau, whose mass-loss rate is much
  higher than that of \mbox{W~Hya}. Also, the HCN emission can depend on the
  variability phase. Imaging of HCN lines with higher $J$ at different
  epochs at higher angular resolutions will be important to address this
  point.

An alternative model for the presence of HCN in oxygen-rich stars 
is that HCN forms as a product of photochemical reactions in the 
circumstellar envelope due to the penetration of interstellar UV photons 
(Charnley et al. \cite{charnley95}). 
In this model, HCN is not expected to form in high abundance at the innermost
radii, and the emission would be seen as a hollow shell.
Van de Sande et al. (\cite{vandesande18}) investigated the effects of
clumping and porosity on the circumstellar chemistry due to 
deeper penetration of the interstellar UV radiation and enhanced
clump density.
Their models show that 
the HCN abundance is generally more enhanced in the inner wind for
low mass-loss rates. The predicted HCN abundances increase
significantly beyond a few $\times 10^{14}$~cm ($\sim$10~\mbox{$R_{\star}$}) as
seen in their Figs. 4--5. 
Ag\'undez et al. (\cite{agundez10}) obtained similar results (their Fig.~3).
These results are not consistent with the presence of HCN close to the star, down to $\sim$1.4~\mbox{$R_{\star}$}, as indicated by the integrated intensity map of the HCN $\varv_2=1^{1e}$ line.

The modeling with the UV radiation from the AGB star (Van de Sande et al.
\cite{vandesande19}) also shows that HCN (and also SO in some cases) is
destroyed in the innermost
regions (their Figs.~4F and 4D), which does not agree with
the observation of the HCN $\varv_2=1^{1e}$ line.
The models with the UV radiation from a companion by 
Van de Sande et al. (\cite{vandesande22}) show the depletion 
of SO in the innermost region (their Fig. 3), which is not seen in our ALMA
data of the SO line (Fig.~\ref{channelmap_so_251}). In addition, we are not
aware of evidence of a companion of \mbox{W~Hya}\ in the literature and in our
ALMA data.
Therefore, the presence of HCN in the inner wind of W~Hya is better
explained by the shock-induced chemistry model rather than the
photochemistry model. 
However, photochemistry may play an important role for understanding
the very extended distribution of the HCN $\varv=0$ line as described below. 

Figure~D.3 shows the HCN emission in a larger
field of view of 2.4\arcsec $\times$ 2.4\arcsec. 
On a large scale, the emission morphology appears to be irregular with an
apparent elongation in the E-W direction. 
Across most of the channels with the extended emission, we can recognize 
clumpy, asymmetric emission within $\sim$0\farcs5 and complex filamentary,
spider-like structures up to a radius of $\sim$1\arcsec\
(corresponding to $\sim$33~\mbox{$R_{\rm cont}$}\ = 48~\mbox{$R_{\star}$}). 
However,
given the largest recoverable scale of $\sim$190~mas, 
a significant amount of flux density 
from diffuse, smooth emission on large angular scales is considered to be
missing. 
In fact, in the aforementioned SMA images obtained by Muller et al. 
(\cite{muller08}), 
which recovered $>$70\% of the total intensity within 28\arcsec,
the HCN emission appears to be smooth and roughly
spherically symmetric out to a radius of 1\farcs0--1\farcs5. 
Therefore, the filamentary features at the large radii simply represent
localized regions with relatively high intensities. For this reason,
we are unable to conclude from our images alone if there is indeed 
a global deviation from spherical symmetry of the HCN emission. 

Muller et al. (\cite{muller08}) detected a velocity gradient over
0\farcs5 in the NW-SE direction and suggested that it might be due to
a bipolar outflow or a rotating envelope. 
Our ALMA images do not show such a velocity gradient, and the
emission within the largest recoverable scale (i.e., within a radius of
$\sim$95~mas) is dominated by the inhomogeneous structures.
However, the missing diffuse, smooth emission in our ALMA data does not
allow us to study the kinematical structure on larger scales, as in the
case of the SO line discussed in Sect.~\ref{subsect_res_so_251}.

Hoai et al. (\cite{hoai22}) report enhanced emission in 
the NNW in \mbox{W~Hya}\
extending to an angular distance of $\sim$150~mas ($\sim$7.2~\mbox{$R_{\star}$})
at blueshifted velocities,
which led the authors to interpret it as an outflowing blob
in front of the star. 
With the distance of 98~pc and a maximum projected velocity of 5~\mbox{km s$^{-1}$},
which corresponds to the wind terminal velocity, 
the blob could have traveled an angular distance of 27~mas in 
2.5~years between the observations reported in Hoai et al. (\cite{hoai22})
and ours. Therefore, the blob should be located at between 150 and 177~mas
from the stellar center at the time of our ALMA observations,
depending on the projection effect. 

The intensity of most of the lines identified in our ALMA data is not
sufficient at the angular distance of 150~mas to detect the possible
presence of this blob. 
Only the 265~GHz HCN and 251~GHz SO (Sect.~\ref{subsect_res_so_251}) lines
show detectable emission
-- albeit still rather weak -- at 150--200~mas due to their intrinsically
very extended nature. 
The channel maps of the SO line show very faint emission at the tip of the
northern arm at an angular distance of 180~mas at $-7.5$~\mbox{km s$^{-1}$}\
(Fig.~\ref{channelmap_so_251}), and
the HCN channel maps show a blob in the north from \mbox{$V_{\rm rel}$}\ = $-9$ to
$-7.5$~\mbox{km s$^{-1}$}\ at angular distances of 150--200~mas
(Fig.~\ref{channelmap_hcn_265}). These features may correspond to the
blueshifted blob reported by Hoai et al. (\cite{hoai22}).

\subsection{AlO and AlOH}
\label{subsect_res_alo_aloh}

We detect the AlO $N$ = 7 -- 6 line near 267.9~GHz ($E_u$ = 51--52~K). 
Its channel maps, shown in Fig.~E.1, 
reveal that 
the signature of AlO is present across a wide velocity range from
\mbox{$V_{\rm rel}$}\ = $-30$ to $+30$~\mbox{km s$^{-1}$}\ due to the presence of multiple hyperfine
components of similar strengths between 267.912 and 267.961~GHz.
We also detect seven individual hyperfine components that belong to the
same transition outside the velocity range shown in the figure.
Although their
channel maps appear to be too noisy, the intensity maps integrated
from \mbox{$V_{\rm rel}$}\ = $-5$ to 5~\mbox{km s$^{-1}$}\ show
the signals of AlO more clearly. We present
the images of these weak hyperfine components in Fig.~E.2 
for completeness.

The AlO emission extends to a radius of 50--80~mas (1.7--2.7~\mbox{$R_{\rm cont}$}\ =
2.4--3.9~\mbox{$R_{\star}$}).
It is worth noting that the channel maps at all velocities between \mbox{$V_{\rm rel}$}\
= $-30$ and 30~\mbox{km s$^{-1}$}\ show emission -- albeit weak -- over the stellar disk.
No other line is found at the observed velocity
shifts in the Splatalogue, CDMS, JPL, and HITRAN catalogs. Therefore, 
we interpret the emission over the stellar disk as of the nonthermal origin 
-- suprathermal or weak maser action -- as in the case of the SiO and \mbox{SO$_{2}$}\
lines described above. 

\begin{figure}
\resizebox{\hsize}{!}{\rotatebox{0}{\includegraphics{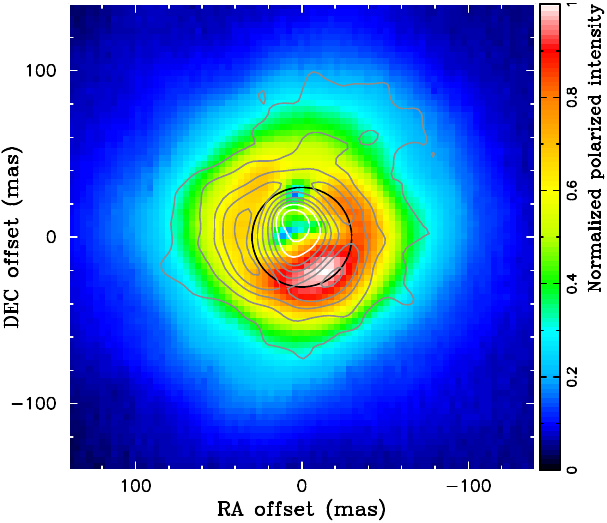}}}
\caption{
  Comparison between the continuum-subtracted integrated intensity map of
  the 267.93~GHz AlO line (contours) and
  the polarized intensity map (color map) obtained at 820~nm.
  The AlO intensity map was obtained by integrating across a velocity range
  from \mbox{$V_{\rm rel}$}\ = $-36$ to $36$~\mbox{km s$^{-1}$}. 
  The black circle represents the millimeter continuum diameter. 
  The gray contours correspond to 0.045 (3$\sigma$), 0.15 (10$\sigma$),
  0.3 (20$\sigma$), 0.45 (30$\sigma$), and 0.675 (45$\sigma$) mJy/beam~\mbox{km s$^{-1}$},
  while the white contours correspond to
  $-0.15$ ($-10 \sigma$) and $-0.045$ ($-$3$\sigma$) mJy/beam~\mbox{km s$^{-1}$}. 
  The polarized intensity is normalized to 1 at its maximum.
}
\label{zimpol_alo_267}
\end{figure}

The spatial extension of the AlO emission in our ALMA data is similar to
that of the AlO $N$ = 8 -- 9 line ($E_u$ $\approx$ 80~K) at 344.4~GHz 
observed in 2015 (Takigawa et al. \cite{takigawa17}). 
However, there are noticeable differences. 
Their AlO image shows an incomplete ring with a radius of $\sim$50~mas
with particularly strong emission in its northern part. 
The integrated intensity map from our data
(Figs.~E.3a and \ref{zimpol_alo_267}) also shows an
incomplete ring-like structure, but the enhanced emission is
now seen in the east and SW. Also, the emission in the NNW plume
did not appear in the AlO image of Takigawa et al. (\cite{takigawa17}). 
The difference in the morphology is likely due to time variations
in the density and temperature of the extended atmosphere 
and hence the excitation of the observed lines.

Figure~E.3 
shows spatially resolved spectra of AlO
extracted at the same five positions over the stellar disk as for
other lines described above as well as at four positions where the
emission is prominent, as marked in Fig.~E.3a. 
The spectra obtained over the stellar disk
(Figs.~E.3b--E.3f)
show not only weak absorption but also the aforementioned emission. 
The spectra extracted from the emission on and off the stellar limb show 
clearly the contribution of the hyperfine components
(Figs.~E.3g--E.3j). 
We can see differences in their contribution at different positions.
While the emission peaks of the individual hyperfine components are prominent
in the spectrum at position 28, those at positions 27 and 29 are less
pronounced. The individual hyperfine components are more smeared out
in the spectrum at position 30. 
This suggests that the AlO gas motion
along the line of sight includes various velocity components,
resulting in larger velocity shifts of each hyperfine component.

 Takigawa et al. (\cite{takigawa17}) show that
AlO emission observed in December 2015 is in good agreement with the
spatial distribution of clumpy dust clouds observed within 150 days
by Ohnaka et al. (\cite{ohnaka16}, \cite{ohnaka17}) using visible
polarimetric imaging.
In Fig.~\ref{zimpol_alo_267}, 
we compare our AlO integrated intensity map with the 820~nm polarized
intensity map taken contemporaneously with the ALMA data reported in Paper~I. 
The AlO emission and the
polarized light from the clumpy dust clouds are in excellent agreement.
The enhanced AlO emission in the east and SW corresponds to two major 
dust clouds seen in the polarized intensity map. The AlO emission 
in the NNW plume also has corresponding signals in the dust-scattered light. 
Therefore, our ALMA imaging confirms the co-location of the AlO emission
and the clumpy dust cloud formation as Takigawa et al. (\cite{takigawa17})
present, this time at another epoch based on the contemporaneous ALMA and
visible polarimetric observations. 
Furthermore, the spatial distributions of the AlO gas and dust clouds appear
to have changed in a similar manner over $\sim$3.5 years.
The agreement of the AlO emission and dust clouds 
indicates that the formation of both AlO and dust are promoted in the same
regions, suggesting that the enhanced AlO leads to higher dust formation. 
However, as Kami\'nski (\cite{kaminski19})
discusses, it is not clear if all AlO depletes onto dust, because gas-phase
chemistry may convert it to other Al-bearing molecules. 

We detect the AlOH $J$ = 8 -- 7 line at 251.79~GHz, and 
its channel maps are shown in Fig.~E.4 
(the maps obtained at $\mbox{$V_{\rm rel}$}\ < -3.0$~\mbox{km s$^{-1}$}\ show the signals from
TiO as described below).
The AlOH emission is much weaker and more compact than that of AlO.
The most prominent feature is the bright emission just off the eastern limb
of the stellar disk detected at \mbox{$V_{\rm rel}$}\ = $-3.0$ to 0.0~\mbox{km s$^{-1}$}. Weaker emission
can be seen in the SW at \mbox{$V_{\rm rel}$}\ = $-1.5$ and 0.0~\mbox{km s$^{-1}$}. 
The NNW plume and the extended elongated atmosphere are not detected. 
The AlOH line at 251.79~GHz consists of multiple hyperfine
components with the strongest ones located at 251.794442--251.794886~GHz.
These strong components are located so close to one another that
they cannot be spectrally resolved.
We do not detect two isolated, much weaker hyperfine components
redshifted by $\sim$8 and 10~\mbox{km s$^{-1}$}\ with respect to the strongest components. 
There are another two weak hyperfine components blueshifted by 5 and 10~\mbox{km s$^{-1}$}\
(blend with the TiO line described below) 
with respect to the strongest ones.
However, their contribution is considered to be negligible, because 
their transition probabilities are similar to the undetected
weak components at $\sim$8 and 10~\mbox{km s$^{-1}$}.

Wallstr\"om et al. (\cite{wallstroem24}) detected both AlO ($N$ = 7 -- 6) and
AlOH ($J$ = 8 -- 7) in three AGB stars (GY~Aql, IRC+10011, and U~Her). 
For U~Her, the authors derived a spatial extent of 4~\mbox{$R_{\star}$}\ and 
an upper limit of 2.8~\mbox{$R_{\star}$}\ for the AlOH and AlO emission, respectively. 
For GY~Aql and IRC+10011, the upper limit of the spatial extent of AlO and
AlOH was derived to be 1.2--1.3~\mbox{$R_{\star}$}\ and 3.7--4.0~\mbox{$R_{\star}$}, respectively.
These results suggest the emission of both AlO and AlOH 
originates from a region within $\sim$4~\mbox{$R_{\star}$}, which is in agreement with
our results of \mbox{W~Hya}.

Also, Decin et al. (\cite{decin17}) presented ALMA observations of the
AlO ($N$ = 9 -- 8) and AlOH ($J$ = 11 -- 10) lines at 345--346~GHz for 
R~Dor, whose mass-loss rate is similar to that of \mbox{W~Hya}\
(a few $\times 10^{-7}$~\mbox{$M_{\sun}$~yr$^{-1}$}). 
Most of the 345~GHz AlO emission in R~Dor originates
from a region within a radius of $\sim$150~mas ($\sim$6~\mbox{$R_{\star}$}\ with an
angular diameter of 51.2~mas from Ohnaka et al. \cite{ohnaka19}), and 
the AlOH emission is much weaker. 
This is qualitatively consistent with our results of \mbox{W~Hya}.
Kami\'nski et al. (\cite{kaminski16}) found that the excitation temperature 
of AlOH (1960~K) is significantly higher than that of AlO (330~K) in
the oxygen-rich AGB star $o$~Cet, which has a mass-loss rate comparable
to \mbox{W~Hya}. This can be interpreted if AlOH forms close to the star,
while the AlO emission originates from a more extended region as our ALMA
data of \mbox{W~Hya}\ show. 
However, it is possible that the AlO and AlOH emission in \mbox{W~Hya}\ depends
on the variability phase. It is important to observe the AlO
and AlOH emission at different
variability phases to better understand the variability-phase dependence
of these Al-bearing molecules.

\begin{figure}
\resizebox{\hsize}{!}{\rotatebox{0}{\includegraphics{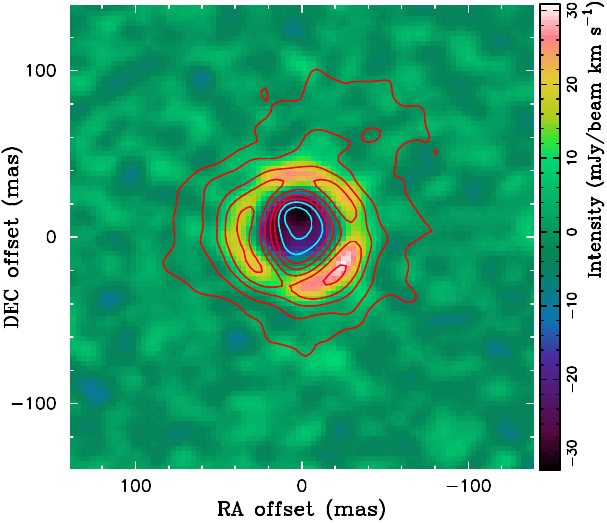}}}
\caption{
  Comparison of the continuum-subtracted integrated intensity maps of the
  lines of TiO ($\varv = 1$, $^3\Delta_1$, $J$ = 8 -- 7)
  and AlO ($N$ = 7 -- 6). The color map and contours represent 
  the TiO and AlO lines, respectively. The TiO intensity map was obtained
  by integrating from \mbox{$V_{\rm rel}$}\ = $-10$ to 6~\mbox{km s$^{-1}$}, avoiding the AlOH line. 
  The red and light blue contours
  correspond to the same positive and negative intensity levels, respectively,
  as shown in Fig.~\ref{zimpol_alo_267}. 
}
\label{integmap_tio_alo}
\end{figure}

\begin{figure}
\resizebox{\hsize}{!}{\rotatebox{0}{\includegraphics{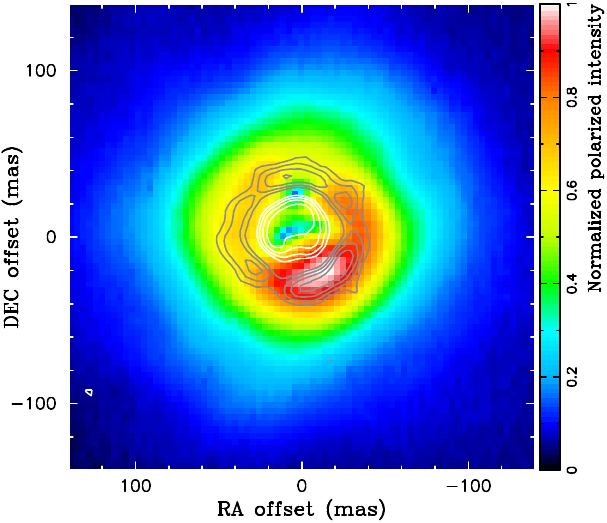}}}
\caption{
  Comparison of the continuum-subtracted integrated intensity maps of the 
  vibrationally excited TiO line
  ($\varv = 1$, $^3\Delta_1$, $J$ = 8 -- 7)
  and the polarized intensity map obtained at 820~nm. 
  The color map and contours represent 
  the polarized intensity and the TiO line,
  respectively. The gray and white contours
  correspond to positive values (0.01, 0.015, 0.02, and 0.025~mJy/beam \mbox{km s$^{-1}$})
  and negative values ($-0.025$, $-0.02$, $-0.015$, and
  $-0.01$~mJy/beam \mbox{km s$^{-1}$}), respectively. 
}
\label{integmap_tio_zimpol}
\end{figure}

\subsection{TiO and \mbox{TiO$_{2}$}}
\label{subsect_res_tio_tio2}

The TiO and \mbox{TiO$_{2}$}\ molecules are considered to be potentially important
for dust formation, because they can provide nucleation seeds, on which
other grain species can grow. 
We detect a vibrationally excited TiO line
($\varv = 1$, $^3\Delta_1$, $J$ = 8 -- 7,
$E_u$ = 1492~K) at 251.802917~GHz, a $^{50}$TiO line 
($\varv = 0$, $^3\Delta_2$, $J$ = 8 -- 7, $E_u$ = 192~K) at 
253.591920~GHz, and a $^{49}$TiO line ($\varv=0$, $^3\Delta_1$, $J$ = 8 -- 7,
$E_u$ = 53~K) at 251.957928~GHz. 
Figure~F.1 
shows the channel maps for the TiO
$\varv=1$ line at 251.802917~GHz, 
which overlaps with the AlOH line described above at $\mbox{$V_{\rm rel}$} \ga 6$~\mbox{km s$^{-1}$}. 
The emission is ring-like just outside the limb of the stellar
disk, extending only to $\sim$40~mas (1.3~\mbox{$R_{\rm cont}$}\ = 1.9~\mbox{$R_{\star}$}).
Enhanced emission can be seen in the
NNW and SE (\mbox{$V_{\rm rel}$}\ = $+1.5$ to $+3.0$~\mbox{km s$^{-1}$}) as well as in the SW
(\mbox{$V_{\rm rel}$}\ = $-3$ to 0.0~\mbox{km s$^{-1}$}).
The channel maps of the $^{50}$TiO line, shown in Fig.~F.2, 
are of a lower S/N ratio of $\sim$5.
Nevertheless, it can be recognized that the emission features are located
along or just outside the limb of the stellar disk. 
The channel maps of the $^{49}$TiO line are of even lower S/N, and therefore,
we only present the integrated intensity map in Fig.~F.3 
for completeness.

TiO emission has been observed at a much greater radius in the visible. 
The continuum-subtracted image of \mbox{W~Hya}\ obtained at 717~nm 
(Ohnaka et al. \cite{ohnaka17}), corresponding to the $\gamma$ bands of the 
TiO's electronic transition, shows emission extending to 100--150~mas about
four years prior to the ALMA observations. Our TiO $\varv=1$ emission is
compact because the transition probability is much smaller than those of
the $\gamma$ bands in the visible. 

Figure~\ref{integmap_tio_alo} shows a comparison of the integrated intensity
maps of the TiO line and the AlO $N$ = 7 -- 6 line. The positions of the
enhanced TiO emission agree very well with that of the AlO line with the three
major emission clumps in the east, SW, and NNW, although the extended
NNW plume is absent in the TiO line. 
Figure~\ref{integmap_tio_zimpol} shows a comparison of the TiO integrated
intensity map and the polarized intensity map at 820~nm reported in Paper~I. 
The TiO emission overlaps well with the scattered light from the
clumpy dust clouds. 
We discuss the interpretation of the overlap in Sect.~\ref{sect_discuss},
together with other molecular lines. 

We detect five \mbox{TiO$_{2}$}\ lines. While the emission is not sufficiently strong
to be seen in the channel maps, the integrated
intensity maps confirm the detection as shown in Fig.~F.4. 
The image of the \mbox{TiO$_{2}$}\ line \#54 at 251.708056~GHz (Fig.~F.4a) 
shows the emission most clearly. It extends up to $\sim$100~mas (4.8~\mbox{$R_{\star}$}) 
in the north and east. The line \#51 also shows emission to $\sim$80~mas,
although the emission is less clear. 
This shows that \mbox{TiO$_{2}$}\ is not entirely lock up in dust grains up to
4--5~\mbox{$R_{\star}$}. 
The other \mbox{TiO$_{2}$}\ lines show emission just outside the limb of the stellar
disk due to their higher excitation energies or a smaller Einstein
coefficient.

We stacked the images of five detected \mbox{TiO$_{2}$}\ lines to better visualize
the spatial distribution of \mbox{TiO$_{2}$}. Figure~F.5 shows
a comparison of the stacked \mbox{TiO$_{2}$}\ image (color map) and the 251.80~GHz TiO
line (contours). The \mbox{TiO$_{2}$}\ emission coincides rather well, if not perfectly,
with the TiO image, with the most pronounced emission located in the east,
SW, and NW. As discussed in Sect.~\ref{sect_discuss}, 
\mbox{TiO$_{2}$}\ is another molecular species, together with some other molecules, 
whose formation is enhanced in the regions 
where shock-induced chemistry is active and dust formation takes place. 

Kami\'nski et al. (\cite{kaminski17}) detect \mbox{TiO$_{2}$}\ toward $o$~Cet
and 
find that the \mbox{TiO$_{2}$}\ emission extends to 2.7--5.5~\mbox{$R_{\star}$}, more extended
than that of TiO ($<$4~\mbox{$R_{\star}$}). Decin et al. (\cite{decin18})
present similar results for R~Dor and IK~Tau. The observed extension of the 
\mbox{TiO$_{2}$}\ lines in \mbox{W~Hya}\ is in qualitative agreement with these results. 
However, we cannot conclude that the abundance distribution of \mbox{TiO$_{2}$}\
is more extended than that of TiO from our data alone, owing to the
differences in the excitation energies. 

\subsection{High excitation OH lines}
\label{subsect_res_oh}

We detect two OH lines ($\varv=0$, $N_j$ = 18$_{35/2}$,
$F$ = 18$^{+}$ -- 18$^{-}$ and $F$ = 17$^{+}$ -- 17$^{-}$) at 265.734659
and 265.765323~GHz, which correspond to the $\Lambda$-doubling
hyperfine components.
Because the S/N ratios of the individual channel maps are quite low,
we present in Fig.~G.1 
the intensity maps integrated from \mbox{$V_{\rm rel}$}\ = $-10$ to 10~\mbox{km s$^{-1}$}. 
Due to their high upper level energy of
8860~K,
these OH lines form in layers close to the star,
resulting in the compact emission. 
The figure shows that the emission is seen in the north and SW along the limb
of the stellar disk.
Interestingly, while the northern
emission is seen just outside the stellar limb, the SW emission is detected
on and inside the limb.

Khouri et al. (\cite{khouri19}) report the
detection of the $\Lambda$-doubling of high excitation OH lines from 107
to 352~GHz in three AGB stars, \mbox{W~Hya}, R~Dor, and IK~Tau, 
including the same 266~GHz pair toward \mbox{W~Hya}\ 
taken at a lower angular resolution of $\sim$110$\times$190~mas
on 2017 July 8 (phase 0.74). 
Their analysis of the OH lines with $E_u$ spanning from 4818 to 12756~K 
shows a rotational temperature of
$2472\pm 184$~K for the emission within a radius of 44~mas.
The temperature is comparable to the continuum brightness temperatures
of 2500--2650~K that they obtained, which can explain the OH emission
seen close to the star in our images.
Baudry et al. (\cite{baudry23}) detect one of the hyperfine components
($F$ = 18$^{+}$ -- 18$^{-}$) only in R~Hya and tentatively in Mira ($o$~Cet)
among their sample of 17 AGB stars and RSGs. 

Khouri et al. (\cite{khouri19}) 
also present high-angular resolution images of \mbox{W~Hya}\ in Band 4 
OH lines at $\sim$130~GHz with a restoring beam of
$\sim$30~mas and detect emission partially over the stellar disk.
Our images obtained at a higher angular resolution reveal that the emission
is primarily located along the limb of the stellar disk. 
Based on the observed flux ratio of the $\Lambda$-doubling lines, 
Khouri et al. (\cite{khouri19}) suggest possible maser action in the OH
lines at 265.7~GHz. 
While the weak OH emission over the stellar disk detected in our data
may also be masers, we need data with higher S/N to confirm this possibility.

\subsection{Unidentified lines}
\label{subsect_res_uid}

There are seven lines that could not be identified.
Figures~H.1, H.2, H.3, and H.4
show the channel maps obtained for U253913, U250652, U252586, and U266902,
respectively. 
The rest frequency of U253913 is close to the vibrationally excited
\mbox{SO$_{2}$}\ line $\varv_2=2$ $\varv_3=1$ $29_{3,26}$--$29_{2,27}$ at 253.915158~GHz
listed in HITRAN. However, its excitation energy $E_u$ = 3867~K seems to
be too high, and the frequency is also uncertain (H. M\"uller, 
  priv. comm.). Therefore, we do not identify the line as \mbox{SO$_{2}$}\ but 
mark is as unidentified.

In the cases of U250652 and U252586,
emission is clearly seen in the east just off
the limb of the stellar disk at \mbox{$V_{\rm rel}$}\ = $-1.5$ and 0~\mbox{km s$^{-1}$}.
The positions of the emission as well as its velocities are similar to that
of the AlOH line discussed in Sect.~\ref{subsect_res_alo_aloh}, although
no AlOH line is present in the list of Splatalogue, CDMS, JPL, and HITRAN.
In the case of U266902, we note that a weak, unidentified line at the same
rest frequency is also detected in VX~Sgr among a sample of 17 O-rich AGB
stars and RSGs (Wallstr\"om et al. \cite{wallstroem24}). U266902 is close
to a direct $\ell$-type transition of H$^{13}$CN line at
266.904987~GHz ($\varv_2=1$, $J$ = 35, $l$ = $1^{e}$--$1^{f}$).
However,
its Einstein coefficient is very small ($\log A_{ul} = -5.818$), and
its upper level energy is relatively high ($E_u$ = 3632~K). 
As shown in Fig.~D.1, 
the vibrationally excited HCN line appears very weak in spite of
its larger Einstein coefficient ($\log A_{ul} = -3.142$), lower
upper level energy ($E_u$ = 1050~K), and its status as the main isotopologue
(i.e., H$^{12}$CN).
Furthermore, direct $l$-type HCN transitions in carbon stars have only been
found in the main isotopologue (Thorwirth et al. \cite{thorwirth03};
Cernicharo et al. \cite{cernicharo11}). 
Therefore, it is unlikely that U266902 is due to H$^{13}$CN. 

For U250797, U253724, and U268099, we only show 
the integrated intensity maps in Fig.~H.5, 
because the individual channel maps are of low S/N ratios. 
We find lines due to AlS (250.802235~GHz) and $^{17}$OH (250.796230~GHz)
near U250797. However, we detect no signatures
of another AlS line at 250.841643~GHz with approximately the same upper level
energy and Einstein coefficient. Neither did we see a trace of two AlS lines
at 250.930853 and 250.936712~GHz with higher Einstein coefficients. Therefore,
we deem the identification as AlS to be unlikely. 
The $^{17}$OH line has a very small Einstein coefficient of
$\log A_{ul} = -7.981$, which also makes it unlikely to be the identification
of U250797.

For U268099 and U253724, we find SO$^{18}$O lines at nearby frequencies: 
$\varv=0$, $24_{4,21}$--$25_{1,24}$ at 268.0964242~GHz listed in the CDMS
catalog for U268099 and
$\varv_2 = 1$, $27_{8,19}$--$28_{7,22}$ at 253.723621~GHz listed in HITRAN 
for U253724. 
However, while there are a number of ground-state SO$^{18}$O lines in the
observed spectral windows, none of them is detected. Therefore, the
identification of U268099 and U253724 as SO$^{18}$O is unlikely. 
Another candidate for U268099 is the vibrationally excited \mbox{SO$_{2}$}\ line
($\varv_2=2$, $32_{9,23}$--$33_{8.26}$, $E_u$ = 729~K) at
268.106284~GHz listed in HITRAN. 
The rest frequency is blueshifted by 8.1~\mbox{km s$^{-1}$}\ with respect to the observed
line. 
Given that the uncertainty in the frequency of vibrationally excited
transitions is often
large, and the velocity shift is caused by the atmospheric dynamics, 
we cannot exclude this line for the identification of U268099.

\section{Discussion}
\label{sect_discuss}

In Paper~I, we find striking agreement between the spatial distributions
of the clumpy dust clouds seen in the 820~nm polarized intensity map and
the nonthermal, likely maser emission of the vibrationally excited \mbox{H$_2$O}\
line ($\varv_2=2$ $6_{5,2}$--$7_{4,3}$) at 268~GHz. 
In the complete data reported in the present paper, we reveal 
similar good agreement between the dust clouds distribution and the lines of 
AlO (Fig.~\ref{zimpol_alo_267}) and TiO (Fig.~\ref{integmap_tio_zimpol}). 
The emission of the vibrationally excited HCN line ($\varv_2=1^{1e}$) also
overlaps with the signals of the dust clouds but extends farther out. 
Figure~I.1 
shows that the emission of more molecular lines
coincides with the dust clouds: \mbox{$^{30}$SiO}\ ($\varv=1$, line \#4),
\mbox{Si$^{17}$O}\ ($\varv=0$, line \#5), 
\mbox{TiO$_{2}$}\ (stacked image of lines \#51--\#55, Sect.~\ref{subsect_res_tio_tio2}), 
SO ($\varv=0$, $N_J$ = $6_5$--$5_4$, line \#41), and two \mbox{SO$_{2}$}\ lines
(lines \#9 and \#20). The emission of more \mbox{SO$_{2}$}\ lines shows noticeable
overlap with the dust clouds, and two lines are presented as examples. 
The emission of these lines traces the paisley-like dust cloud in the SW
quadrant, another cloud with moderate signals in the east, and the
weaker dust signals in the NNW plume. 

The observed overlap between the dust clouds and molecular line
emission indicates not only AlO and TiO but also some other molecules
are involved in the chemical network including dust formation. 
We propose in Paper~I that the 268~GHz \mbox{H$_2$O}\ emission and dust both trace
cool, dense pockets. Moreover, \mbox{H$_2$O}\ is an important oxidizing
agent for the formation of a number of molecules and molecular clusters 
that are precursors of dust, and SiO also plays an important role in these
chemical reactions (Goumans \& Bromley \cite{goumans12};
Gobrecht et al. \cite{gobrecht16}; Andersson et al. \cite{andersson23}).
AlO is also considered to be important as a precursor 
of seed particles (Gobrecht et al. \cite{gobrecht22}). 
Therefore, the agreement between the emission of \mbox{H$_2$O}, SiO, and AlO 
and the dust clouds can be interpreted as that these molecules directly
take part in the chemical network leading to dust formation.

An alternative interpretation of 
the overlap of these molecular species with dust
emission is that their formation may be favored in regions where the dust is
located. Examples of these regions include dense pockets created by convection
and/or pulsation.
For example, it is possible that some Si is processed back
to the gas phase from the dust grains and forms the SiO molecule.
As discussed in Sect.~\ref{subsect_res_sio_model}, 
the present data do not provide tight constraints on the Si abundance
in the gas phase, and multiline observations including $^{28}$SiO are 
needed to address this point. 

Gobrecht et al. (\cite{gobrecht16}) also show that there are
molecules that form in post-shock gas-phase chemistry but are not
directly associated with grain nucleation. SO and \mbox{SO$_{2}$}, which forms
from SO, are among such molecules. In general, sulfur is not significantly
depleted onto dust grains (Danilovich et al. \cite{danilovich20} and
references therein). On the other hand, grain nucleation also starts behind 
shock fronts, where the density is enhanced 
(e.g., Liljegren et al. \cite{liljegren17}). Therefore, the observed
agreement between the SO and \mbox{SO$_{2}$}\ emission and dust clouds lends
support to the idea that the formation of these molecular species and dust grains is
both associated with shocks in the same localized regions. 
As discussed in Sect.~\ref{subsect_res_hcn_265}, HCN also likely forms in
shock-induced chemistry, and the spatial distribution of the HCN
$\varv_2=1^{1e}$ emission indeed covers the dust cloud formation regions.
However, the HCN emission is much more extended than the dust clouds,
suggesting that this molecular species is not actively involved in or depleted 
by grain formation. The differences in the distribution of \mbox{SO$_{2}$} /SO
and HCN may be due to the differences in their formation reactions and
excitation, which depend on the ambient conditions.

While \mbox{TiO$_{2}$}\ is also considered to be a precursor of dust
(Gail \& Sedlmayr \cite{gail98}; Jeong et al. \cite{jeong03}), 
the recent observations do not confirm its importance 
in dust formation (Kami\'nski et al. \cite{kaminski17};
Decin et al. \cite{decin18}). 
The shock-chemistry models of Gobrecht et al. (\cite{gobrecht16}) also show
that only a modest amount of \mbox{TiO$_{2}$}\ can form from TiO in shocks 
and conclude that \mbox{TiO$_{2}$}\ does not play an important role
as seed particles.
Therefore, the agreement of the TiO and \mbox{TiO$_{2}$}\ emission with the dust clouds
does not necessarily indicate that these molecules actively participate
in grain nucleation but that they form in shock-induced chemistry 
in the same localized regions as the dust clouds, as in the cases of SO and 
\mbox{SO$_{2}$}. 

The infall within $\sim$2.5~\mbox{$R_{\rm cont}$}\ ($\sim$3.6~\mbox{$R_{\star}$}) derived from
the SiO lines (Sect.~\ref{subsect_res_sio_model})
and inferred from the 268~GHz \mbox{H$_2$O}\ line
(Paper~I) means that the dust grains are forming in the infalling
material if we assume that the gas and dust grains are kinematically coupled.
As we point out in Paper~I, this picture agrees with the 1D
dynamical models, which show that shocks develop where the infalling
material collides with the upwelling gas, and dust formation takes place 
behind shocks, where the density is enhanced
(see Fig.~5 of Liljegren et al. \cite{liljegren17}). 
As discussed in Sect.~\ref{subsect_res_sio_model}, the velocity profile
derived from our modeling is consistent with this picture. 
The recent 3D models (Freytag \& H\"ofner \cite{freytag23}) also show that
there are dust clouds infalling up to $\sim$25~\mbox{km s$^{-1}$}, because the grain
size is not large enough for the radiation pressure to accelerate the clouds
outward. The observed infall velocities are within the prediction of
the 3D models.

\section{Conclusions}
\label{sect_concl}

We identify 57 spectral lines of SiO, \mbox{H$_2$O}, \mbox{SO$_{2}$}, AlO, AlOH, TiO,
\mbox{TiO$_{2}$}, OH, SO, and HCN as well as their isotopologues in \mbox{W~Hya}\ 
between 250 and 268~GHz. The angular resolution of $\sim$20~mas of our
ALMA data and the large angular size of \mbox{W~Hya}\ of $\sim$60~mas in the
millimeter continuum have enabled us to spatially
resolve the stellar disk and atmosphere within several stellar radii.
We detect a plume in the NNW and a tail-like structure in the SSW,
both extending to $\sim$100~mas ($\sim$4.8~\mbox{$R_{\star}$}). The atmosphere is
elongated in the ENE-WSW direction with semimajor and semiminor axes of
$\sim$70 and $\sim$40~mas, respectively. 

About 2/3 of the identified lines appear in absorption over the stellar disk
as expected for thermal excitation. The absorption is inhomogeneous over the
stellar disk.
However, about 1/3 of the identified lines -- some lines of SiO, \mbox{H$_2$O},
\mbox{SO$_{2}$}, and AlO --
show emission on top of the continuum
over the stellar disk, which cannot be explained by material warmer
than the millimeter continuum-forming layer.
This indicates nonthermal emission -- either suprathermal
emission or maser action. Particularly, we detect clear maser emission from
the \mbox{$^{29}$SiO}\ line ($\varv=2$, $J$ = 6 -- 5),
in addition to the likely maser emission of the vibrationally excited \mbox{H$_2$O}\
line at 268~GHz ($\varv_2=2$, $J_{K_a,K_c}$ = 6$_{5,2}$ -- 7$_{4,3}$) over the
stellar disk reported in Paper~I. 
The emission over the stellar disk complicates the interpretation of the
data in terms of kinematics.
Nevertheless, our modeling of two SiO lines without signatures of
nonthermal emission suggests an accelerating infall toward the
continuum-forming layer, from $\sim$0~\mbox{km s$^{-1}$}\ at $\sim$75~mas
($\sim$3.6~\mbox{$R_{\star}$}) to $\sim$11~\mbox{km s$^{-1}$}\ at $\sim$50~mas ($\sim$2.4~\mbox{$R_{\star}$}), and
outflow at up to 10~\mbox{km s$^{-1}$}\ in deeper layers.
The dynamics in this region are dominated by
the pulsation and/or convection; they remain gravitationally bound, and the
material is not yet outflowing in a stationary wind.
  A detailed non-LTE analysis is needed to derive the abundances of the 
  different molecular species, which will allow us to better understand
  the chemical processes involving molecules and dust as well as their
  interplay with stellar pulsation and convection. 

The emission of the SiO and AlO lines as well as the 268~GHz
\mbox{H$_2$O}\ line from Paper~I 
shows noticeable agreement with the spatial distribution of the clumpy dust
clouds seen in the visible polarimetric images contemporaneously taken with
our ALMA data.
This is consistent with the chemical models that directly associate
SiO, \mbox{H$_2$O}, and AlO with the grain nucleation process. 
On the other hand, the overlap of the SO and \mbox{SO$_{2}$}\ emission with the
dust clouds implies that these molecular species and dust form behind
shock fronts,
even though SO and \mbox{SO$_{2}$}\ are not expected to be associated with dust
formation.  
This can also be the case for TiO and \mbox{TiO$_{2}$}. 

We detect the vibrationally excited HCN line emission very close to the
star, down to the millimeter continuum radius of $\sim$30~mas
($\sim$1.4~\mbox{$R_{\star}$}). The HCN emission overlaps with the dust clouds but it is
more extended, which implies that the inner HCN is likely produced by
shock-induced chemistry but not actively involved in dust formation.

The emission of the HCN ($\varv$ = 0, $J$ = 3 -- 2) and SO ($N_J$ = $6_5$ -- 
$5_4$) lines is much more extended up to $\sim$1\arcsec. 
We detect the filamentary, spider-like structures, which likely correspond
to localized regions with high intensities.
However, diffuse, smooth emission on scales larger than $\sim$190~mas 
is missing due to the lack of short baselines, which prevented us
from drawing a conclusion on the large-scale kinematics.
This issue can only be solved with observations combining ALMA’s extended and
compact configurations, preferably simultaneously, given the variabilities
in line intensities and spatial distributions.
As discussed above, it is
also important to carry out high-resolution observations at other variability
phases to understand the phase dependence of the molecular emission in the
inner winds, which will be tested against chemical and hydrodynamical
models. A frequent access to both long and short baselines (on a timescale of
months) is essential; this is currently unavailable but may be considered
for a future upgrade of ALMA.
\vspace*{3ex}

\noindent
{\large \sffamily \bfseries Data availability}
\vspace*{1.3ex}

\noindent
The figures in Appendices are available on Zenodo \url{https://doi.org/10.5281/zenodo.17118092}
\vspace*{3ex}

\begin{acknowledgement}
This paper makes use of the following ALMA data:
ADS/JAO.ALMA\#2018.1.01239.S. ALMA is a partnership of ESO (representing its
member states), NSF (USA) and NINS (Japan), together with NRC (Canada), NSTC
and ASIAA (Taiwan), and KASI (Republic of Korea), in cooperation with the
Republic of Chile. The Joint ALMA Observatory is operated by ESO, AUI/NRAO and
NAOJ.
Data reduction was carried out at the IRAM ARC node.
We are grateful to Holger M\"uller for the discussion about the line
identification.
We also thank the anonymous referee, whose careful and constructive comments
helped us improve the manuscript. 
K.O. acknowledges the support of the Agencia Nacional de 
Investigaci\'on Cient\'ifica y Desarrollo (ANID) through
the FONDECYT Regular grant 1240301. 
K.T.W. acknowledges support from the European Research Council (ERC) under
the European Union's Horizon 2020 research and innovation programme
(Grant agreement no. 883867, project EXWINGS).
This research made use of the \mbox{SIMBAD} database, 
operated at the CDS, Strasbourg, France. 

\end{acknowledgement}

\end{document}